\documentclass[journal]{vgtc}
  \usepackage[nocompress]{cite}
  \usepackage{cite}
  \usepackage[pdftex]{graphicx}
  \DeclareGraphicsExtensions{.pdf,.jpeg,.png}
\usepackage{amsmath}
\usepackage{amsfonts}
\usepackage[ruled,vlined,linesnumbered,noend]{algorithm2e}
\usepackage[normalem]{ulem}

\SetAlgoCaptionSeparator{}

\usepackage{xspace}
\usepackage[colorlinks=true,urlcolor=blue]{hyperref}
\usepackage{comment}

\usepackage{amsmath}
\usepackage{booktabs}
\usepackage{makecell}
\usepackage{tabularx}

\usepackage{todonotes}
\usepackage{xcolor,colortbl}

\usepackage{float}
\restylefloat{table}

\usepackage{cite}  

\usepackage[style=base]{caption}
\usepackage{subcaption}
\usepackage[ruled,vlined]{algorithm2e}
\SetKwInput{KwInput}{Input}
\SetKwInput{KwOutput}{Output}
\SetKwProg{Fn}{Function}{:}{}

\usepackage[T1]{fontenc}
\usepackage{tgbonum}
\usepackage{cancel}
\usepackage{soul}

\usepackage{multirow}

\newcommand{\blue}[1]{\textcolor{black}{#1}}

\newcommand{\yedCircular}{CIR\xspace}
\newcommand{\sfdpWithPrism}{sfdp+p\xspace}
\newcommand{\reyansAlgo}{Desirable\-Edge\-Length\-guided (DELG) Algorithm\xspace}
\newcommand{\reyansAlgoAcronym}{RT\_L\xspace}
\newcommand{\mingweisAlgo}{Compactness\-guided (CG) Algorithm\xspace}
\newcommand{\mingweisAlgoAcronym}{RT\_C\xspace}
\newcommand{\khaledsAlgoAcronymDelg}{PRT\_L\xspace}
\newcommand{\khaledsAlgoAcronymCG}{PRT\_C\xspace}
\newcommand{\khaledsAlgoAcronym}{PRT\xspace}

\newcommand{\desiredEdgeLengthAcronym}{DEL\xspace}
\newcommand{\compactnessAcronym}{CM\xspace}
\newcommand{\reyansInit}{{\fontfamily{lmtt}\selectfont Edge-Length-Initialization}\xspace}
\newcommand{\mingweisInit}{{\fontfamily{lmtt}\selectfont Compact-Initialization}\xspace}
\newcommand{\khaledsAlgoAcronymDelgStar}{PRT\_L$^*$\xspace}
\newcommand{\khaledsAlgoAcronymCgStar}{PRT\_C$^*$\xspace}
\begin{document}
%
\title{A Scalable Method for Readable Tree Layouts}
%
%
%
%

\author{Kathryn Gray, Mingwei Li, Reyan Ahmed, Md. Khaledur Rahman, Ariful Azad, Stephen Kobourov, Katy B\"{o}rner}

%
%

\markboth{IEEE Transactions on Visualization and Computer Graphics}%
{Gray \MakeLowercase{\textit{et al.}}: A Scalable Method for Readable Tree Layouts}
%



\let\oldtwocolumn\twocolumn
\renewcommand\twocolumn[1][]{%
    \oldtwocolumn[{#1}{
    \begin{center}
           \includegraphics[width=\textwidth]{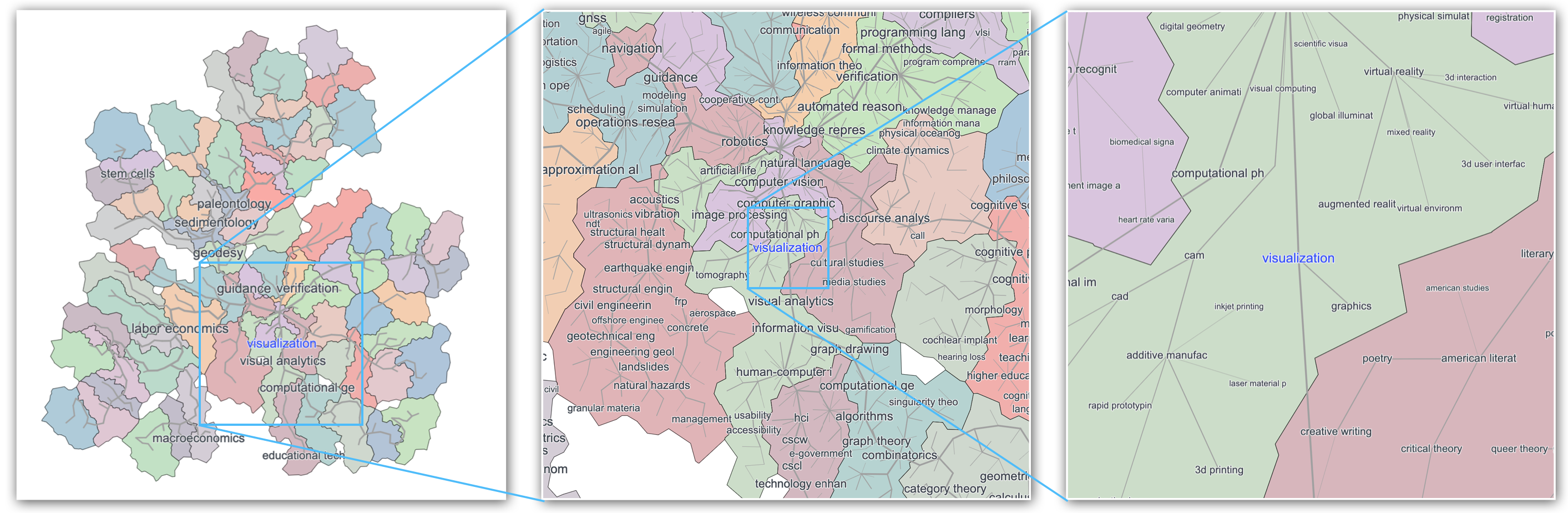}
           \captionof{figure}{
  {A map of a  real-world research topics network with over 5,000 nodes that provides semantic zooming, generated with the scalable method for readable tree layout. The visualization provides an overview of the dataset, showing the high-level structure, including important nodes and edges. Zooming into a particular area of interest provides more details. 
  The layout obtained by the proposed algorithm is compact, there are no edge crossings, and there are no label overlaps.}
  }
           \label{fig:teaser}
        \end{center}
    }]
}

\abstract{
Large tree structures are ubiquitous and real-world relational datasets often have information associated with nodes (e.g., labels or other attributes) and edges (e.g., weights or distances) that need to be communicated to the viewers. Yet, scalable, easy to read tree layouts are difficult to achieve. 
We consider tree layouts to be readable if they meet some basic requirements: node labels should not overlap, edges should not cross, 
edge lengths should be preserved, and the output should be compact. 
There are many algorithms for drawing trees, although very few take node labels or edge lengths into account, and none optimizes all requirements above. With this in mind, we propose a new scalable method for readable tree layouts. The algorithm guarantees that the layout has no edge crossings and no label overlaps, and optimizes one of the remaining aspects: desired edge lengths and compactness. 
We evaluate the performance of the new algorithm by comparison with related earlier approaches using several real-world datasets, ranging from a few thousand nodes to hundreds of thousands of nodes.
Tree layout algorithms can  be used to visualize large general graphs, by extracting a hierarchy of progressively larger trees. We illustrate this functionality by presenting several map-like visualizations generated by the new tree layout algorithm.
}


\maketitle


\vspace{4cm}
\section{Introduction}\label{sec:introduction}

Many real-world datasets can be represented by a network where each node represents an object and each link represents a relationship between objects; e.g., the tree of life captures the evolutionary connections between species and a research topics network captures relationships between 
research areas; see Fig.~\ref{fig:teaser}. Abstract networks can be modeled by node-link diagrams, with points representing nodes and segments/curves representing the edges.  However, real-world datasets have labels associated with nodes and attributes such as edge lengths that are not captured in node-link diagrams. For example, the tree of life has species names as node labels and evolutionary distance between the corresponding species as edge data. These networks would benefit from a visualization that shows the labels and captures the desired edge lengths. However, just adding labels to an existing layout will result in an unreadable visualization with many overlapping labels. One could scale the layout to remove such overlaps, but this would blow up the drawing area to an unmanageable size. Label overlaps could also be removed with specialized overlap removal algorithms but result in layout changes: not just changing edge lengths, but also the topology (e.g., breaking up clusters, or introducing new edge crossings).

 \begin{figure*}[thp]
	\centering
	\subfloat[\yedCircular]{\label{fig:t8_direct}\includegraphics[width=.31\columnwidth]{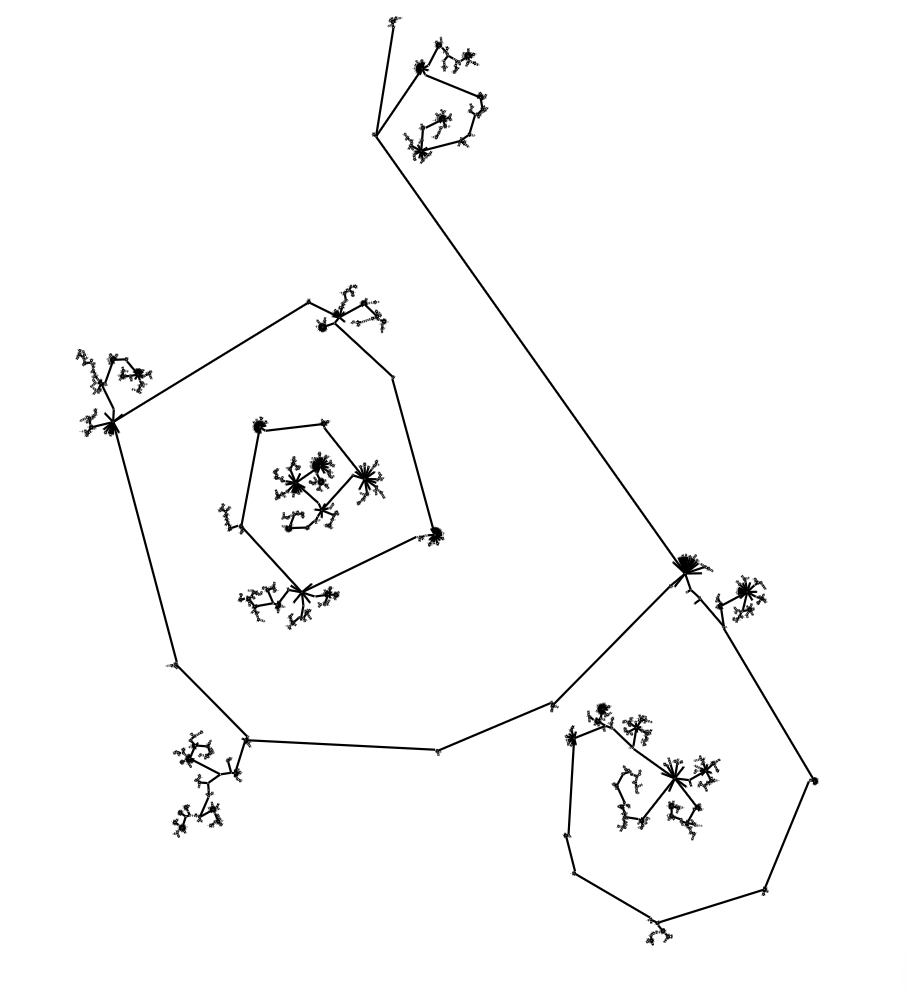}}
	\hfill
 	\subfloat[\sfdpWithPrism]{\label{fig:t8_prism}\includegraphics[width=.42\columnwidth]{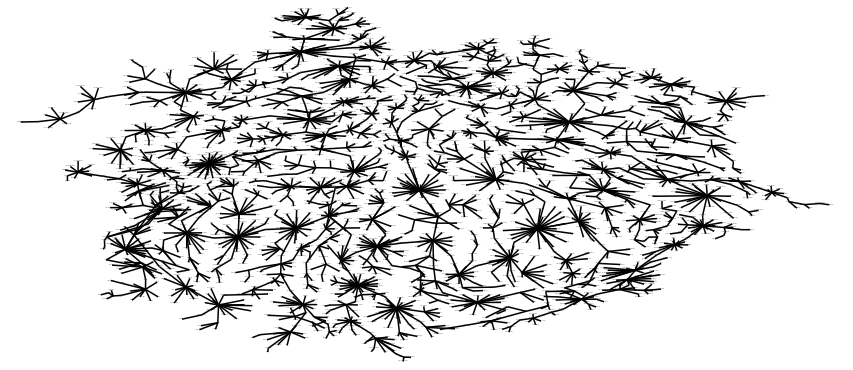}}
	\hfill
	\subfloat[\reyansAlgoAcronym]{\label{fig:t8_angular_lastfm}\includegraphics[width=.24\columnwidth]{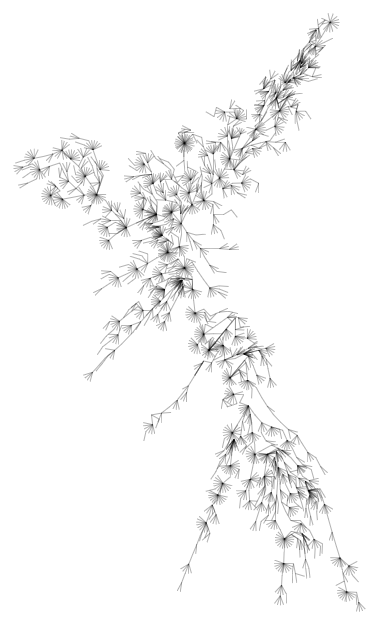}}
	\hfill
	\subfloat[\mingweisAlgoAcronym]{\label{fig:t8mw_lastfm}\includegraphics[width=.33\columnwidth]{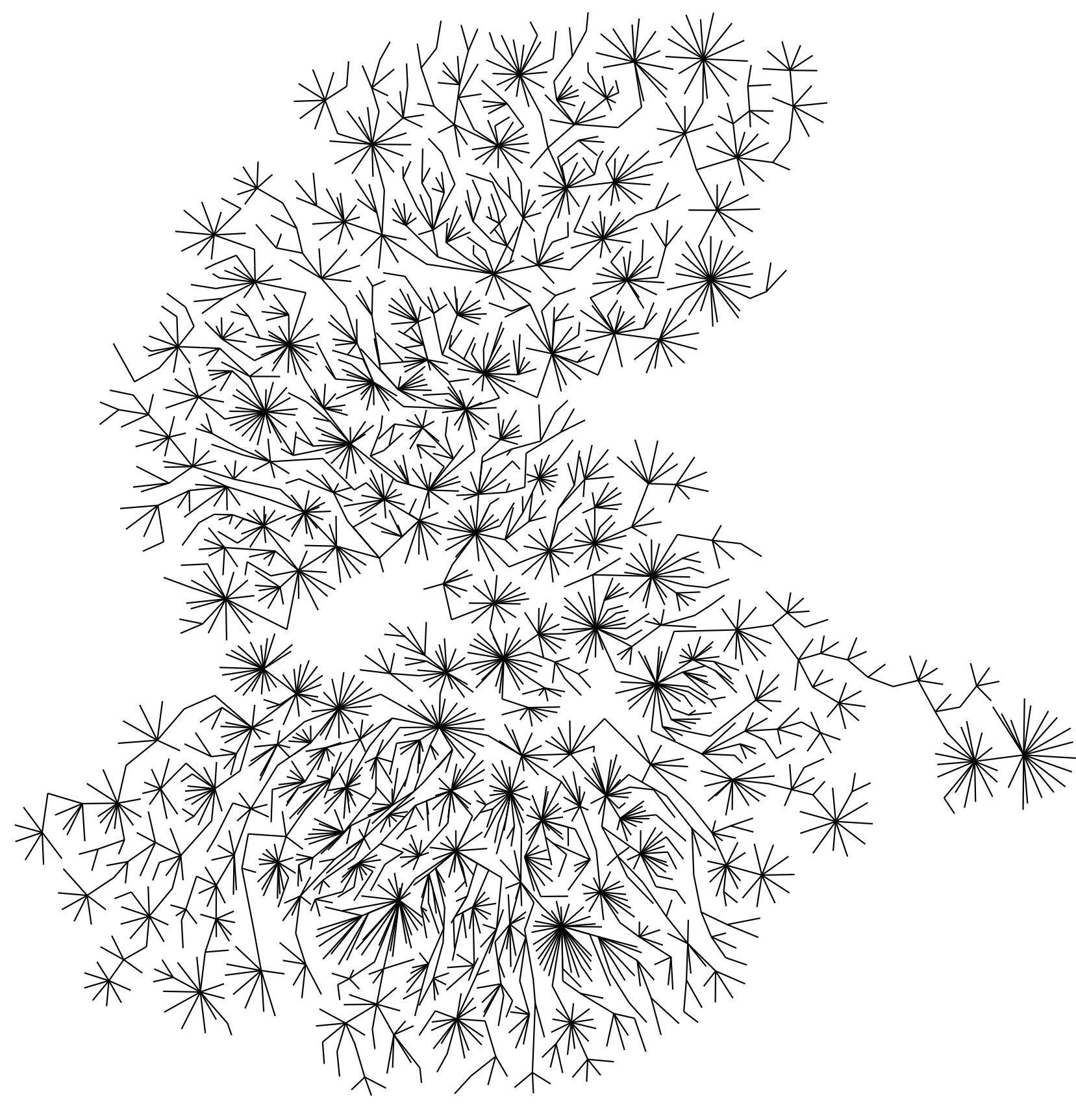}}
	\hfill
 	\subfloat[\khaledsAlgoAcronymDelg]{\label{fig:t8kh_lastfm_delg}\includegraphics[width=.33\columnwidth]{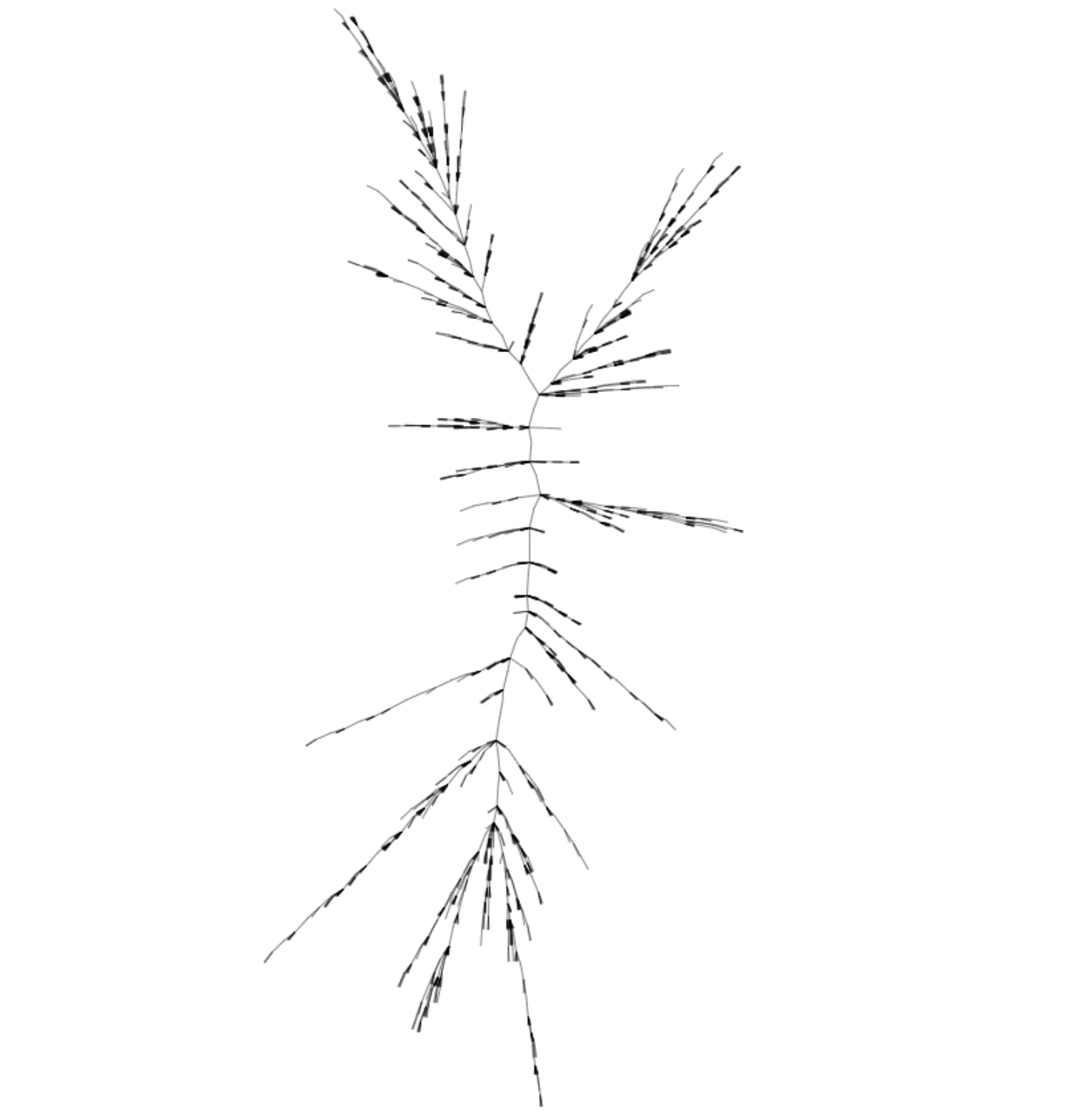}}
 	\hfill
 	\subfloat[\khaledsAlgoAcronymCG]{\label{fig:t8kh_lastfm_cg}\includegraphics[width=.30\columnwidth]{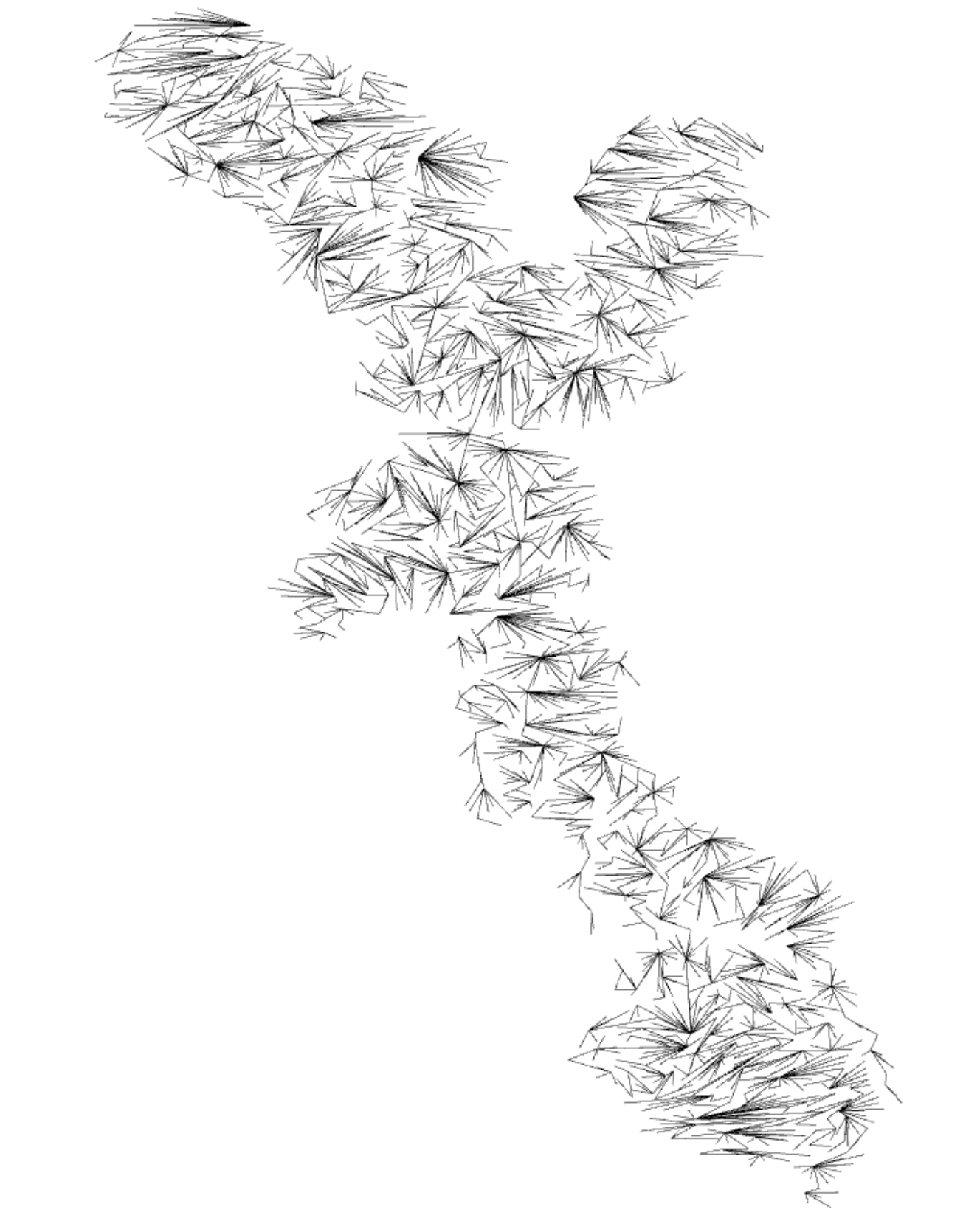}}

 	\subfloat[\yedCircular]{\label{fig:t8mw_lastfm_zoom_in_cir}\includegraphics[width=.31\columnwidth]{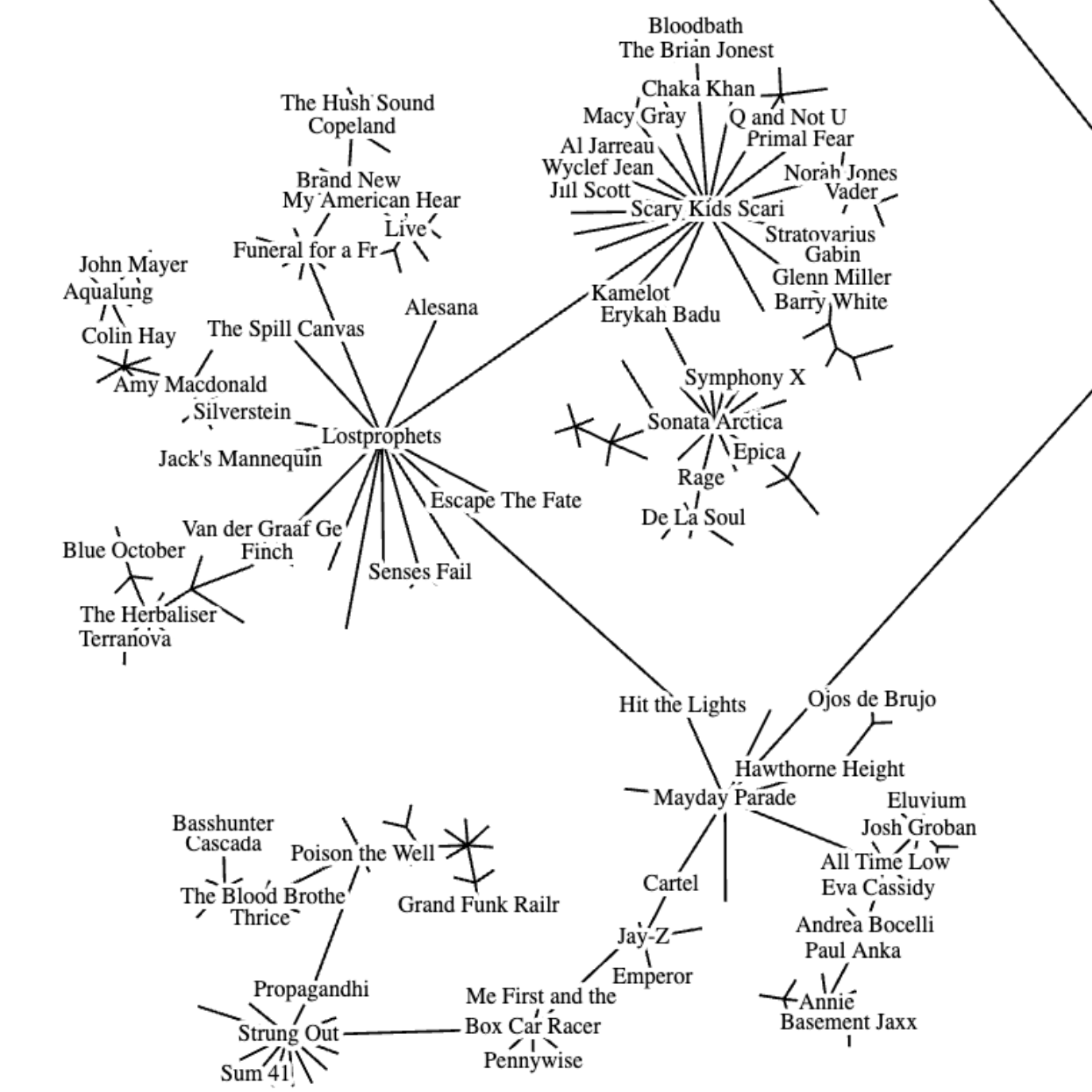}}
 	\hfill
 	\subfloat[\sfdpWithPrism]{\label{fig:t8mw_lastfm_zoom_in_sfdp}\includegraphics[width=.31\columnwidth]{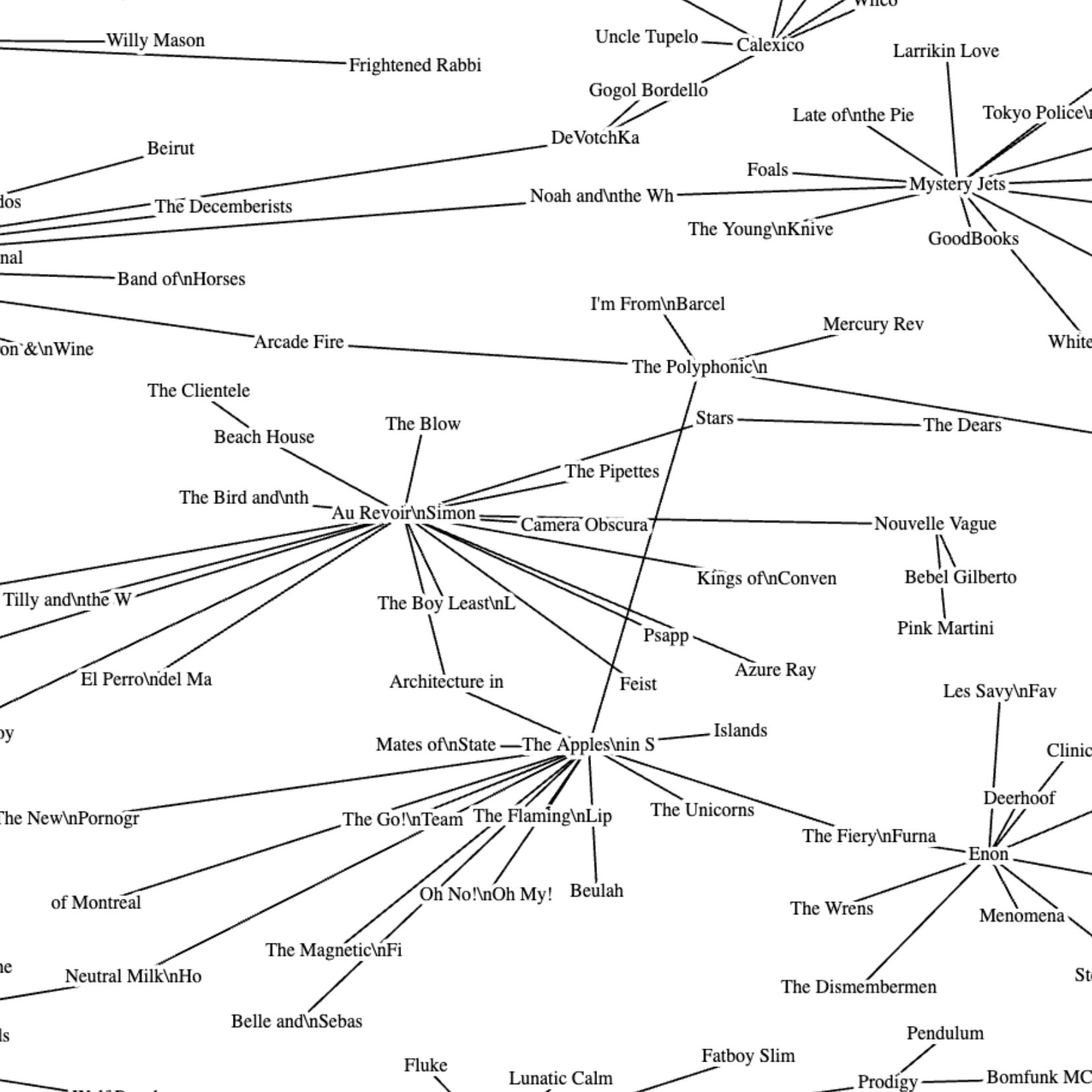}}
 	\hfill
 	\subfloat[\reyansAlgoAcronym]{\label{fig:t8mw_lastfm_zoom_in_delg}\includegraphics[width=.31\columnwidth]{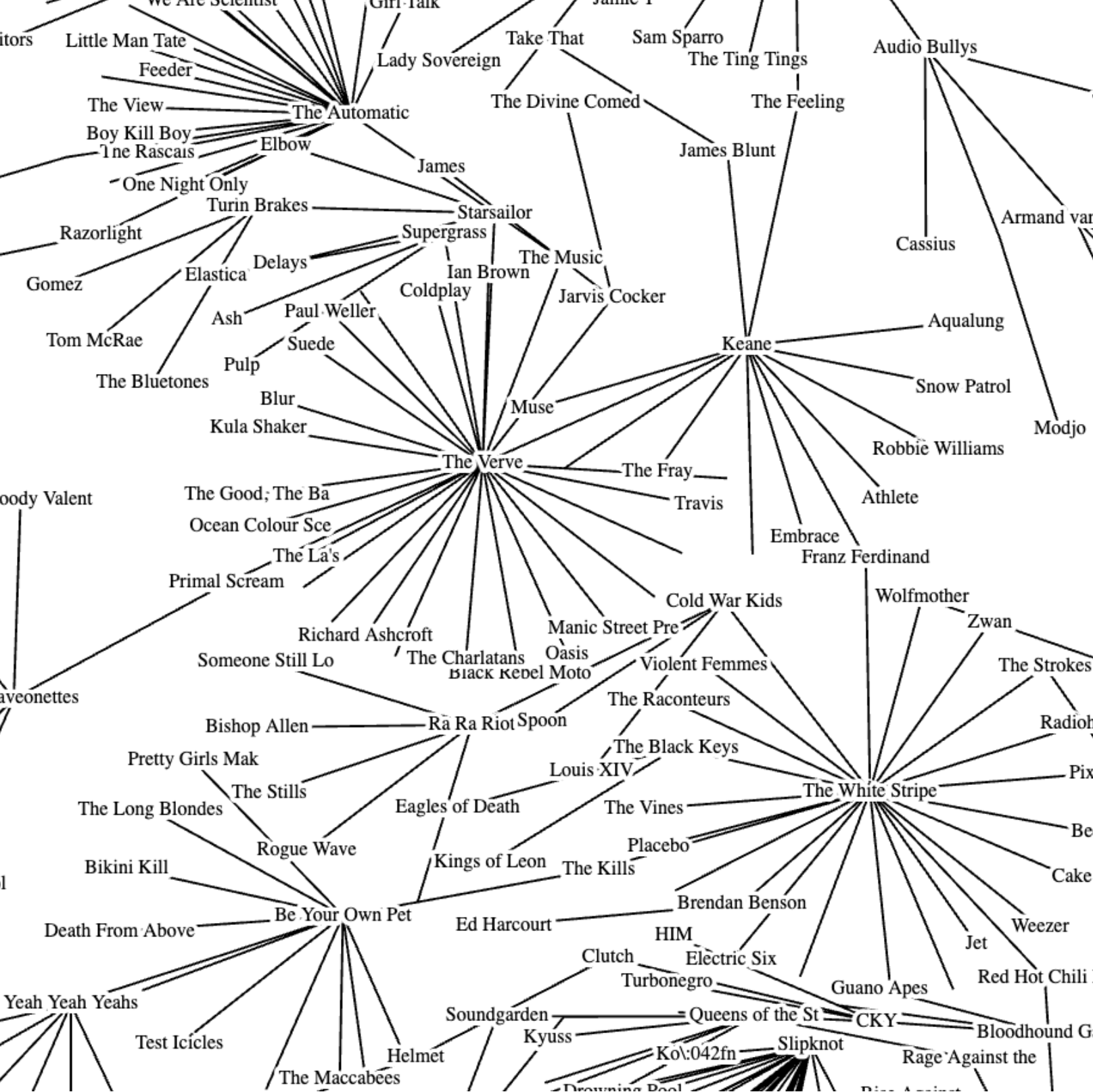}}
 	\hfill
 	\subfloat[\mingweisAlgoAcronym]{\label{fig:t8mw_lastfm_zoom_in_cg}\includegraphics[width=.31\columnwidth]{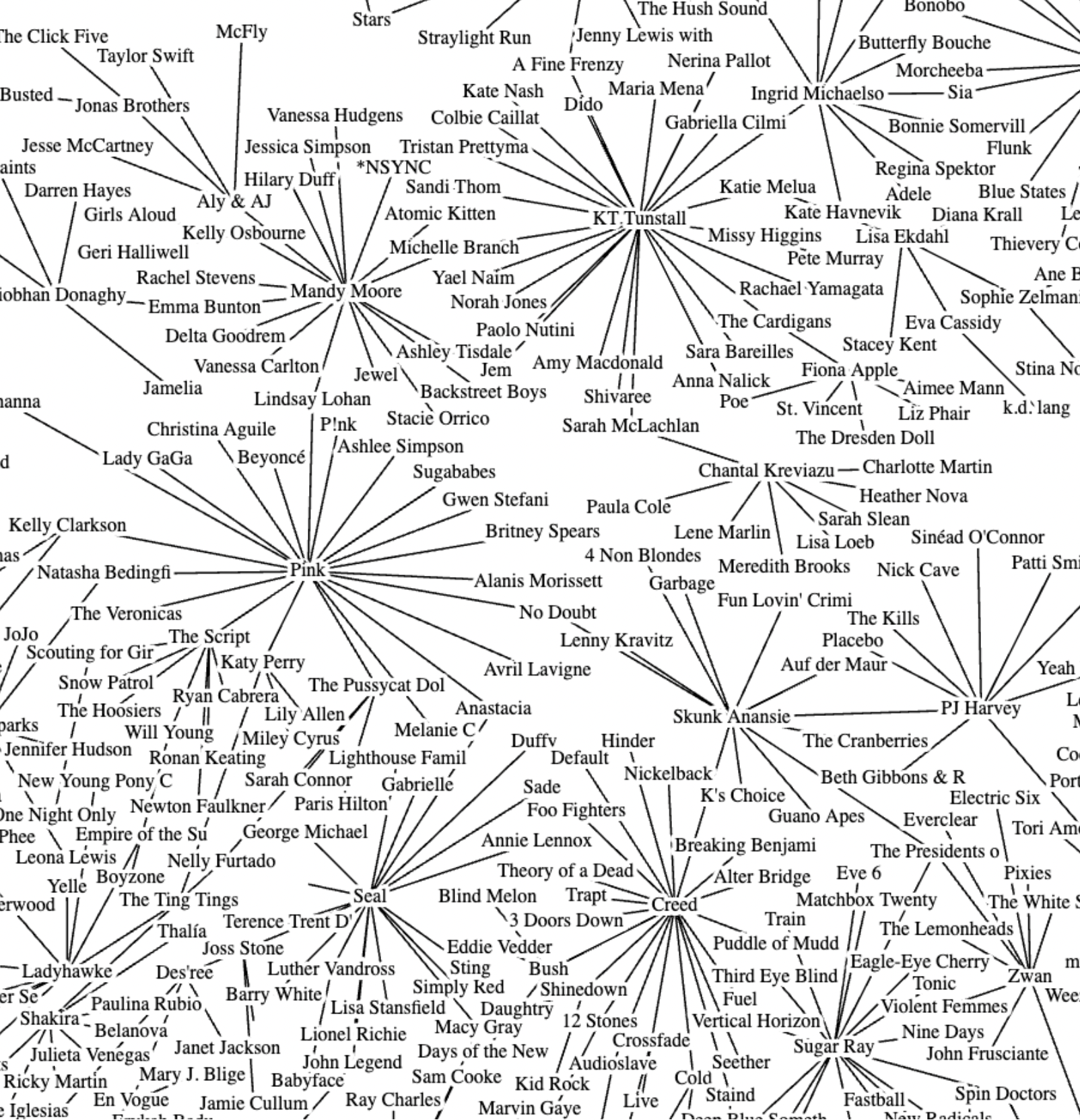}}
 	\hfill
 	\subfloat[\khaledsAlgoAcronymDelg]{\label{fig:t8mw_lastfm_zoom_in_pdelg}\includegraphics[width=.31\columnwidth]{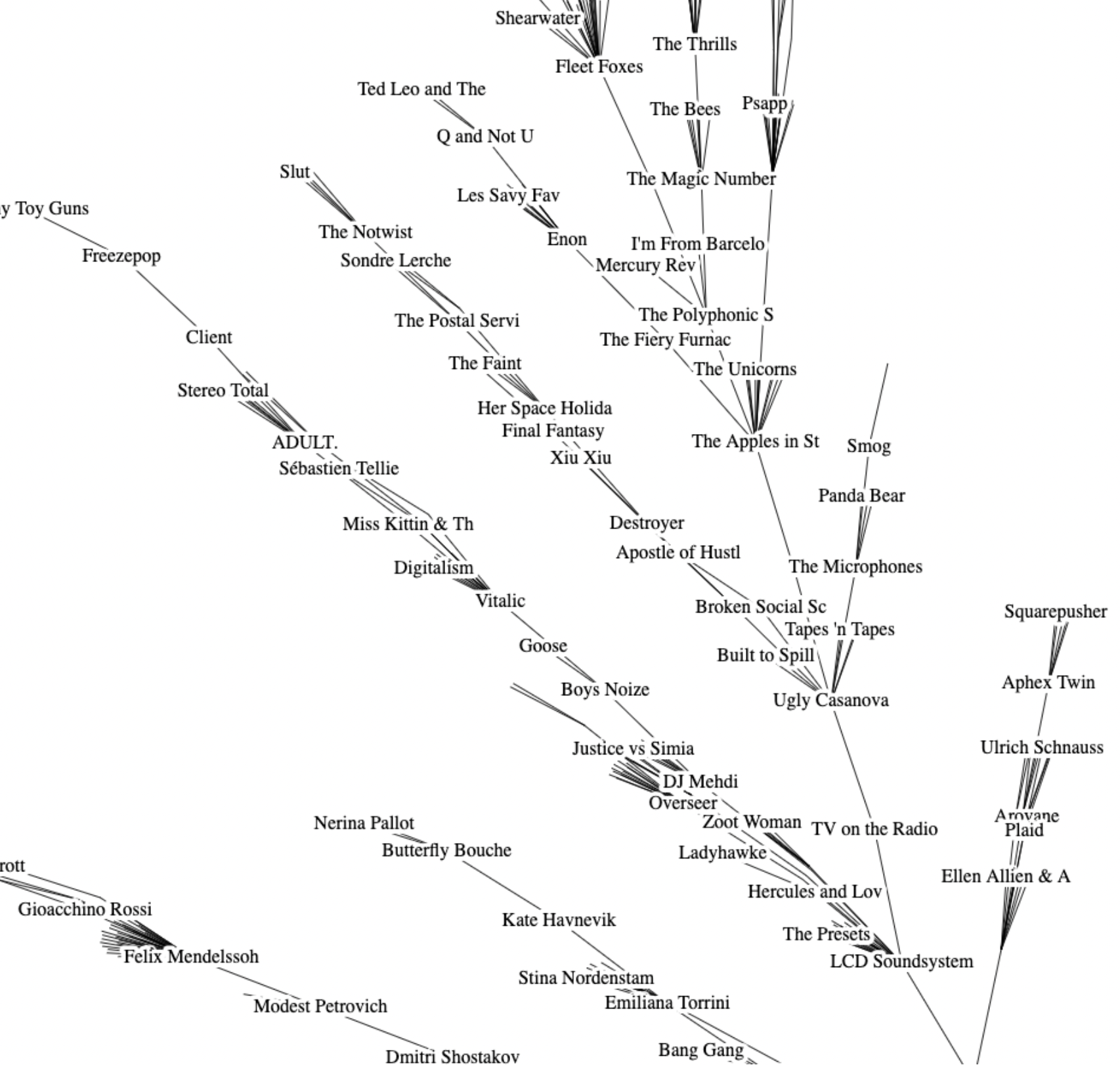}}
 	 \hfill
 	\subfloat[\khaledsAlgoAcronymCG]{\label{fig:t8mw_lastfm_zoom_in_pcg}\includegraphics[width=.31\columnwidth]{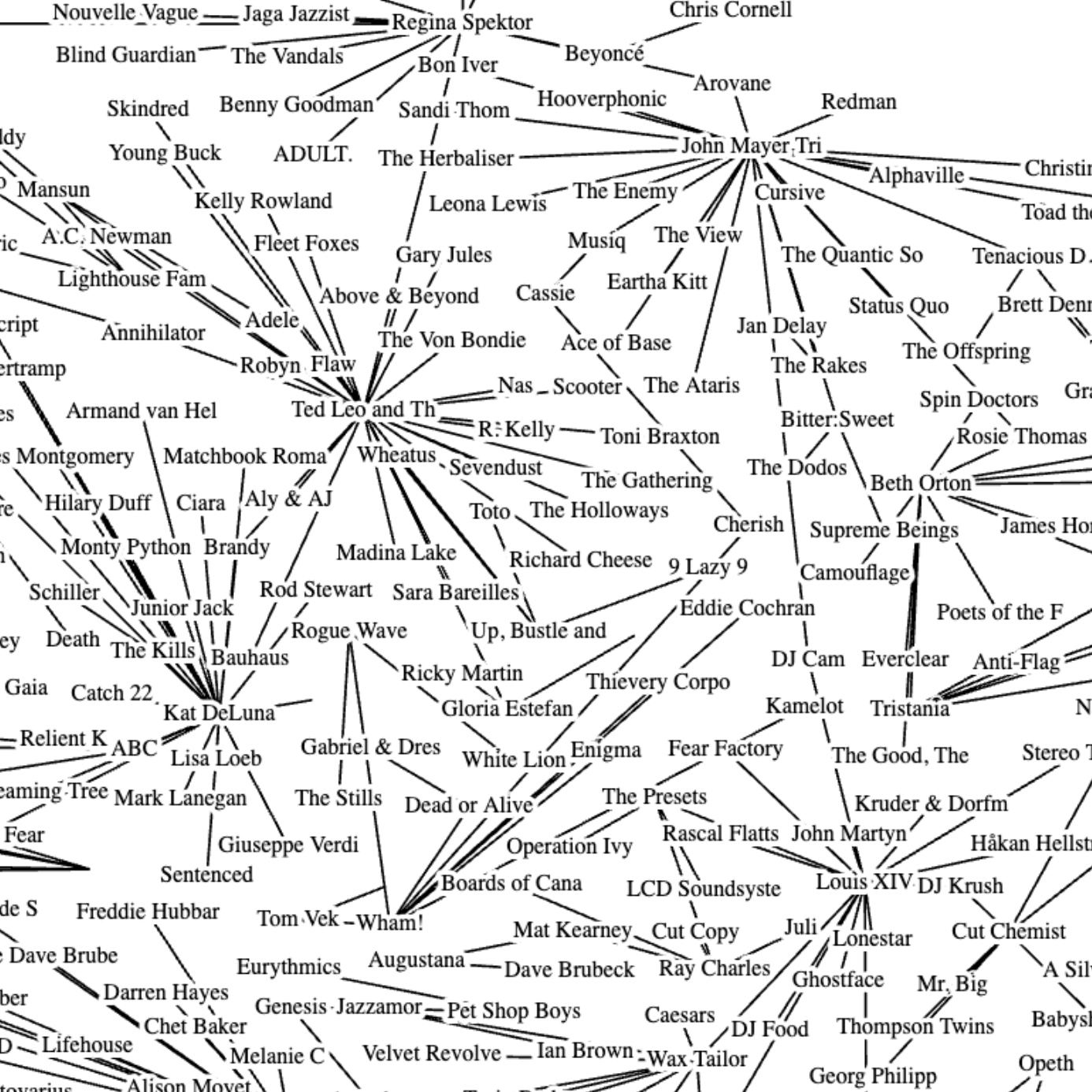}}
 	
	\caption{Comparison of the layout of the Last.fm graph with uniform edge lengths.
 The \yedCircular layout produces edges with very different lengths, while the \sfdpWithPrism has crossings, and neither of them captures the structure of the underlying tree. On the other hand, our algorithms generate layouts that show the structure of the input network. In an overview such as this one, several of the results look similar, but in these zoomed-out layouts we cannot see all the crossings, label overlaps, and area used; we provide a detailed quantitative evaluation in  Table~\ref{tab:eval-all}, which shows differences as large as orders of magnitude.}

	\label{fig:compare-lastfm}
\end{figure*}


We consider tree layouts to be {\em readable} if they meet some basic requirements: node labels
should not overlap, edges should not cross, edge lengths should be preserved, and the output should be compact. 
This gives us two hard {\em constraints}: (C1) No edge crossings, (C2) No label overlaps. We also consider two additional desirable properties that the algorithm {\em optimizes}: (O1) desired edge lengths  and (O2) compactness of the drawing area. Finally, to efficiently handle large networks, we also parallelize the computation. 

Preserving pre-specified, desired edge lengths is important to many real-world datasets, but is not taken into account by most network and tree layout algorithms. Keeping track of the drawing area required (e.g., by comparing to the sum of  areas of all labels), makes it clear that simply scaling up a given layout until labels do not overlap results in unusable layouts with areas that are 4-6 orders of magnitude greater than needed. Finally, the scalability of the algorithm is important when dealing with larger datasets containing hundreds of thousands of nodes.

Despite there being more than 300 algorithms for drawing trees~\cite{shulz2011treevis}, none can guarantee the two constraints (no crossings, no overlaps), while also optimizing desired edge lengths and area. With this in mind, we propose a scalable method that can guarantee both constraints while optimizing desired edge lengths and area. 
We evaluate four variants of the proposed method for Readable Tree layout ($RT_L$, $RT_C$, $PRT_L$, $PRT_C$) with 4 different real-world datasets of different sizes: from trees with 2,588 nodes, up to trees with 100,347 nodes; see Fig.~\ref{fig:compare-lastfm}.
We further compare our method against state-of-the-art general network layout and tree layout algorithms, by relaxing some of the constraints. We experimented with half a dozen prior methods, but none of them are directly comparable (as discussed in detail in Sec.~\ref{sect:related}). In this paper we report the results obtained by two of these prior methods: 
(spfd$+$p) the scalable force directed placement~\cite{hu2005efficient} algorithm together with label overlap removal via the PRoxImity Stress Model in GraphViz~\cite{ellson2001graphviz}, and (CIR) the CIRcular tree layout algorithm from yED~\cite{wiese2004yfiles}.

We also show the utility of the proposed readable tree layouts method in visualizing general networks. There are many algorithms for extracting {\em important} trees from a given network: minimum spanning trees,  maximum spanning trees, network backbone trees, etc. Motivated by people's familiarity with maps~\cite{doi:10.1177/1473871615594652}, we use a multi-level Steiner tree algorithm~\cite{Ahmed-JEA-19} to create a level-of-detail representation of an underlying general network, which underlies an interactive map-like representation that provides semantic zooming; see Fig.~\ref{fig:teaser}.

\label{sec:alg_framework}

We propose a new scalable method for visualizing large, labeled trees. The method maintains the two hard constraints (C1) No edge crossings and (C2) No label overlaps, while optimizing two desirable properties (O1) desired edge lengths and (O2) compactness of the drawing area.
The two constraints and two optimization goals underlie {\em readable} layouts, as shown in prior work:

Edge crossings are known to make network layouts less readable~\cite{compactness}. Since trees are planar, it is possible to create layouts without crossings, justifying the first constraint (C1).
\blue{Overlapping labels detract from readability and are often a metric for the usability of the labeled network layouts~\cite{overlap,kittivorawongfast}. This supports the second constraint (C2).} 

Preserving desired edge lengths is a standard requirement in instances where edge lengths capture important information, e.g.,  evolutionary time in the tree of life, and 
phylogenetic trees in general~\cite{Bachmaier2005,phylodraw,itol,blanch2015dendrogramix,ballen2017walking,hug2016new}.
\blue{ Edge-length preservation is also used for sensor network reconstruction~\cite{efrat2010} and when representing similarity matrices~\cite{zager2008graph}. This property is also needed in the final step of creating map-like visualization: semantic zooming depends on linearly increasing edge lengths.}
Uniform edge lengths (a special case where all desired edge lengths are the same) are preferable in cases where all edges represent the same notion of connectivity~\cite{overlap,compactness}. 
These reasons validate the first optimization goal (O1). Note that simultaneously ensuring C1 and O1 is an NP-hard problem~\cite{delnphard}, which is why O1 is optimized, rather than guaranteed.

Layout compactness is an important feature for providing an effective overview of the underlying network~\cite{compactness} as a lot of white space can be detrimental to readability~\cite{INKA}. This justifies the second optimization goal O2.



\begin{figure*}[t]
    \centering
    \includegraphics[width=.98\linewidth]{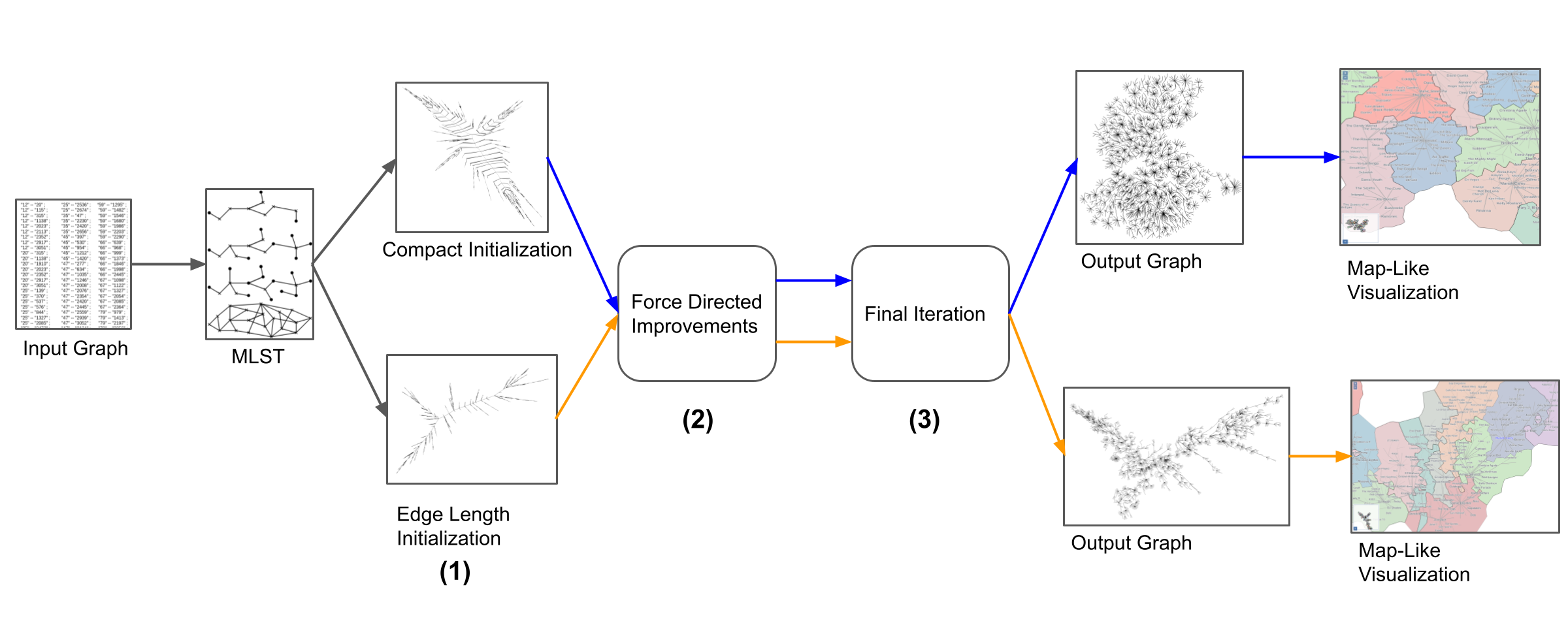}
\caption{\blue{
Overview of the readable tree (RT) method. 
The input is a node-labeled tree with pre-specified edge lengths from which a multi-level Steiner tree (MLST) is computed.
(1) RT initializes with a crossing-free layout, with options targeting compactness or edge length preservation. 
(2) A force-directed improvement removes label overlaps, preserves desired edge lengths, and minimizes the drawing area. 
(3) Remaining label overlaps are removed through resizing and position fine tuning. Note that we have two options for prioritizing either compactness or edge length preservation, matching the corresponding initialization. The tree layout together with the MLST drive a Map-like visualizations with semantic zooming.}}
    \label{fig:framework}
\end{figure*}

Fig.~\ref{fig:framework} shows the workflow of the proposed readable tree (RT) layout method. The input is a node-labeled tree with pre-specified edge lengths from which a multi-level Steiner tree is computed to provide semantic zooming.
The algorithm next computes a crossing-free initial layout (C1) of the tree and maintains this property in every subsequent step.
In the iterative refinement step, the algorithm employs force-directed layout improvement, tailored to remove label overlaps, preserve desired edge lengths (O1), and minimize the drawing area (O2). \blue{In the final iteration step any remaining overlaps are removed, thus enforcing (C2).
The output is used to provide a map-like representation with semantic zooming, which utilizes linearly increasing edge lengths for the different levels of detail.}

For consistency in the experimental evaluation, we use the same methodology to assign desirable edge lengths to all datasets. Specifically, we extract multi-level Steiner trees (described in Sec.~\ref{sect:mlst}) and assign desired lengths proportional to the level where the edge first appears.

\begin{figure}[bht]
	\centering
        \subfloat[]{
    	\label{fig:opposition_delg}
    	\includegraphics[width=0.45\columnwidth]{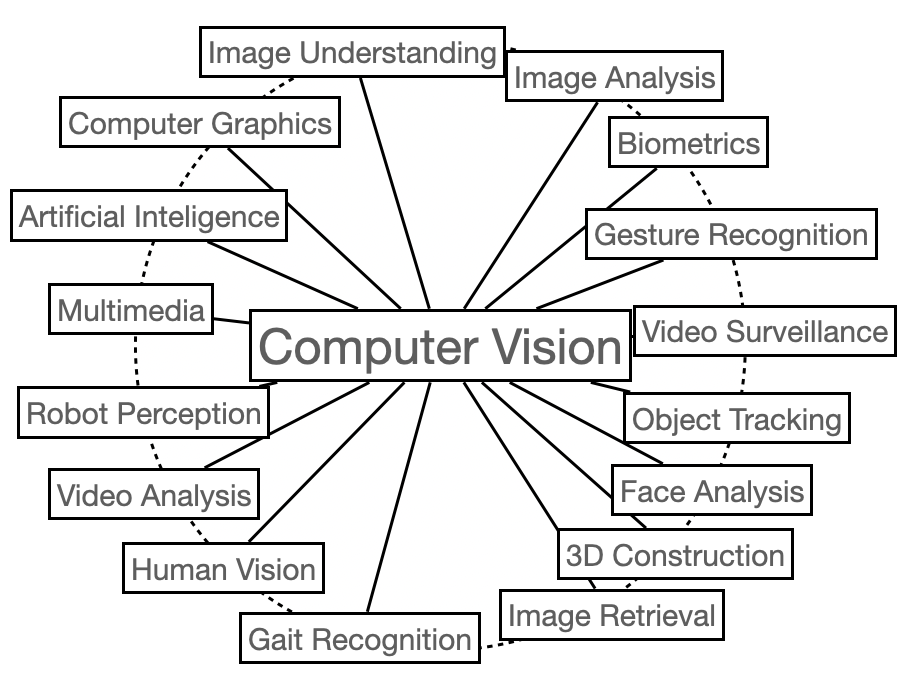}
        }
	\subfloat[]{
            \label{fig:opposition_cg}
            \includegraphics[width=0.5\columnwidth]{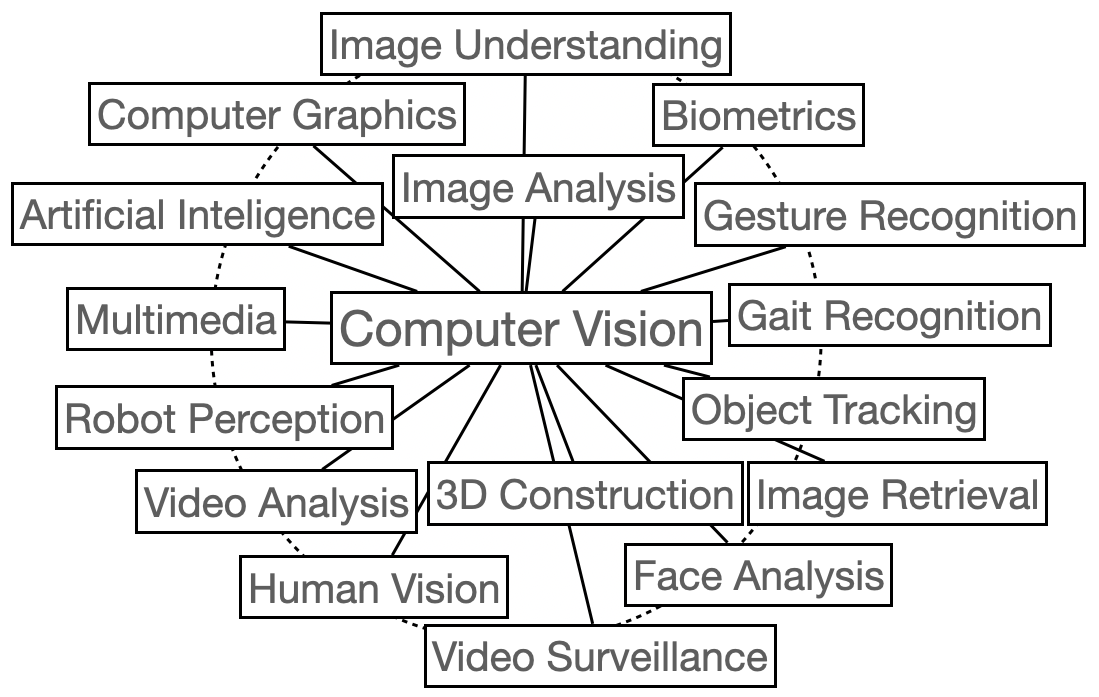}
        }
    \caption{\blue{Preserving desired edge lengths and maintaining compactness can be contradictory goals, especially when maintaining no label overlaps. In this example, all edge lengths should be the same. A layout algorithm can either (a) preserve edge lengths at the expense of layout compactness; or (b) optimize compactness at the expense of edge lengths.
    } 
    }
    \label{fig:MG_uniform_with_labels}
\end{figure}

Note that simultaneously improving the desirable properties is a non-trivial task, as the individual properties could require contradictory layout changes.
For example, consider a high-degree node with all adjacent edges having the same desirable length. It is not possible to preserve the edge lengths and obtain a compact layout simultaneously (with no label overlaps). To preserve edge lengths without overlapping labels we need a larger drawing area. To obtain a compact layout with no label overlaps we need to distort edge lengths; see Fig.~\ref{fig:MG_uniform_with_labels}.



With this in mind, our proposed method can emphasize desired edge length preservation or compactness. This is determined by selecting one of the two initial layouts. \reyansInit preserves desired edge lengths but allows label overlaps. 
\mingweisInit prioritizes compactness but does not preserve desired edge lengths. 
In the following steps, (iterative refinement and final iteration) these initial layouts are modified to ensure C1 and C2 and optimize O1 and O2. 


 We provide two implementations of the proposed method: one that requires parallel hardware and one that does not. 
The d3.js~\cite{bostock2011d3} version works well for smaller instances.
The parallel variant of our method, using OpenMP~\cite{chandra2001parallel},  is 1-2 orders of magnitude faster and can handle trees with hundreds of thousands of nodes. Note that the interactive map-like visualization with semantic zooming relies on the readable tree layout algorithm which is run just once per dataset as a pre-processing step, and we consider runtime in minutes to be acceptable.

We quantitatively evaluate our algorithms by measuring compactness, desired edge length preservation, and runtime. For comparison, we use two state-of-the-art layout methods. Note that the two prior methods do not take desired edge lengths into account, and to make the comparison somewhat fair we use uniform edge lengths. Also, one of the two methods can produce edge crossings.

Since even the smallest tree we work with has more than 2500 labeled nodes, we also make the results accessible via interactive, map-like visualizations. Specifically, using the multi-level Steiner tree hierarchy extracted from the input, we provide semantic zooming functionality that allows us to see the global structure (high level) and local details (low level). The interactive visualization is accessible here: \url{https://tiga1231.github.io/zmlt/demo/overview.html} 

\section{Related Work}
\label{sect:related}
\noindent \textbf{Tree and Network Layout Algorithms:} Drawing trees has a rich history: 
 Treevis.net~\cite{shulz2011treevis} contains over three hundred different types of visualizations.  Here we briefly review algorithms for 2D node-link representations, starting with arguably the best-known one by 
Reingold and Tilford\cite{Reingold1981}. This and other early variants draw trees recursively in a bottom-up sweep.
These methods produce crossings-free layouts but do not consider node labels or edge lengths. 
We will now broaden our scope to general graphs.  While general graphs are not necessarily planar, the layout techniques and ideas can be applied to trees.  Most general network layout algorithms use a force-directed~\cite{eades1984heuristic,fruchterman1991graph} or stress model~\cite{brandes2007eigensolver,koren2002ace} and provide a single static drawing.
The force-directed model works well for small networks but does not scale to large networks. Speedup techniques employ a multi-scale variant~\cite{hh-msadg-99,GGK04} or use parallel and distributed computing architecture such as 
VxOrd~\cite{bkk-mbs-05},
BatchLayout~\cite{rahman2020batchlayout},
and MULTI-GILA~\cite{arleo2018distributed}. 
Libraries such as GraphViz~\cite{ellson2001graphviz} and  OGDF~\cite{chimani2011} 
provide many general network layouts, but may not support interactions. Whereas visualization toolkits such as 
Gephi~\cite{bastian2009gephi} and yEd~\cite{wiese2004yfiles} support visual network manipulation, and while they can handle large networks,
the amount of information rendered statically on the screen makes the visualization difficult to use for large networks.

\smallskip\noindent \textbf{Overlap Removal and Topology Preservation:}\label{sec:topo_pres} 
In theory, nodes can be treated as points, but in practice, nodes are labeled and these labels must be shown in the layout~\cite{nobre2019state,hadlak2015survey}. 
Overlapping labels pose a major problem for most layout generation algorithms, and severely affect the usability of the resulting visualizations. 
A simple solution to remove overlaps is to scale the drawing until the labels no longer overlap. This approach is straightforward, although it may result in an exponential increase in the drawing area. 
Marriott \textit{et al.}~\cite{marriott2003removing} proposed to scale the layout using different scaling factors for the $x$ and $y$ coordinates. This reduces the overall blowup in size but may result in a poor aspect ratio.
Gansner and North~\cite{gansner1998improved},  Gansner and Hu~\cite{gansner2009efficient}, and Nachmanson \textit{et al.}~\cite{nachmanson2017node} describe overlap-removal techniques with better aspect ratio and modest additional area. However, these approaches can and do introduce edge-crossings, even when starting with a crossings-free input. 
Placing labeled nodes without overlaps has also been studied~\cite{luboschik2008particle,mote2007fast,kittivorawongfast}, but these approaches also cannot guarantee crossings-free layouts.

\smallskip\noindent {\bf The need for new algorithms:} 
While there exist many algorithms for generating tree and network layouts, to the best of our knowledge, no existing algorithm considers the four aspects of the readability of labeled tree layouts: no edge crossings, no node overlaps, compact drawing area, and preserved desired edge lengths.
For example, one of the most frequently used network visualization systems,   GraphViz~\cite{ellson2001graphviz},  has an efficient layout algorithm based on the scalable force directed placement (sfdp) algorithm~\cite{hu2005efficient} and can remove label overlaps via the PRoxImity Stress Model (PRISM)~\cite{gansner2009efficient}. The output, however, does not optimize the given edge lengths and cannot ensure the crossing constraint; examples in this paper contain 100-1000 crossings. 

The popular visualization library d3.js 
provides a link-force feature to optimize desired edge lengths, but cannot ensure the crossing constraint and cannot remove label overlaps without blowing up the drawing area. Another excellent visualization toolkit, yEd~\cite{wiese2004yfiles}, provides a method that can draw trees without edge crossings and optimize compactness. However, none of the methods available in yED can preserve the desired edge lengths and one cannot remove label overlap without blowing up the drawing area.
Nguyen and Huang~\cite{space_optimized} describe an algorithm for compact tree layouts (note that we use a similar initialization step), however, their approach is not concerned with edge lengths and node labels.

Different dimensionality reduction techniques such as t-SNE~\cite{van2008visualizing} and its variants~\cite{kobak2019art,leow19GraphTSNE} are hard to use to visualize tree networks since they do not guarantee crossing-free and label overlap-free drawing; in one of our experiments, we observed that t-SNE can generate more than 50 crossings in a small tree of 100 nodes. 
We believe our paper fills this gap in the literature by developing and making available a scalable method for readable tree layouts.

\section{Readable Tree Layout Algorithm}


Our algorithm has several parts: multi-level Steiner tree extraction, an initial layout, a force-directed layout improvement, a final iteration ensuring no label overlaps, and a map-like visualization; see Fig.~\ref{fig:framework}. \blue{ We provide two options for the initialization, which results in two different potential outputs.} Since we rely on existing techniques for the first step and the last steps, we briefly discuss them in Sec.~\ref{se:visualizationtool}. In this section, we discuss the remaining three steps in detail.

\subsection{Initialization}

As mentioned in  Sec.~\ref{sec:introduction}, desired edge length preservation and compactness are contradictory optimization goals. Hence, our algorithm can produce two different types of layouts: one emphasizing the desired edge length preservation and the other emphasizing compactness. The two initialization steps below influence the end results, these are discussed in the evaluation section, Sec.~\ref{sec:evaluation}.


\smallskip\noindent {\bf Edge Length Initialization: }

\begin{figure}[thp]
	\centering
	\subfloat[]{
	\label{fig:balanced}
	\includegraphics[width=.24\columnwidth]{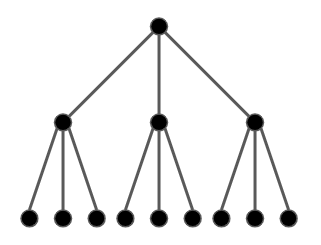}}
	\subfloat[]{\label{fig:symmetric}\includegraphics[width=.24\columnwidth]{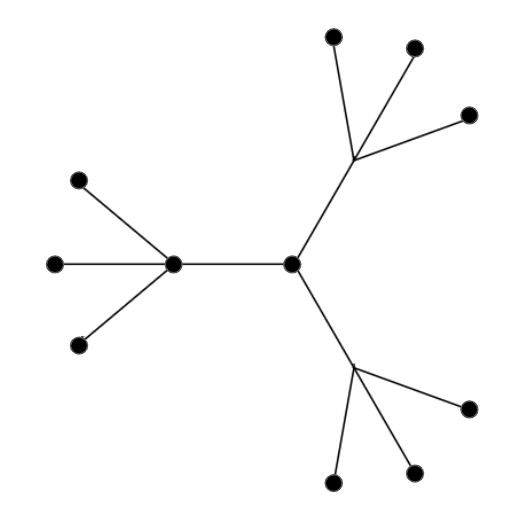}}
	\subfloat[]{\label{fig:non_balanced}\includegraphics[width=.24\columnwidth]{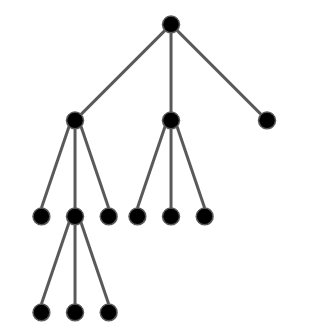}}
	\subfloat[]{\label{fig:non_symmetric}\includegraphics[width=.24\columnwidth]{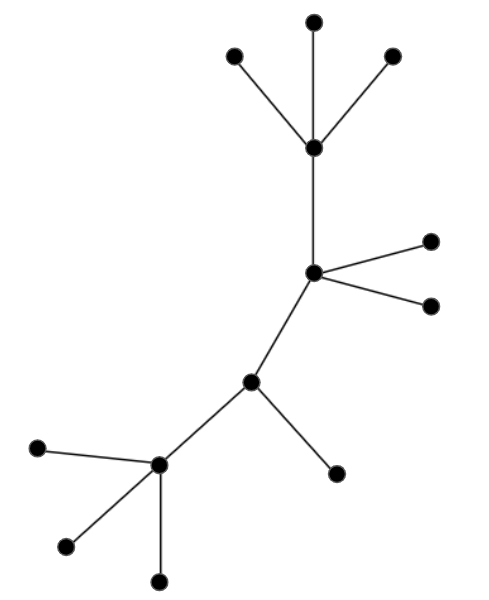}}
	\caption{Illustrating the output layout of \reyansInit w.r.t.~different types of inputs. If the tree is a) balanced then the output is b) symmetric. If the tree is c) not balanced then the output is d) not symmetric.}
	\label{fig:balanced_vs_non}
\end{figure}

\begin{figure}[t]
    \centering
    \includegraphics[width=0.35\columnwidth]{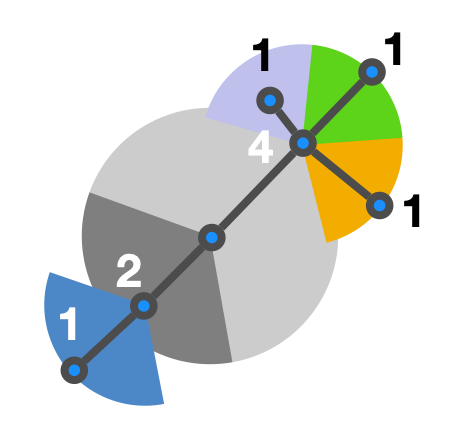}
    \includegraphics[width=0.55\columnwidth]{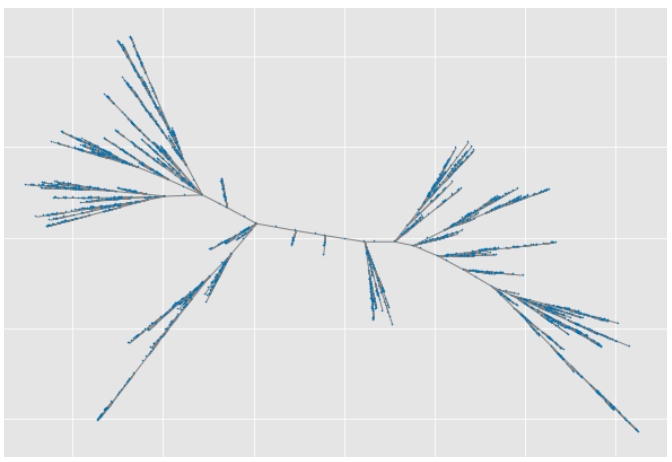}
    \caption{\textbf{Left:} Illustration of initial radial layout of \reyansAlgoAcronym. The numbers indicate the proportion of wedge sectors determined by the size of induced subtrees of child nodes. \textbf{Right:} Initial layout of the last.FM network.}
    \label{fig:init-DELG}
\end{figure}

The first initialization creates a layout that is crossing-free and preserves all edge lengths, although it may have label overlaps. Our crossings-free initialization is similar to a prior work~\cite{Bachmaier2005}.
We select a root node and assign a wedge region (sector) to every child. 
Each child is then placed along an angle bisector of the assigned wedge, away from its parent by the desired edge length.
We continue this process until the coordinates for each node have been computed; see pseudocode Alg. \reyansInit. 
We traverse the nodes using breadth-first search (BFS) which visits parents before children.
Since the angular regions are unbounded, the algorithm can preserve all the desired edge lengths exactly. 
When we assign wedge regions to child nodes, the angles are  proportional to the size of the subtree rooted from the child; see Fig.~\ref{fig:init-DELG}. 
When the input is a balanced tree this algorithm computes a symmetric layout; see Fig.~\ref{fig:balanced_vs_non}.

\begin{algorithm}[t]
\caption{\reyansInit}
 \label{alg:angular}
\KwInput{
    $G = (V, E)$ \tcp{The tree network}
    $root \in V$ \tcp{root of the tree}
    $\{\dots, n_{v}, \dots\}$ \tcp{Size of subtree rooted from $v \in V$}
    $\{\dots, DEL_{uv}, \dots\}$ \tcp{Desired edge length from $u$ to $v \in V$}
}
\KwOutput{
    $X$ \tcp{Crossing-free initial layout for RT\_L}
}
\Fn{Initial\_Layout($G, root$)}{
    $X_{root} \leftarrow (0,0)$\;
    $W_{root} \leftarrow wedge(center=root,$ \\
        $\quad radius=DEL_{root}, angle\_range=[0, 2\pi))$\;
    \For {parent node $p \in BFS(G, root)$}{
        $ \{c_1, c_2 \dots c_{|C|}\} \leftarrow children(p)$\;
        $angle\_ranges = \{\dots A_{c_i} \dots \} \leftarrow partition(angle\_range(W_p); n_{c_1}, n_{c_2}, \dots)$\;
        \For {$c_i \in C$}{
            $W_{c_i} \leftarrow wedge(center=p,$ \\
                $\quad radius=DEL_{c_i}, angle\_range=A_{c_i})$\;
            $X_{c_i} \leftarrow midpoint(arc(W_{c_i}))$\;
        }
    }
    return $X$\;
}
\end{algorithm}

\smallskip\noindent {\bf Compact Initialization:}

\begin{figure}[t]
    \centering
    \includegraphics[width=0.32\columnwidth]{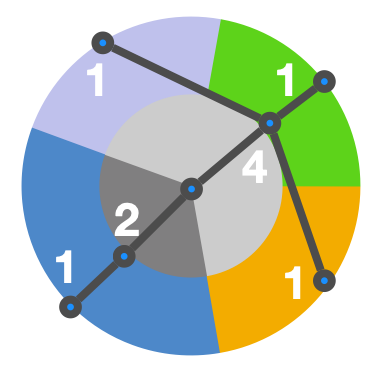}
    \includegraphics[width=0.38\columnwidth]{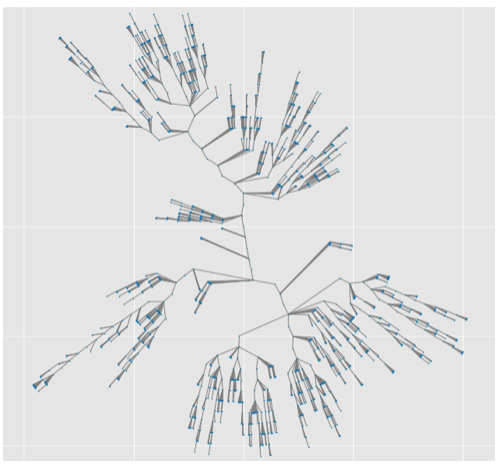}
    \caption{\textbf{Left:} Illustration of initial radial layout of \mingweisAlgoAcronym. The number indicates the proportion of wedge sectors, determined by the size of induced subtrees of child nodes. \textbf{Right:} Initial layout of the last.FM network by \mingweisAlgoAcronym.
    }
    \label{fig:init-CG}
\end{figure}

The second initialization also creates a crossing-free layout that is compact, although desired edge lengths might be distorted.
Instead of assigning wedge regions to children of a node, we assign the entire fan area to the children and place children along the arc of the fan; see pseudocode Alg. \mingweisInit. 
This results in a wider spread of nodes, but does not preserve the desired edge lengths; see Fig.~\ref{fig:init-CG} and compare it with Fig.~\ref{fig:init-DELG}. 

\begin{algorithm}[t]
\caption{\mingweisInit}
\KwInput{
    $G = (V, E)$ \tcp{The tree network}
    $root \in V$ \tcp{root of the tree}
    $\{\dots, n_{v}, \dots\}$ \tcp{Size of subtree rooted from $v \in V$}
    $\{\dots, d_{v}, \dots\}$ \tcp{Number of hops from the root to $v \in V$}
}
\KwOutput{
    $X$ \tcp{Crossing-free initial layout for RT\_C}
}
\Fn{Initial\_Layout($G, root$)}{
    $X_{root} \leftarrow (0,0)$\;
    $W_{root} \leftarrow wedge(center=root,$ \\
        $\quad radius=1, angle\_range=[0, 2\pi))$\;
    \For {parent node $p \in BFS(G, root)$}{
        $ \{c_1, c_2 \dots c_{|C|}\} \leftarrow reorder(children(p), $\\
        $\quad mode=\text{'centralize heavy subtrees'})$\;
        $angle\_ranges = \{\dots A_{c_i} \dots \} \leftarrow partition(angle\_range(W_p); n_{c_1}, n_{c_2}, \dots)$\;
        \For {$c_i \in C$}{
            $W_{c_i} \leftarrow wedge(center=root,$ \\
                $\quad radius=d_{c_i}, angle\_range=A_{c_i})$\;
            $X_{c_i} \leftarrow midpoint(arc(W_{c_i}))$\;
        }
    }
    return $X$\;
}
\end{algorithm}

\subsection{Force Directed Improvements}\label{sec:force-directed-optimization}

The next step is to use a force directed algorithm to optimize our soft constraints. Here we describe  all of the forces used, as well as how we prevent edge crossings at every step of the force directed improvement.

\subsubsection{Maintaining the Crossings-Free Constraint}


The force-directed algorithm improves the layout by applying different forces while ensuring that there are no edge crossings introduced in any iteration of the algorithm. The algorithm starts with a layout computed in the previous initialization step. In each iteration of the algorithm, it computes different forces for each node as discussed in the next sections. Then for each node $v$, it computes the movement $T_v$ applied by the forces. The algorithm combines the forces linearly, with scaling factors discussed in Sec.~\ref{sec:alg_params}. If $T_v$ introduces any edge crossings, then the algorithm does not apply the movement. Otherwise, it computes the new coordinate of $v$ according to $T_v$. The algorithm continues this step until a maximum number of iterations is reached; see pseudocode Alg.~{\fontfamily{lmtt}\selectfont Force-Directed-Improvement}.

\begin{algorithm}[t]
\SetAlgoLined
\KwInput{
    $G = (V, E)$ \tcp{The tree network}
    $X$ \tcp{Crossing-free initial layout}
    $niter$ \tcp{Number of iterations}
}
\KwOutput{
    $X$ \tcp{Improved crossing-free layout}
}
\Fn{Force\_Directed\_Improvement($G, X$)}{
\For{$i = 1, 2, \ldots, niter$}
{
    \For{\textit{each node} $u\in V$}
    {
        \tcp{Label overlap force}
        \For{\textit{each node} $v \in \text{collision\_region(u)}$}{
            $T_v \leftarrow \Delta T_v + f_c(X_v, X_{u})$ \;
        }
        \tcp{Edge length force}
        \For{\textit{each neighbor} $v$ \textit{of} $u$}
        {
            \If{$length(X_u, X_v) > l_{uv}$}
            {
                $T_v\leftarrow T_v + f_a(X_u, X_v)$\;
            }
            \Else 
            {
                $T_v\leftarrow T_v - f_r(X_u, X_v)$\;
            }
        }
        \tcp{Distribution force}
        \For{\textit{each node} $v$ \textit{of} $u$}
        {
            $T_v\leftarrow T_v - f_{d}(X_u, X_v)$ \;
        }
        \tcp{Node-edge force}
        \For{\textit{each neighbor} $v$ \textit{of} $u$}
        {
            $T_v\leftarrow T_v - f_{node-edge}(X_u, X_v)$ \;
        }
        \tcp{Maintaining no edge crossings}
        \If{$T_v$ \textit{does not introduce edge-crossing}}
        {
            $X_v \leftarrow X_v + T_v$\;
        }
    }
}
    return $X$\;
}
\caption{{\fontfamily{lmtt}\selectfont Force-Directed-Improvement}}
\label{alg:crossing-check}
\end{algorithm}

\subsubsection{Label Overlap Force}\label{sect:collision-force}


We use an elliptical force by modifying the traditional collision force to help remove the label overlaps.
Since labels are typically wider than they are tall, a circular collision region potentially wastes space above and below the labels. We build an elliptical force out of a circular collision force by stretching the $y$-coordinate by a constant factor $b$ (e.g., by default we use $b=3$) before a circular collision force is applied, and restoring the coordinates after the force is applied.  The velocity computed by the collision force is processed in a similar manner, with a reciprocal scaling factor. 
Formally, let $X_v$ denote the coordinate of node $v$. 
We specify a different collision radius depending on label size, denoted by $r_v$, for every node $v$. 
Note that the collision radius depends on both the font size of a label and the number of characters in the label.
A circular collision force first calculates a movement $T'_v$, and then the elliptical movement $T_v$ is computed by stretching the $x$-coordinate and compressing the $y$-coordinate, e.g., $T_v.x = T'_v.x \times b, T_v.y = T'_v/y$.  Then we update the coordinate $X_v$ by adding $T_v$.

\subsubsection{Edge Length Force}\label{sect:edge-length-force}

The edge length force is designed to maintain the desired edge lengths.  
For every edge, we apply either a repulsive force $f^e_r = K/d \cdot I_{d < l_e}(d)$ (when the edge is compressed) or an attractive force $f^e_a = K d \cdot I_{d > l_e}(d)$ (when the edge is stretched), determined by the indicator function.
The force is proportional/reciprocal to  distance $d$.

\subsubsection{Distribution Force}\label{sect:distribution-force}
We define a global node distribution force as a repulsive force between every pair of nodes.
We set the repulsive force between two nodes  inversely proportional to the squared distance in the current layout; similar to an electrical charge between nodes:
$|f_{d}(u, v)| = s(u,v) / ||X_u - X_v||^2,$
where $s(u,v)$ denotes the strength of the force between nodes and depends on the longest desired edge length adjacent to $u$ and $v$. 
We set
$s(u,v) = max_{w \in V, (u,w) \in E}\; \{l_{uw}\} * max_{w \in V, (v,w) \in E} \; \{l_{vw}\}.$
\subsubsection{Node-Edge Force}\label{sect:nodes-and-edges-force}
Finally, we define a force between nodes and edges.  This improves readability by reducing the number of instances where labels are placed over edges.  This force is inversely proportional to the distance between the node and edge, acting orthogonally from the edge (evaluating to zero if the node does not project onto the edge segment or is too far from the edge):
$|f_{node-edge}(v, e)| = c / d(v, e),$
where $c$ is a constant across all pairs and $d(v, e)$ denotes the Euclidean distance between node $v$ and edge $e$.

\subsection{Final Iteration}
\label{sec:delgpostprocessing}
The final iteration is needed to ensure any remaining overlaps are removed.
It is possible that this step will have no work to do, but this check is necessary in order to enforce our hard constraint (C2).

In this step, we go over all pairs of overlapping nodes and move them until the overlap is repaired. 
To do this we check whether we can move one of the overlapping nodes so that the distance between two nodes increases without introducing any crossing and label overlap.
Specifically, for each node $v$ of the pair of nodes, we consider a square bounding box that has a small area. We denote the set of nodes in that bounding box by $V'$. We then sample some random points from that bounding box. For each of these sample points, we check whether we have an overlap-free and crossing-free drawing. If we find such a point, then we move $v$ to that point and consider the next label overlap; see pseudocode Alg.~{\fontfamily{lmtt}\selectfont Final-Iteration}. 

Note that in some cases this step is not needed at all, as all overlaps are removed during the force-directed layout improvement step. In other cases (e.g., larger input instances with denser subtrees) a handful of final iteration steps are needed to remove remaining overlaps.

\begin{algorithm}[ht]
\SetAlgoLined
\KwInput{
    $steps$ \tcp{Sample size}
    $size$ \tcp{Width of sample area}
}
\KwOutput{
    $X$ \tcp{A crossing-free layout with reduced label overlaps}
}
\Fn{Final\_Overlap\_Removal($steps, size$)}{
 \For{$u, v \in V \times V$ s.t. $u$ and $v$ overlaps}{
    \For{$k = 1, 2, \cdots, steps$}{
      $r = random(0, 1)$\;
      $\Delta X_u = (r * size) - (size/2)$\;
      \If{$\Delta X_u$ does not introduce crossing and new overlap}{
        $X_u \leftarrow X_u + \Delta X_u$
      }
      \If{$u$ and $v$ overlaps}{
        $\Delta X_v = (r * size) - (size/2)$\;
        \If{$\Delta X_u$ does not introduce crossing and new overlap}{
          $X_v \leftarrow X_v + \Delta X_v$
        }
      }
  }
 }
 return $X$;
}
 \caption{{\fontfamily{lmtt}\selectfont Final-Iteration}}
 \label{alg:delgpostprocessing}
\end{algorithm}

\section{Parallel Readable Tree Drawing}
Here we describe the parallel version of the algorithm (PRT), which again maintains the hard constraints, but now also adds scalability to handle larger trees. The previous algorithm, RT, generates good-quality layouts that satisfy all the hard constraints and optimize the soft constraints well.  Additionally, RT does not require any specialized equipment and can be run on any computer. However, it is sequential in nature and does not work well for networks with more than about 5000 nodes. 
The parallel tree version takes advantage of opportunities to speed up some of the necessary computations. As before, there are two variants: Edge Length Initialized Parallel Readable Tree (\khaledsAlgoAcronymDelg) algorithm and Compactness Initialized Parallel Readable Tree (\khaledsAlgoAcronymCG) algorithm that emphasize the preservation of desired edge lengths and compactness.

Note that force calculations in the
PRT 
algorithm has an inherent dependence on neighbors and non-neighbors (i.e., collision/edge forces for a node depend on the coordinates of other nodes). If such forces are computed in different threads, they cannot be seamlessly integrated. 
Instead of running the entire algorithm in parallel, we use the mini-batch approach, similar to BatchLayout~\cite{rahman2020batchlayout}. The mini-batch technique is commonly used for parallel Stochastic Gradient Descent (SGD) training, where one gradient update has a dependency on other gradient updates \cite{goyal2017accurate}.

The PRT algorithm follows the same workflow shown in Fig.~\ref{fig:framework}:
a crossings-free initial layout, followed by customized force-directed improvement, and a final iteration that enforces the overlaps constraint. 

\begin{algorithm}[!h]
\caption{{\fontfamily{lmtt}\selectfont PRT-Center-Node}}
\label{alg:centernode}
\DontPrintSemicolon
\SetAlgoLined
\KwInput{
    $G= (V, E)$ \tcp{The tree network}
}
\KwOutput{
    $n_c$ \tcp{center node id $n_c$}
}
\Fn{PRT\_Center\_Node($G$)}{
    $C_u \leftarrow 0.0$, $\forall u\in V$\;
    \For{\textit{each node} $u\in V$ \textbf{in parallel}}
    {
        $\mathcal{D} \leftarrow \sum_{v\in V} dist(u, v)$\;
        $C_u \leftarrow \frac{|V|}{\mathcal{D}}$
    }
    $n_c \leftarrow \underset{u}{\mathrm{arg\;max}}\; C_u$\;
    return $n_c$\;
}
\end{algorithm}

\medskip\noindent{\bf Initialization: } We create an initial layout of the tree with no edge crossings using parallelized versions of Alg.~{\fontfamily{lmtt}\selectfont Edge-Length-Initialization} and Alg.~{\fontfamily{lmtt}\selectfont Compact-Initialization}. The most time-consuming part here is determining the center node that serves as \emph{root}. The time complexity of this step is $O(n^2)$, where $n$ is the number of vertices in $G$. Thus, we fully parallelize Alg.~{\fontfamily{lmtt}\selectfont PRT-Center-Node}, where normalized closeness centrality~\cite{bavelas1950communication} is computed for each vertex. Then, we take the node with the maximum score
as the root. Since $G$ is a tree, we can apply BFS to compute the distance between vertices $u$ and $v$ as shown in line 3 of 
 Alg.~{\fontfamily{lmtt}\selectfont PRT-Center-Node}.

We place the center node at the origin of the Cartesian coordinate system and iteratively traverse the tree in a BFS fashion, placing nodes so that they do not introduce edge-crossings. The runtime of this part of the algorithm is $O(n)$. Thus, finding the center node is the slowest step of Initialization, and that has been effectively parallelized so that it has minimal effect on the overall runtime.


\begin{algorithm}[!htb]
\caption{{\fontfamily{lmtt}\selectfont PRT-Force-Directed-Improvement}}
\label{algo:btforce}
\DontPrintSemicolon
\SetAlgoLined

\KwInput{
    $G = (V, E)$ \tcp{The tree network}
    $X$ \tcp{Crossing-free initial layout}
    $batch$ \tcp{No. of vertex batches form V}
    $samples$ \tcp{Sample size}
    $niter$ \tcp{Number of iterations}
}
\KwOutput{
    $X$ \tcp{Improved crossing-free layout}
}


\Fn{PRT\_Force\_Directed\_Improvement($G, X$)}{
    \For{$i = 1, 2, \ldots, niter$}
    {
        $T \leftarrow \{0\}^{|V|\times 2}$\;
        Partition $V$ into $B = \lceil\frac{|V|}{batch}\rceil$ batches\; 
        \For{\textit{each batch} $B\in V$}
        {
            \For{\textit{each node} $u\in B$ \textbf{in parallel}}
            {
                \tcp{Label overlap force}
                \For{\textit{each node} $v \in \text{collision\_region(u)}$}{
                    $T_v \leftarrow \Delta T_v + f_c(X_v, X_{u})$ \;
                }
                \tcp{Edge length force}
                \For{\textit{each neighbor} $v$ \textit{of} $u$}
                {
                    \If{$length(X_u, X_v) > l_{uv}$}
                    {
                        $T_u\leftarrow T_u + f_a(X_u, X_v)$\;
                    }
                    \Else 
                    {
                        $T_u\leftarrow T_u - f_r(X_u, X_v)$\;
                    }
                }
                \tcp{Distribution force}
                \For{\textit{a random node} $w$ \textit{upto} \textbf{\textit{samples}} \textit{times}}
                {
                    $T_u\leftarrow T_u - f_{d}(X_u, X_{w})$ \;
                }
                \tcp{Node-edge force}
                \For{\textit{each neighbor} $v$ \textit{of} $u$}
                {
                    $T_u\leftarrow T_u - f_{node-edge}(X_u, X_v)$ \;
                }
                \tcp{Maintaining no edge crossings}
                \If{$T_u$ \textit{does not introduce edge-crossing}}
                {
                    $X_u \leftarrow X_u + T_u$\;
                }
            }
        }
    }
    return $X$;
}
\end{algorithm}

\medskip\noindent{\bf Parallel Force-directed Improvement: }
We describe the parallel force-directed improvement of layout in Alg.~{\fontfamily{lmtt}\selectfont PRT-Force-Directed-Improvement}. In each iteration of the algorithm, we select a batch $B$ from the set of nodes $V$ and compute attractive and repulsive forces in parallel similar to the BatchLayout method \cite{rahman2020batchlayout}. 
 We apply label overlap forces, edge length forces, and node-edge forces to each node and edge of $B$ in a similar way described in Sec.~\ref{sec:force-directed-optimization}.
To compute repulsive distribution forces with respect to the non-neighboring nodes, we select $sample$ nodes at random, to speed up the process by approximate repulsive force computation~\cite{rahman2020force2vec}. For each random node $w$, we compute repulsive force $f_{d}(X_u, X_{w})$ and update the temporary coordinates $T_u$. Note that this force computation for nodes within the same batch is independent and thus we can run it in parallel. 
Before updating the coordinates of a batch, we check whether it introduces edge-crossings (line 22).
Even though an increased number of batches exposes more parallelism, the quality of the layout may be negatively impacted, as observed in stochastic gradient descent (SGD)~\cite{rahman2020force2vec}. 
We found that a batch size of 128 or 256 gives a good balance between speed and quality~\cite{rahman2020batchlayout}.
Since the batch size is small compared to the size of the tree, we perform sequential updates in line 23.  
However, we perform the edge-crossing check in parallel.

\medskip\noindent{\bf Parallel Final Iteration: }
This step checks for any remaining label overlaps and repairs them in parallel. 
The underlying edge crossings check method is similar to the parallel force computation in Alg.~{\fontfamily{lmtt}\selectfont PRT-Force-Directed-Improvement}. 
Other than the parallel edge crossings check, the algorithm is similar to Alg.~{\fontfamily{lmtt}\selectfont Final-Iteration}. For each overlapping pair of nodes, we sample some random points from a square bounding box that has a small area. For each of these sample points, we check whether we have an overlap-free and crossing-free drawing. When we find such a point, we move the node there.


\section{Algorithm Parameters}
\label{sec:alg_params}

Like many force-directed algorithms, our RT and PRT methods depend on several parameters. We set some default parameter values based on prior work. For example, we select effective approximation algorithms to compute multi-level Steiner trees and the number of levels in the trees proportional to the size of the underlying data based on prior work~\cite{Ahmed-JEA-19}. We set the batch size of Alg.~{\fontfamily{lmtt}\selectfont PRT-Force-Directed-Improvement} equal to 128-256 as in BatchLayout~\cite{rahman2020batchlayout}. Repulsive force parameters are based on those in Force2Vec~\cite{rahman2020force2vec}. 

We performed a small-scale parameter search for most of the remaining parameters.
To determine good values for these parameters we extracted samples with 2,000 nodes from each of our datasets and analyzed the effect of  parameter modification.
We do this by setting default values from pilot experiments and modifying one parameter at a time while fixing the remaining ones.

\subsection{Different Forces}
In Alg.~{\fontfamily{lmtt}\selectfont Force-Directed-Improvement} (lines 6-16) and Alg.~{\fontfamily{lmtt}\selectfont PRT-Force-Directed-Improvement} (lines 11-21), we combine several different forces acting on each node of the input network. 
Specifically, there are four types of forces: the label overlap force $F_c$ to remove label overlaps, the edge length force $F_l$ to achieve the desired edge lengths, the distribution force $F_{d}$ to distribute the nodes in the drawing area uniformly, and the node-edge force $F_{ne}$ to keep adjacent nodes closer. Each of these forces has a strength, or scaling factor, that indicates its impact on the overall force. We denote the strengths of the label overlap force, edge length force, distribution force, and node-edge force by $S_c, S_l, S_{d}$ and $S_{ne}$, respectively. The range of the strength values is $[0-1]$. Then the total force $F(u)$ applied on a node $u$ is calculated by the following equation:
$$F(u) = S_c \cdot F_c(u) + S_l \cdot F_l(u) + S_{d} \cdot F_{d}(u) + S_{ne} \cdot F_{ne}(u)$$

\medskip\noindent{\bf Edge Length Force: }The edge length force, $F_l$, plays a very important role in all RT and PRT variants as it corresponds to one of the two optimization goals (O1). With this in mind, we use the highest strength for this force: $S_l = 1$.

\begin{figure}[th]
	\centering
	\subfloat[]{\label{fig:force_overlaps}\includegraphics[width=.43\columnwidth]{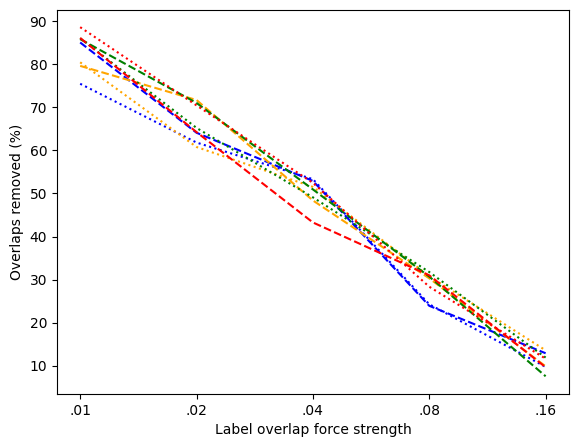}}
	\subfloat[]{\label{fig:force_DEL}\includegraphics[width=.44\columnwidth]{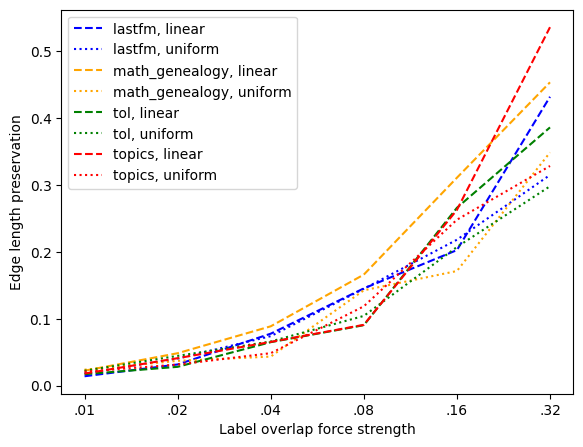}}
	\caption{Illustrating the impact of the label overlap force strength. (a) shows the percentage of label overlaps with respect to the label overlap force.  (b) shows the edge length preservation with respect to the label overlap force.
	}
	\label{fig:collision_force_strength}
\end{figure}

\medskip\noindent{\bf Label Overlap Force: }
The initialization based on edge length preservation consistently results in more overlaps than the compactness-based one.
Hence to determine an appropriate strength for the label overlap force, we use Alg.~{\fontfamily{lmtt}\selectfont Edge-Length-Initialization} in the parameter search.
 Fig.~\ref{fig:force_overlaps} shows the percentage of label overlaps of all networks, with respect to the initial overlaps, after 50 iterations of Alg.~{\fontfamily{lmtt}\selectfont Force-Directed-Improvement}.
 As we increase the strength of this force from 0 to 1, overlaps decrease, while edge length preservation decreases,  as illustrated in Fig.~\ref{fig:force_DEL}. To balance this, we set the strength of label overlap force to 0.16, leaving  around 10\% overlaps (which are removed in the final step), while providing good edge length preservation.

\begin{figure}[th]
	\centering
	\subfloat[]{\label{fig:ellipse-edge-1}\includegraphics[width=.45\columnwidth]{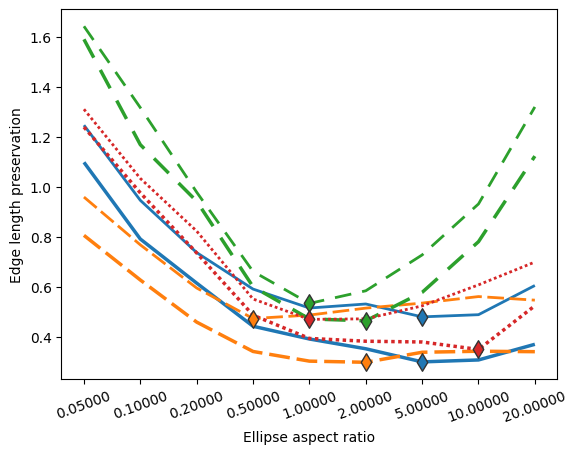}}
	\subfloat[]{\label{fig:ellipse-edge-2}\includegraphics[width=.45\columnwidth]{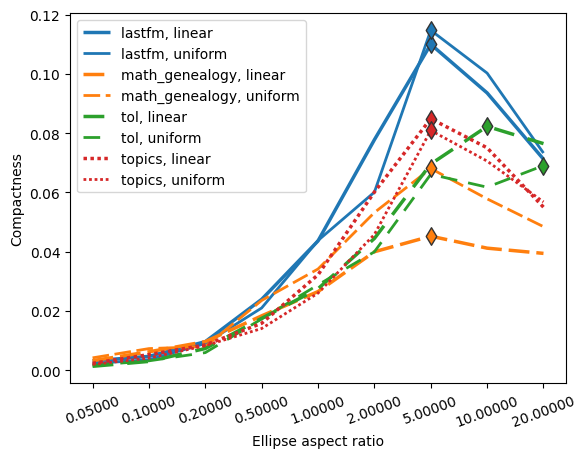}}
	\caption{Impact of label overlap force directions (ellipse aspect ratio) on the two quality metrics: Edge length (a) and compactness (b). Diamonds mark the best parameter setup for the specific network.
	}
	\label{fig:ellipse}
\end{figure}

    We also explore the aspect ratio of the ellipse in the label overlap force. The aspect ratio, or the parameter $b$ defined in Sec.~\ref{sect:collision-force}), denotes a wide elliptical collision force when $b > 1$ and a tall one when $0 < b < 1$.
    As shown in Fig.~\ref{fig:ellipse}, the edge lengths can be preserved with a wide range of ellipse aspect ratios with the best compaction achieved when the aspect ratio is set to 5.

\begin{figure}[th]
	\centering
	\subfloat[]{\label{fig:charge-edge}\includegraphics[width=.45\columnwidth]{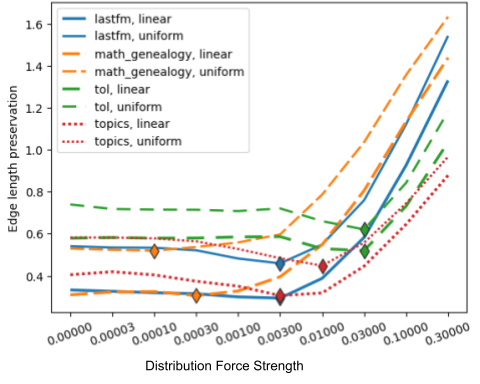}}
	\subfloat[]{\label{fig:charge-compactness}\includegraphics[width=.45\columnwidth]{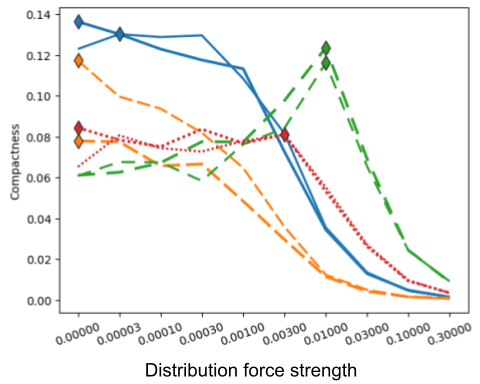}}
	\caption{Impact of distribution force on the two quality metrics: Edge length (a) and compactness (b). Diamonds mark the best parameter setup for the specific network.
	}
	\label{fig:charge}
\end{figure}

\medskip\noindent{\bf Distribution and Node-Edge Forces: }
The distribution force and the node-edge force play a larger role in the compactness initialization, and we experimentally determine suitable strengths with Alg.~\mingweisInit.
As shown in Fig.~\ref{fig:charge}, smaller force strength leads to better edge length preservation and better compactness for all datasets, and so we set $S_{d}=0.003$ 

\begin{figure}[th]
	\centering
	\subfloat[]{\label{fig:nodeedge-edge}\includegraphics[width=.45\columnwidth]{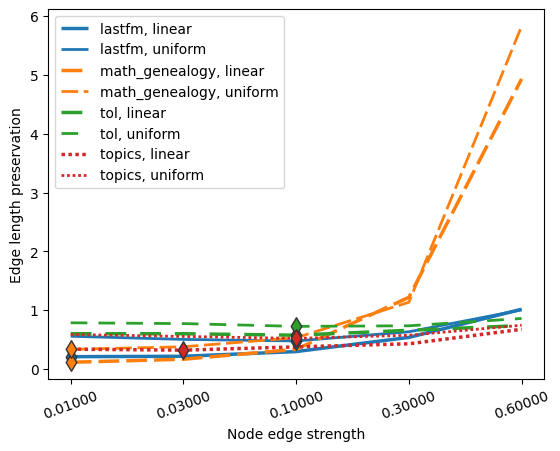}}
	\subfloat[]{\label{fig:nodeedge-compactness}\includegraphics[width=.45\columnwidth]{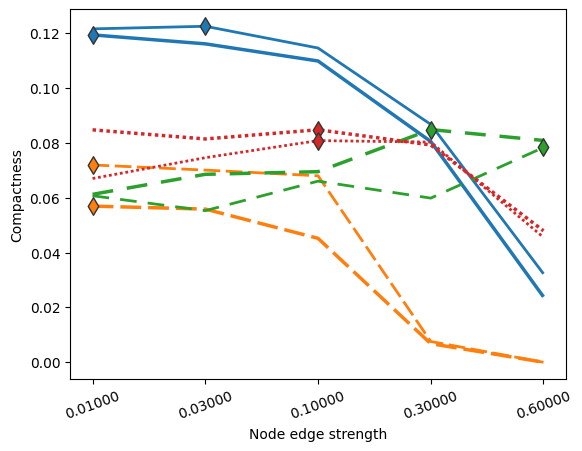}}
	\caption{Impact of node-edge force on the two quality metrics: Edge length (a) and compactness (b). Diamonds mark the best parameter setup for the specific network.
	}
	\label{fig:nodeedge}
\end{figure}

We also explore the strength of node-edge force defined in Sec.~\ref{sect:nodes-and-edges-force}. 
As shown in Fig.~\ref{fig:nodeedge}, both edge lengths preservation and compactness are better with a weak node-edge force. Setting node-edge force strength $S_ne= 0.1$ seems to provide a reasonable balance between the two metrics.

\subsection{Number of Iterations}
The force-directed improvement step of the algorithm and the final iteration both use iterative refinements; the number of iterations of each of these steps impacts the quality of the layouts and the runtime of the overall algorithm.

\begin{figure}[th]
	\centering
	\subfloat[]{\label{fig:iterations_overlaps}\includegraphics[width=.45\columnwidth]{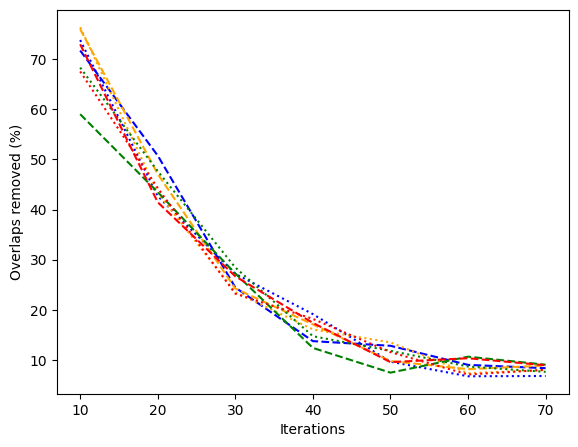}}
	\subfloat[]{\label{fig:iterations_time}\includegraphics[width=.45\columnwidth]{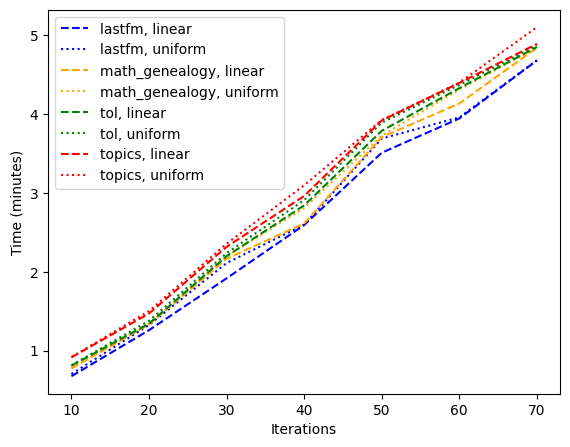}}
	\caption{Illustrating the impact of the number of iterations on overlaps and time.
	(a) shows the percentage of overlaps by iterations and (b) shows the time by the number of iterations.  We choose a value to balance the two considerations.}
	\label{fig:force_iterations}
\end{figure}

\medskip\noindent{\bf Number of Force-directed Iterations: }
The force-directed algorithm converges relatively quickly when initialized using Alg.~\mingweisInit. On the other hand, Alg.~{\fontfamily{lmtt}\selectfont Edge-Length-Initialization} needs more force-directed iterations to remove all label overlaps.
Hence we analyze the impact of the number of iterations when initializing with Alg.~{\fontfamily{lmtt}\selectfont Edge-Length-Initialization}. 
As illustrated in Fig.~\ref{fig:iterations_overlaps}, early iterations remove many overlaps, with diminishing returns after 40-50 iterations. On the other hand, the running time increases with the number of iterations as shown in Fig.~\ref{fig:iterations_time}. Hence, we set the number of iterations equal to 50.

\begin{figure}[th]
	\centering
	\subfloat[]{\label{fig:sample_time}\includegraphics[width=.45\columnwidth]{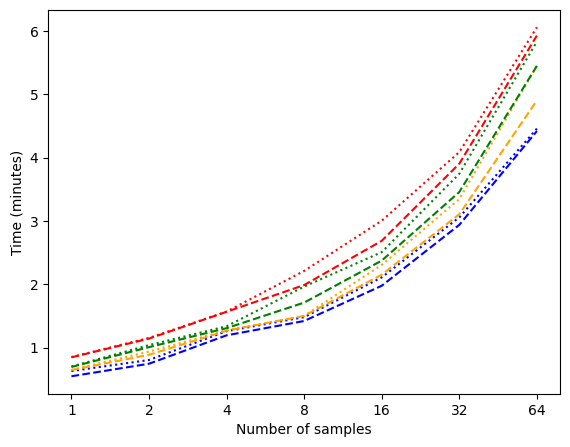}}
	\subfloat[]{\label{fig:width_samples}\includegraphics[width=.45\columnwidth]{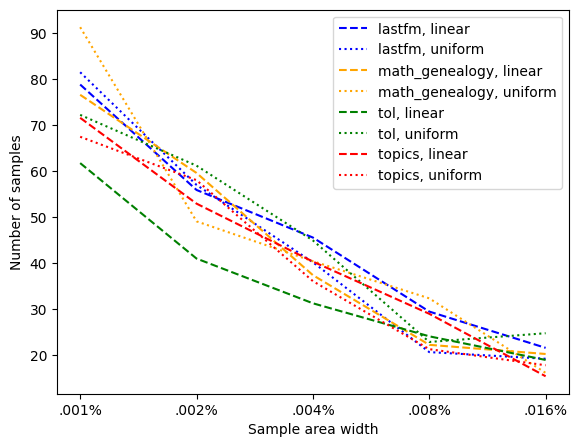}}
	\caption{Illustrating the impact of the number of samples and sample area width of the final overlap removal algorithm. (a) shows the running time by the number of samples.  (b) shows the width of the sample area for the number of samples needed.  We choose a value that allows for fewer samples to reduce the runtime.
	}
	\label{fig:postprocessing_params}
\end{figure}

\medskip\noindent{\bf Final Iteration: }
Since the force-directed improvement step terminates after 50 iterations, it may not have removed all overlaps. The final iteration step enforces the no-overlaps constraint (C2) by removing any remaining overlaps. Its performance depends on two  parameters: the number of samples and the width of the square sample area. 
As illustrated in Fig.~\ref{fig:sample_time}, the running time increases as the number of samples increases. We can reduce the number of samples by tuning the width of the sample area. We denote the width of the sample area by the percentage of the minimal square box that contains the drawing. As illustrated in Fig.~\ref{fig:width_samples}, as the width of the sample area increases, the number of needed samples is reduced. Hence, we set the sample width to $0.01-0.02\%$ of the total area and the number of samples to 20.

\section{Evaluation}
\label{sec:evaluation}

In this section, we evaluate different algorithms using four real-world datasets. 
We denote the RT algorithm by \reyansAlgoAcronym and \mingweisAlgoAcronym when initialized by Alg.~\reyansInit and Alg.~\mingweisInit, respectively. Similarly, we denote the PRT algorithm by \khaledsAlgoAcronymDelg and \khaledsAlgoAcronymCG when initialized by Alg.~\reyansInit and Alg. \mingweisInit, respectively.

\medskip\noindent{\bf Prior Methods: }
As there are no prior algorithms that guarantee the two constraints (no crossings, no overlaps), while optimizing desired edge lengths and compactness, it can be somewhat unfair to compare against prior approaches. Nevertheless, with some careful modifications (and clarifications) we can use existing tree/network layout algorithms in a comparison. We chose two such algorithms as described below.

GraphViz~\cite{ellson2001graphviz} can efficiently lay out a given tree or network with sfdp~\cite{hu2005efficient}, label the nodes and then remove overlaps via the PRoxImity Stress Model (PRISM)~\cite{gansner2009efficient}. Note that the output does not optimize given edge lengths and does not guarantee that trees are drawn in a crossings-free manner.
We denote this by \sfdpWithPrism. 

The yED~\cite{wiese2004yfiles} system provides several methods that can draw trees without edge crossings and optimize compactness. Note that yED does not optimize edge lengths and the only way to remove label overlap is to scale the drawing area.
We use the \textit{Circular Layout} (\yedCircular) in \textit{yED} as it produces the most compact layouts.

To provide a fair comparison, we consider two settings for the evaluation: one in which we have different desired edge lengths
and the other one considers uniform edge lengths (to make it possible to compare with \sfdpWithPrism and \yedCircular). 

	

\medskip\noindent{\bf Datasets:}
We extract seven networks from four datasets.

\textbf{Last.FM Network}~\cite{gansner2009putting}, extracted from the last.fm Internet radio station with $2588$ nodes and $28221$ edges.
The nodes are popular musical artists with weights corresponding to the number of listeners. Edges are placed between similar artists, based on listening habits. 

\textbf{Google Topics Network}~\cite{burd2018graph} has as nodes  research topics from Google Scholar's  profiles, with weights corresponding to the number of people working on them. Edges are placed between pairs of topics that co-occur in profiles. We work with two versions of this network: one with $34,741$ nodes and $646,565$ edges, and the other a smaller subset with $5001$ nodes.

\textbf{Tree of Life}\footnote{\url{http://tolweb.org/tree/}}, 
 extracted from the tree of life web project. 
 This dataset contains a node for every species, with an edge between two nodes representing the phylogenetic connection between the two. We work with two versions of this network: 
 one with $35,960$ nodes, and the other with $2,934$ nodes.
 
\textbf{Math Genealogy Network}\footnote{\url{https://genealogy.math.ndsu.nodak.edu/}},  every node represents a mathematician with edges capturing (advisor, advisee) relationship. 
We work with two versions of this network: one with $257,501$ nodes, 
and the other with $3,016$ nodes.

\medskip\noindent{\bf Dataset Processing: }
We compute the multi-level Steiner tree as described in Sec.~\ref{se:visualizationtool}. The details about the terminal selection method, number of levels, and desired edge lengths are provided in that section. We assign edge lengths, increasing linearly as we go from the lowest level to the top. Note that we can equivalently consider such a multi-level tree as a single-level tree with different edge lengths. Specifically, a multi-level tree has a hierarchical structure: all the nodes and edges of a particular level are also present in the lower levels. Hence, for each edge, we consider the highest level where the edge is present. We assign the desired edge length of this edge to the edge length of that level. 

With this in mind, we compare the performance of the four variants of our algorithm using a single-level tree, where all edges are present and have different desired edge lengths. 

Since the \sfdpWithPrism and \yedCircular methods cannot handle different edge lengths when comparing them to our algorithm, we consider the setting where the trees are given as above, but the edge lengths are uniform.


\begin{table*}[t]
\centering
\setlength\tabcolsep{3.5pt}
\resizebox{\textwidth}{!}{%
	\begin{tabular}{|l|r|l|r|r|r|r|r|r|r|r|r|r|r|r|r|r|r|r|r|r|r|r|r|r|}
		\hline
		\multicolumn{3}{|c|}{Network} 
		&\multicolumn{6}{c|}{\texttt{Desired Edge Length} $\downarrow$} 
		&\multicolumn{6}{c|}{\texttt{Compactness} $\uparrow$}
		&\multicolumn{8}{c|}{\texttt{Runtime (sec)} $\downarrow$}
		&\multicolumn{2}{c|}{\texttt{Crossings} $\downarrow$}
		\\ \hline 
		Name & $|V|$ & Edge Length 
		&\texttt{\yedCircular}&\texttt{\sfdpWithPrism}&\texttt{\reyansAlgoAcronym}&\texttt{\mingweisAlgoAcronym}&\texttt{\khaledsAlgoAcronymDelg}&\texttt{\khaledsAlgoAcronymCG}
		&\texttt{\yedCircular}&\texttt{\sfdpWithPrism}&\texttt{\reyansAlgoAcronym}&\texttt{\mingweisAlgoAcronym}&\texttt{\khaledsAlgoAcronymDelg}&\texttt{\khaledsAlgoAcronymCG}
		&\texttt{\yedCircular}&\texttt{\sfdpWithPrism}&\texttt{\reyansAlgoAcronym}&\texttt{\mingweisAlgoAcronym}&\texttt{\khaledsAlgoAcronymDelg} & \texttt{\khaledsAlgoAcronymDelgStar} & \texttt{\khaledsAlgoAcronymCG} & \texttt{\khaledsAlgoAcronymCgStar}
		&\texttt{\sfdpWithPrism}&\texttt{others}
		\\ \hline
\multirow{2}{*}{Last.FM} & \multirow{2}{*}{2,588} 

& uniform & 2.06 & 0.56 & \textbf{0.18} & 0.45 & 0.19 & 0.42
 & 3e-7 & 0.04 & 0.01 & \textbf{0.13} & 0.02 & 0.11
 & \textbf{5} & 8 & 926 & 309 & 54 & 15 & 26 & 7
 & 147 & \textbf{0} \\ \cline{3-25}

&& linear & - & - & \textbf{0.13} & 0.45 & 0.15 & 0.43 
 & - & - & 0.01 & \textbf{0.07} & 0.02 & 0.05 
 & - & - & 1635 & 233 & 55 & 15 & 31 & \textbf{9}
  & - & \textbf{0} \\ \hline 

\multirow{2}{*}{Topics} & \multirow{2}{*}{5,001} 

& uniform & 1.93 & 0.72 & \textbf{0.27} & 0.38 & 0.28 & 0.36
 & 7e-6 & 0.03 & 0.01 & \textbf{0.05} & 0.02 & 0.04
 & 44 & \textbf{22} & 3997 & 1000 & 175 & 32 & 142 & 23 
 & 203 & \textbf{0} \\ \cline{3-25}

&& linear & - & - & \textbf{0.14} & 0.30 & 0.16 & 0.28
 & - & - & 0.01 & \textbf{0.05} & 0.01 & 0.04
 & - & - & 8456 & 1096 & 178 &  32 & 147 & \textbf{25}
 & - & \textbf{0} \\ \hline 

\multirow{2}{*}{Tree of Life} & \multirow{2}{*}{2,934} 

& uniform & 1.24 & 0.53 & 0.43 & 0.49 & \textbf{0.42} & 0.47
 & 6e-6 & \textbf{0.09} & 0.01 & \textbf{0.09} & 0.02 & 0.08
 & \textbf{7} & \textbf{7} & 3157 & 333 & 62 & 36 & 43 & 21
 & 134 & \textbf{0} \\ \cline{3-25}

&& linear & - & - & \textbf{0.45} & 0.47 & 0.46 & 0.47
 & - & - & 0.01 &  \textbf{0.10} & 0.01 & 0.07
 & - & - & 1178 & 698 & 61 & 55 & 49 & \textbf{26}
 & - & \textbf{0} \\ \hline

\multirow{2}{*}{Math Genealogy} & \multirow{2}{*}{3,016} 

& uniform & 3.15 & 0.78 & \textbf{0.21} & 0.40 & 0.23 & 0.34
 & 3e-6 & \textbf{0.13} & 0.02 & 0.02 & 0.02 & 0.02
 & \textbf{5} & 6 & 2067 & 774 & 57  & 14 & 46 & 11
 & 1508 & \textbf{0} \\ \cline{3-25}

&& linear & - & - & \textbf{0.29} & 0.34 & 0.31 & 0.36
 & - & - & 0.02 & \textbf{0.03} & 0.02 & \textbf{0.03}
 & - & - & 1340 & 563 & 64 & 14 & 51 & \textbf{12}
  & - & \textbf{0} \\ \hline

\multirow{1}{*}{Topics (large)} & \multirow{1}{*}{34,758} 

& uniform & 3.88 & \textbf{0.76} & - & - & 0.84 & 0.91
 & 3e-6 & \textbf{0.005} & - & - & 0.001 & 0.004
 & 1259 & \textbf{258} & - & - & 9346 & 2925 & 8375 & 2658
 & 7713 & \textbf{0} \\ \hline

\multirow{1}{*}{Tree of Life (large)} & \multirow{1}{*}{35,960} 

& uniform & 2.17 & \textbf{0.83} & - & - & 1.29 & 1.36
 & 3e-6 & \textbf{0.006} & - & - & 0.001 & \textbf{0.006}
 & 792 & \textbf{455} & - & - & 8780 & 2184 & 7194 & 1856
 & 12088 & \textbf{0} \\ \hline

\multirow{1}{*}{Genealogy (large)} & \multirow{1}{*}{100,347} 

& uniform & - & 0.88 & - & - & \textbf{0.83} & 0.87
 & - & \textbf{0.007} & - & - & 0.002 & 0.006
 & - & \textbf{2041} & - & - & - & 8689 & - & 7675
 & 27478 & \textbf{0} \\ \hline

\end{tabular}
}
\caption{Quantitative algorithmic comparison using desired edge length preservation, compactness, and runtime for different algorithms. The $\downarrow$ next to a criterion indicates lower scores are better and $\uparrow$ indices that higher scores are better. \khaledsAlgoAcronymDelgStar and \khaledsAlgoAcronymCgStar refer to the experiments run on the Skylake server, while \khaledsAlgoAcronymDelg and \khaledsAlgoAcronymCG refer to the experiments run on the laptop.}
\label{tab:eval-all}
\end{table*}

\medskip\noindent{\bf Quantitative Evaluation: }
We measure the optimization goals: desired edge length preservation and
compactness, as well as the runtime.

\smallskip
\noindent \textbf{Desired Edge Length 
 (\desiredEdgeLengthAcronym):}  evaluates the normalized desired edge lengths in each layer. 
 Given the  desired edge lengths $\{l_{ij}: (i,j) \in E\}$, defined in Sec.~\ref{se:visualizationtool}, and coordinates of the nodes $X$ in the computed layout, we evaluate \desiredEdgeLengthAcronym with the following  formula:
    \begin{align}
    \text{\texttt{\desiredEdgeLengthAcronym}} &= \sqrt{ \frac{1}{|E|} \sum\limits_{(i,j) \in E}\;  
    \left(\frac{||X_i - X_j|| - l_{ij}}{l_{ij}}\right)^2} \label{eq:loss-desired-edge-length}
    \end{align}
    This measures the root mean square of the relative error, 
    producing a positive number, with $0$ corresponding to perfect preservation.

\smallskip
\noindent \textbf{Compactness Measure  (\compactnessAcronym):}  measures the ratio between the total areas of labels (the minimum possible area needed to draw all labels without overlaps) and the area of the actual drawing (measured by the area of the smallest bounding rectangle). CM scores are in the range $[0,1]$, where $1$ corresponds to perfect area utilization, this measure is the fourth found in the work by McGuffin and Robert~\cite{compact_measures}.  \begin{align}
    \text{\texttt{\compactnessAcronym}} &= \frac{ \sum\limits_{v \in V} \text{label\_area}(v) }{(X_{max,0}-X_{min,0})(X_{max,1}-X_{min,1})}
     \label{eq:loss-Compactness}
    \end{align}

\medskip\noindent\textbf{\blue{Experimental Environment:}} \blue{We conducted all but one experiment on a laptop, configured with {MacOS}, 2.3 GHz Dual-Core Intel Core i5, 8GB RAM, and 4 logical cores. The server is configured with Linux OS, Intel(R) Xeon(R) Platinum 8160 CPU @ 2.10GHz, 256GB RAM, 2 sockets, and 24 cores per socket. The remaining experiment used a Skylake server (indicated by \khaledsAlgoAcronymDelgStar and \khaledsAlgoAcronymCgStar). 
}

  \medskip\noindent\textbf{Results:}
We evaluate the performance of all algorithms on seven trees extracted from the four datasets above. Our two RT variants (algorithms \reyansAlgoAcronym and \mingweisAlgoAcronym) are applied only to the 4 small trees. The two PRT variants, sfdp+p, and CIR are applied to all 7 trees. Note that for each of the 4 small trees we consider two different desired edge lengths: the {\em linear edge setting} used to drive the interactive, zoomable visualization, and the {\em uniform edge setting} which makes it possible to compare our algorithms to the prior ones (sfdp+p and CIR).
We show all Last.fm trees layouts (with the uniform edge length setting) 
in Fig.~\ref{fig:compare-lastfm}.  
We provide additional layouts in the supplementary materials.
These figures highlight some significant differences which stand out visually, which we discuss in Sec.~\ref{sec:qualitative}. 

We provide all quantitative data in Table~\ref{tab:eval-all}.
From these results, we can see that in the {\em linear edge length setting} \reyansAlgoAcronym does best in all instances. As expected, \khaledsAlgoAcronymDelg provides similar scores. When looking at the compactness measure, \mingweisAlgoAcronym performs best and, again, the parallel variant \khaledsAlgoAcronymCG performs nearly as well. Blank entries in the Desired Edge Length or Compactness scores indicate that the experiment was not performed for this setting (e.g., the linear setting for sfp+p and CIR) or the computation did not terminate by the maximum time limit of 8 hours (e.g., CIR with the large Math Genealogy tree).

Overall, the RT variants provide slightly better results than the PRT variants, both in edge length preservation and in compactness. We believe this due to RT utilizing the fine-tuned force-directed algorithm in d3.js, while the PRT force-directed algorithm is our own, and not as well tuned.

The PRT variants are  usually more than an order of magnitude faster than the RT variants. This is due to parallelizing the underlying computations and implementation in C++, which is faster than d3.js.
Note that we report four parallel runtimes: \khaledsAlgoAcronymDelg, \khaledsAlgoAcronymCG (running on a laptop), and \khaledsAlgoAcronymDelgStar, \khaledsAlgoAcronymCgStar (running on a server). 
Blank entries in the runtime columns indicate that the algorithm did not terminate by the maximum time limit of 8 hours. 

Recall that the {\em uniform edge length setting} for the small networks makes it possible to compare the prior algorithms (sfdp+p and CIR) against PR and PRT. Here either \reyansAlgoAcronym or \khaledsAlgoAcronymDelg performs best in desired edge length preservation and the scores of the two algorithms are similar. On the other hand, \mingweisAlgoAcronym and \khaledsAlgoAcronymCG  performs well in compactness, although \sfdpWithPrism outperforms our algorithms in some instances (at the expense of edge crossings). The \yedCircular method is the fastest in most of the instances (at the expense of both compactness and edge lengths), and the running times of \sfdpWithPrism and {\khaledsAlgoAcronymCgStar} are comparable. 

For the 3 large networks, we only consider the uniform edge setting. The desired edge length scores of \sfdpWithPrism and \khaledsAlgoAcronymDelg are comparable, with \sfdpWithPrism outperforming \khaledsAlgoAcronym in two of the three cases. Similarly, \sfdpWithPrism and \khaledsAlgoAcronymCG are the best  in compactness. \sfdpWithPrism  is the clear winner in speed, at the expense of thousands of edge crossings; see last two columns of Table~\ref{tab:eval-all}.

\section{Multi-Level Interactive Visualization}
\label{se:visualizationtool}
The focus of this paper is the algorithmic framework for creating readable tree layouts. However, on a desktop, laptop, or phone screen, even {\em viewing} a large tree with thousands of labeled nodes, requires more than  a good layout.

With this in mind, we process all trees and their layouts further, to provide an interactive visualization environment. The idea is to create a hierarchy of trees, starting with the input and extracting progressively smaller trees that capture more and more abstract views, recalling  interactive geographic maps. Important nodes (like large cities) and edges (like highways) are present in all representation levels. Less important nodes (like smaller towns) and edges (like smaller roads) appear when zooming in.
We create such a hierarchy using  multi-level Steiner trees and combine the results with the readable tree layout and an interactive map-like environment. Note that this approach is applicable to arbitrary networks and not just trees.

\subsection{Multi-Level Steiner Trees}
\label{sect:mlst}

A Steiner tree  minimizes the total weight of the subtree spanning a given subset of nodes (called the terminals). The multi-level Steiner tree problem is a generalization of the Steiner tree problem where the objective is to minimize the sum of the edge weights at all levels. As both problems are NP-hard, we use  approximation algorithms that have been shown to work well in practice~\cite{Ahmed-JEA-19}. 

Formally, given a node-weighted and edge-weighted network  $G=(V, E)$, we want to visualize $G$ with the aid of a hierarchy of progressively larger trees $T_1=(V_1, E_1) \subset T_2=(V_2, E_2) \subset \dots \subset T_n=(V_n, E_n) \subseteq  G$, such that $V_1 \subset V_2 \subset \dots \subset V_n = V$ and $E_1 \subset E_2 \subset \dots \subset E_n \subset E$. We use multi-level Steiner trees in order to make the hierarchy  representative of the underlying network, based on a node filtration  $V_1 \subset V_2 \subset \dots \subset V_n = V$ with the most important nodes (highest weight) in $V_1$, the next most important nodes added to form $V_2$, and so on. A solution to the multi-level Steiner tree problem then creates the set of progressively larger trees $T_1=(V_1, E_1) \subset T_2=(V_2, E_2) \subset \dots \subset T_n=(V_n, E_n) \subseteq  G$ using the most important (highest weight) edges. 

\subsection{Dataset Processing}
Many real-life networks come with well-defined notions of the importance of nodes and edges. Node importance can be determined by the number of listeners for a specific band in the Last.FM network, or by the number of researchers of a specific topic in the Google Topics network. In the absence of such node information, importance can be computed based on structural properties of the network, such as degree centrality, eigenvector centrality, etc~\cite{bonacich2007some}. Similarly, edge importance can be given (e.g., number of listeners of a pair of bands in the Last.FM network, number of researchers listing a pair of topics in the Google Topics network), or can be computed based on structural properties (e.g., betweenness centrality, PageRank).
\smallskip

\noindent \textbf{Node Weights:}
The Last.FM network and the topics network have node weights (number of listeners and number of researchers, respectively) and we use the heavy nodes in the higher levels and lighter nodes in the lower levels. 
For the tree of life and math genealogy networks, we set the node weight equal to the node degree (degree centrality).

We select the terminals of the multi-level Steiner tree instance according to the node weights: higher-level terminal sets contain  heavy nodes since the larger the weight is the more important the node is. The size of the terminal sets grows linearly. If we have $n$ nodes and $h$ levels then the top terminal set contains the most important $n/h$ nodes. The next terminal set contains all the nodes of the top terminal set, together with the next $n/h$ important nodes. We continue the process until the bottom level when all nodes are present.

\smallskip

\noindent \textbf{Edge Weights:} 
Edge weights in the Last.FM network are given by the number of listeners of the corresponding pair of bands. Similarly, edge weights in the Google Topics network are the number of researchers listing the corresponding pair of topics. 
Edge lengths in the Tree of Life network are given by the phylogenetic distance between the two endpoints.
The math genealogy network is unweighted, and we use uniform edge lengths. 



\smallskip
\noindent \textbf{Desired Edge Length Settings:}
The edge weights described above can be used for desired edge lengths by computing the reciprocal of the original edge weights (converting similarities into dissimilarities). We have used this setting in experimenting with our algorithms but this setting is not what we report in the paper as it does not lend itself to semantic zooming  and prior methods cannot handle desired edge lengths.

The simplest desired edge length setting used in the paper is the {\em uniform edge length setting}. We need such a setting to be able to compare the performance of our algorithms against those of prior methods (sfdp+p and CIR) that do not take edge lengths into account.

We can use the given edge weights and combine them with the multi-level Steiner tree solution to create edge lengths that work well with semantic zooming. This is the {\em linear edge setting} used in some of the experiments discussed in the paper.
While in the uniform setting, edge lengths are set to 200 in the linear setting the edge lengths depend on the level on which the edge first appears in the multi-level Steiner tree (which in turn depends on the underlying edge weights). The desired edge length on the lowest level is $l_{min}=200$ and this value is increased by $l_{add}$ each time we got to a higher level. In our examples, $l_{add}$ varies depending on the number of levels.  
We use a different number of levels for different networks, with more levels for larger networks.
For the smallest dataset, the Last.FM network, we have only 8 levels while for the largest dataset, the math genealogy tree, there are more than 100 levels.

\subsection{Map-like Visualization}
From the tree layout, we generate a map using the GMap system~\cite{gansner2010gmap}. The map generation method depends on a clustering step, and by default we use the MapSets~\cite{kobourov2014visualizing} clustering technique. Our system uses OpenLayers~\cite{openlayers} with
zooming and panning provided via buttons, mouse scrolling, or through the mini-map.
When viewing level $i$, all nodes at this or higher levels are labeled and there are no label overlaps.
Edge widths are determined based on their levels: higher-level edges are thicker, and lower-level edges are thinner.
A search bar allows for direct queries with auto-complete suggestions and clicking on a search result recenters the map on the selected node.
By default, we show labels with at most 16 characters (truncating longer ones) but the full label is shown on a mouse-over event.
Node attributes and edge attributes  are provided when clicking on the node/edge. An example of this visualization can be seen in Fig.~\ref{fig:teaser}.

\section{Discussion}
\label{sec:qualitative}

Comparing our algorithms shows that they  do well in their respective optimization goals. 
Table~\ref{tab:eval-all} confirms that the \reyansAlgoAcronym and \khaledsAlgoAcronymDelg algorithms outperform the other two in desired edge length preservation, \mingweisAlgoAcronym and \khaledsAlgoAcronymCG algorithms outperform the other two in compactness, and \khaledsAlgoAcronymCgStar is the fastest of the variants.

Comparing the new algorithms to the two prior ones (only possible when using uniform edge lengths) also seem encouraging, even though \sfdpWithPrism introduces crossings in all layouts and \yedCircular has compactness scores that are orders of magnitude worse:
\begin{itemize}
    \item \reyansAlgoAcronym and \khaledsAlgoAcronymDelg outperform \sfdpWithPrism w.r.t.  edge lengths on all 4 small networks
    \item \mingweisAlgoAcronym and \khaledsAlgoAcronymCG outperform \sfdpWithPrism w.r.t.  compactness on 2/4 small networks, and they are almost equal in another dataset

    \item \reyansAlgoAcronym and \khaledsAlgoAcronymDelg outperform \yedCircular w.r.t. edge lengths on all 4 small networks
    \item \mingweisAlgoAcronym and \khaledsAlgoAcronymCG outperform \yedCircular w.r.t. compactness on all 4 small networks
    \item \khaledsAlgoAcronymDelgStar and \khaledsAlgoAcronymCgStar are slower than both \sfdpWithPrism and \yedCircular on the large networks, but not by much.
\end{itemize}

\subsection{Qualitative Analysis}

Here we take a closer look at the results returned by the new algorithms and the two prior ones.
%
Fig.~\ref{fig:compare-lastfm} shows the results of 
last.FM networks. 
As the name implies, the circular layout (\yedCircular) wraps branches of the tree into spiraling circles to form a compact layout.
Edges close to the center of the tree are stretched in order to provide large areas for the subtrees, resulting in poor edge length preservation. 
Since leaves are drawn in small regions, the overall layout must be scaled a lot in order to show the labels without overlaps, yielding poor compactness.
The \sfdpWithPrism results are consistently good in compactness, at the expense of many crossings.
Our new methods are better at capturing the global structure. 

\begin{figure}[thp]
	\centering
	\subfloat[\reyansAlgoAcronym]{\label{fig:edge-length-analysis-reyan}\includegraphics[width=.45\columnwidth]{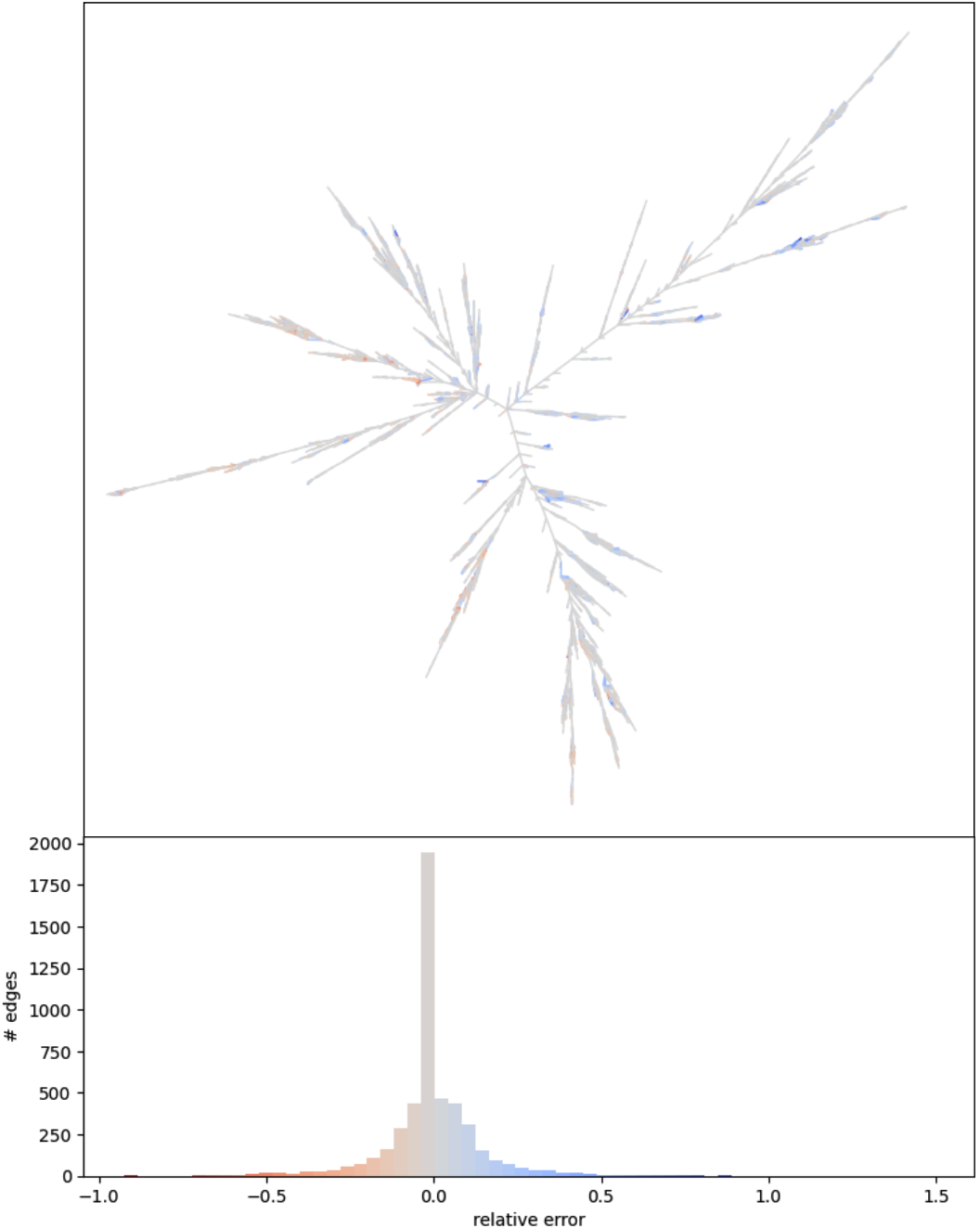}}
	\subfloat[\mingweisAlgoAcronym]{\label{fig:edge-length-analysis-mw}\includegraphics[width=.43\columnwidth]{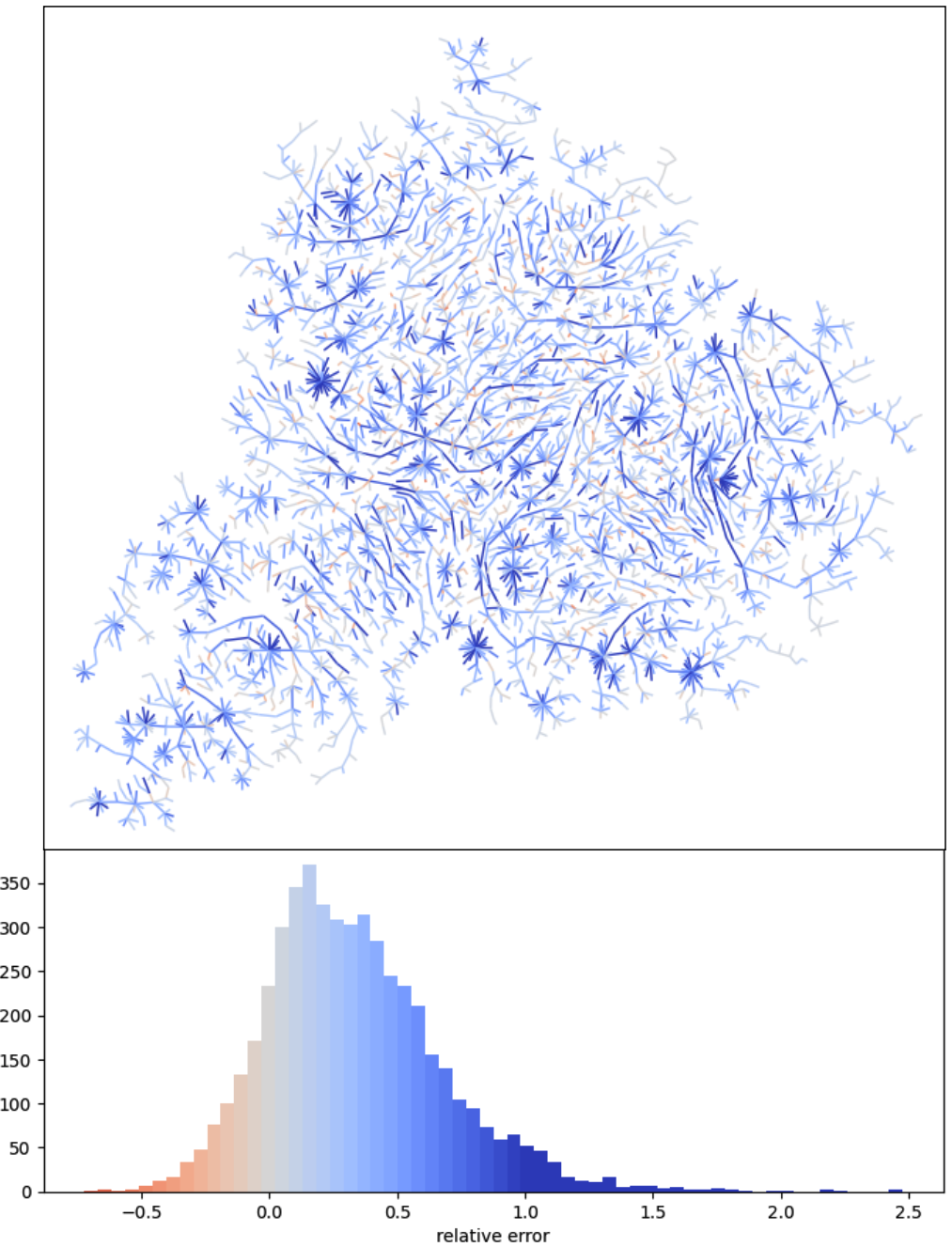}}
	\caption{Analysis of failure modes in desired edge length preservation. \textbf{Top left:} In the layout of the topic network computed by  \reyansAlgoAcronym, most edges are drawn with their desired edge lengths. \textbf{Bottom left:} A histogram of relative errors in desired edge length preservation, colored in the same way as the layout. \textbf{Right}: The same analysis on the \mingweisAlgoAcronym algorithm, showing \mingweisAlgoAcronym stretches (blue) more edges to improve compactness.
	}
	\label{fig:edge-length-analysis}
	\vspace{-10pt}
\end{figure}

Next, we look more closely at the layouts obtained from \reyansAlgoAcronym and \mingweisAlgoAcronym.
First, we focus on desired edge length preservation. Fig.~\ref{fig:edge-length-analysis} colors individual edges in the layout by their relative error. 
Recall that we use the relative error to measure the desired length preservation in Eqn.~\ref{eq:loss-desired-edge-length}.
For each edge $(u,v) \in E$, it measures the discrepancy between the actual edge length $||X_u - X_v||$ in the drawing and the given desired edge length $l_{uv}$:
$\text{relative\_error(u,v)} = (||X_u - X_v|| - l_{uv})/l_{uv}.$

With \reyansAlgoAcronym, we observed from the left layout and histogram in Fig.~\ref{fig:edge-length-analysis} that most edges are drawn with their desired edge lengths due to its edge-length guided initialization. 
We can see a few stretched or compressed edges (colored in blue and red in Fig.~\ref{fig:edge-length-analysis}), but since most of the other edges are drawn near perfectly in terms of edge length, the drawing has a low variation in relative error.
The errors are larger in \mingweisAlgoAcronym,
which seems to be due to the way label overlaps are handled: in dense regions, e.g. around high-degree nodes, the edges are more likely to be stretched, whereas on the periphery the edge lengths are better preserved.

\begin{figure}[thp]
	\centering
	\includegraphics[width=0.90\columnwidth]{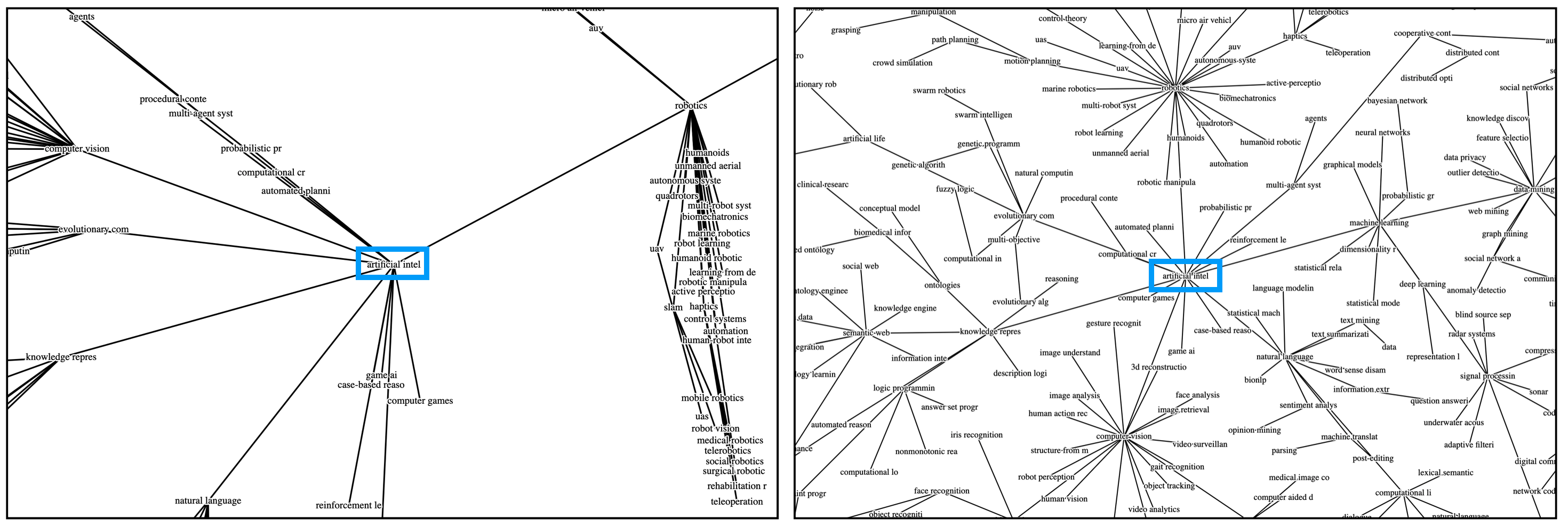}
	\caption{
	    Analysis of failure modes in layout compactness. \textbf{Left:} The topic network layout returned by \reyansAlgoAcronym  has more unused space and fails to distribute heavy subtrees around the `Artificial Intelligence' node (pointed and circled in blue) evenly. \textbf{Right:} \mingweisAlgoAcronym  generates a more balanced layout and is able to distribute subtrees more evenly.
	}\label{fig:compactness-analysis}
\end{figure}

Next, we look at the compactness of the two layouts. 
Consider the difference between the two algorithms in their rendering of the region around the `Artificial Intelligence' node in the maps, as shown in Fig.~\ref{fig:compactness-analysis}. 
From the layout overview in Fig.~\ref{fig:compactness-analysis}, we can already see that \reyansAlgoAcronym  uses space less efficiently, it has more empty, white space. 
The `Artificial Intelligence' node, highlighted in blue in Fig.~\ref{fig:compactness-analysis}, is close to multiple heavy subtrees such as those from the nodes `natural language processing', `machine learning', and `computer vision'. 
\mingweisAlgoAcronym distributes the heavy trees evenly, due to its initial layout.
 \reyansAlgoAcronym, on the other hand, places most heavy branches on the left side, resulting in less efficient use of the drawing area.

\begin{figure}
    \centering
    \includegraphics[width=0.4\linewidth,height=4.0cm]{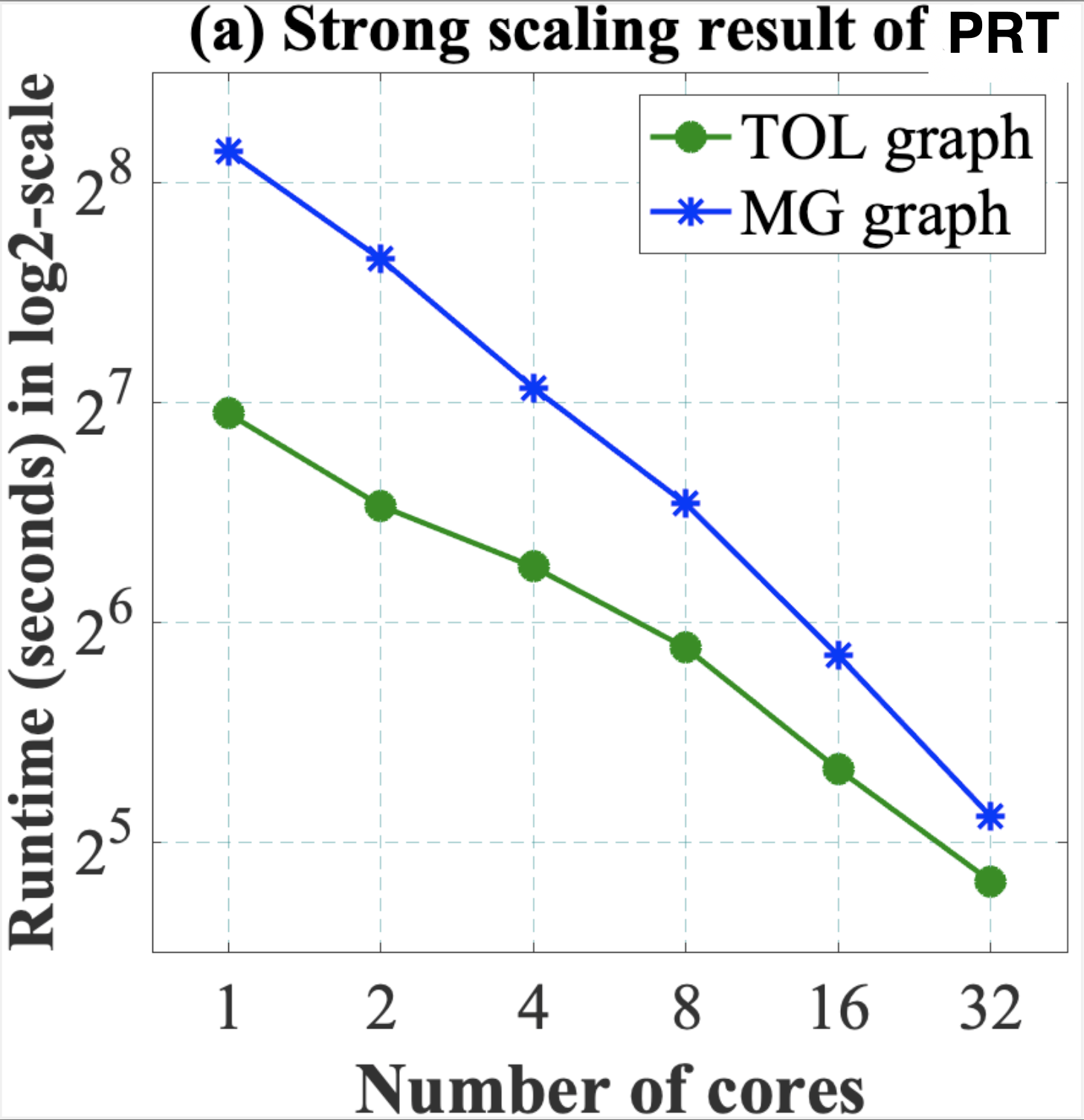}
    \includegraphics[width=0.4\linewidth,height=4.0cm]{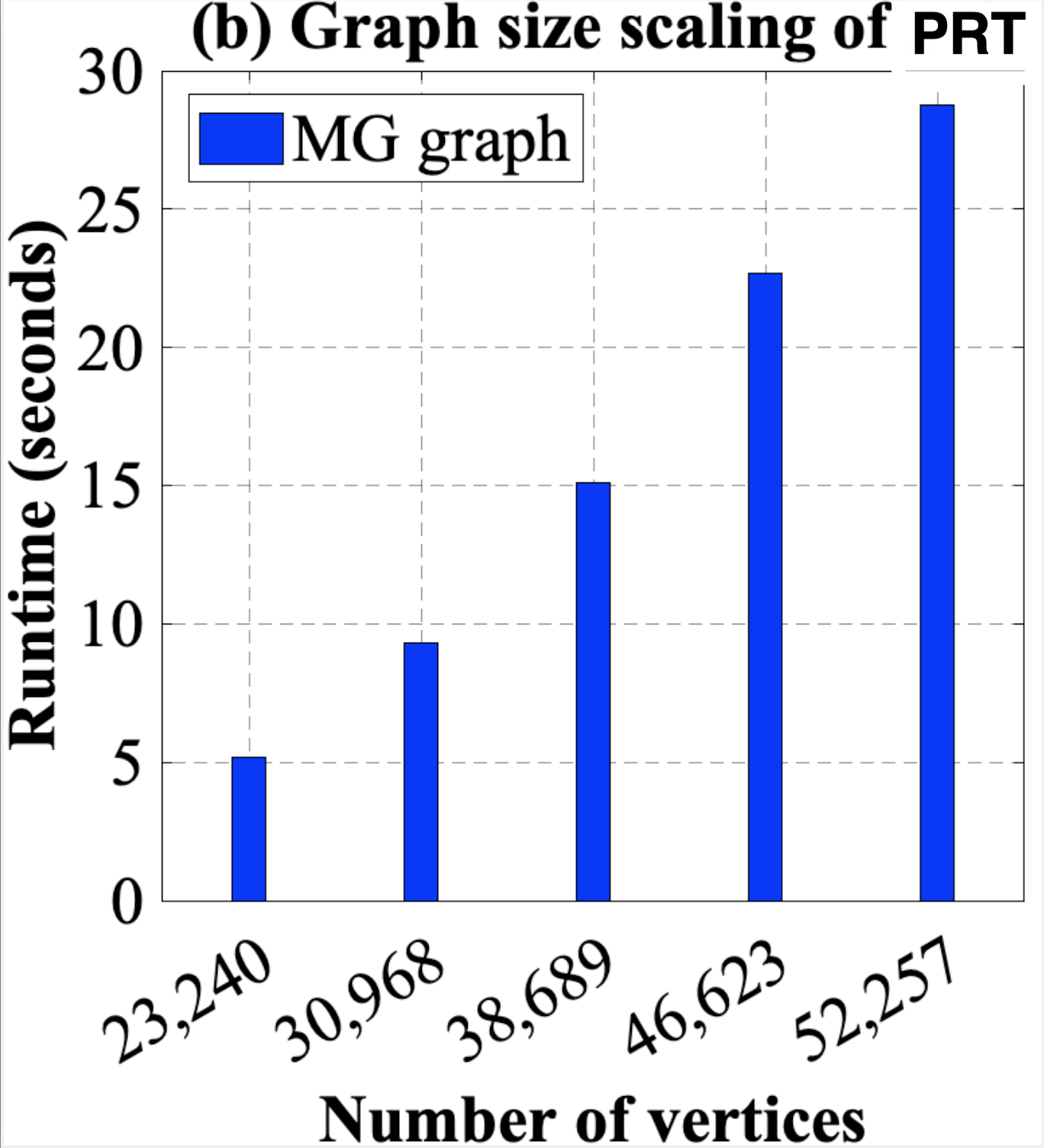}
    \caption{(a) Strong scaling results of Tree of Life (TOL) dataset (35,960 nodes) and Math Genealogy (MG) dataset (52,257 nodes). (b) Network scaling results for different size of MG trees using 48 cores. Only per-iteration runtime is reported.}
    \label{fig:scalability}
\end{figure}

\subsection{Scalability} 

We experimented further with the \khaledsAlgoAcronym algorithm to evaluate how it behaves with a larger number of cores and with a larger number of nodes, shown in Fig.~\ref{fig:scalability}(a) and Fig.~\ref{fig:scalability}(b), respectively. In Fig.~\ref{fig:scalability}(a), we show strong scaling results for the Tree of Life and Math Genealogy datasets. For both datasets, we observe that the runtime decreases almost linearly as the number of cores increases. In Fig.~\ref{fig:scalability}(b), we report the per-iteration runtime on the Math Genealogy tree when increasing the size of the trees. We observe that the runtime increases almost linearly with the size of the tree. This provides support for the scalability of the \khaledsAlgoAcronym algorithm. 
\blue{It is notable that while  \khaledsAlgoAcronym does not outperform RT, \khaledsAlgoAcronym's numbers are very close to the best values; see Table~\ref{tab:eval-all}. For example, we can see that in the {\em linear edge length setting} the desired edge length measure for \khaledsAlgoAcronymDelg is 0.15 and  \reyansAlgoAcronym is 0.13. In general, most \khaledsAlgoAcronymCG and  \mingweisAlgoAcronym values are within a few percentage points.}


\begin{figure}[H]
    \centering
    \fbox{\includegraphics[width=0.8\linewidth]{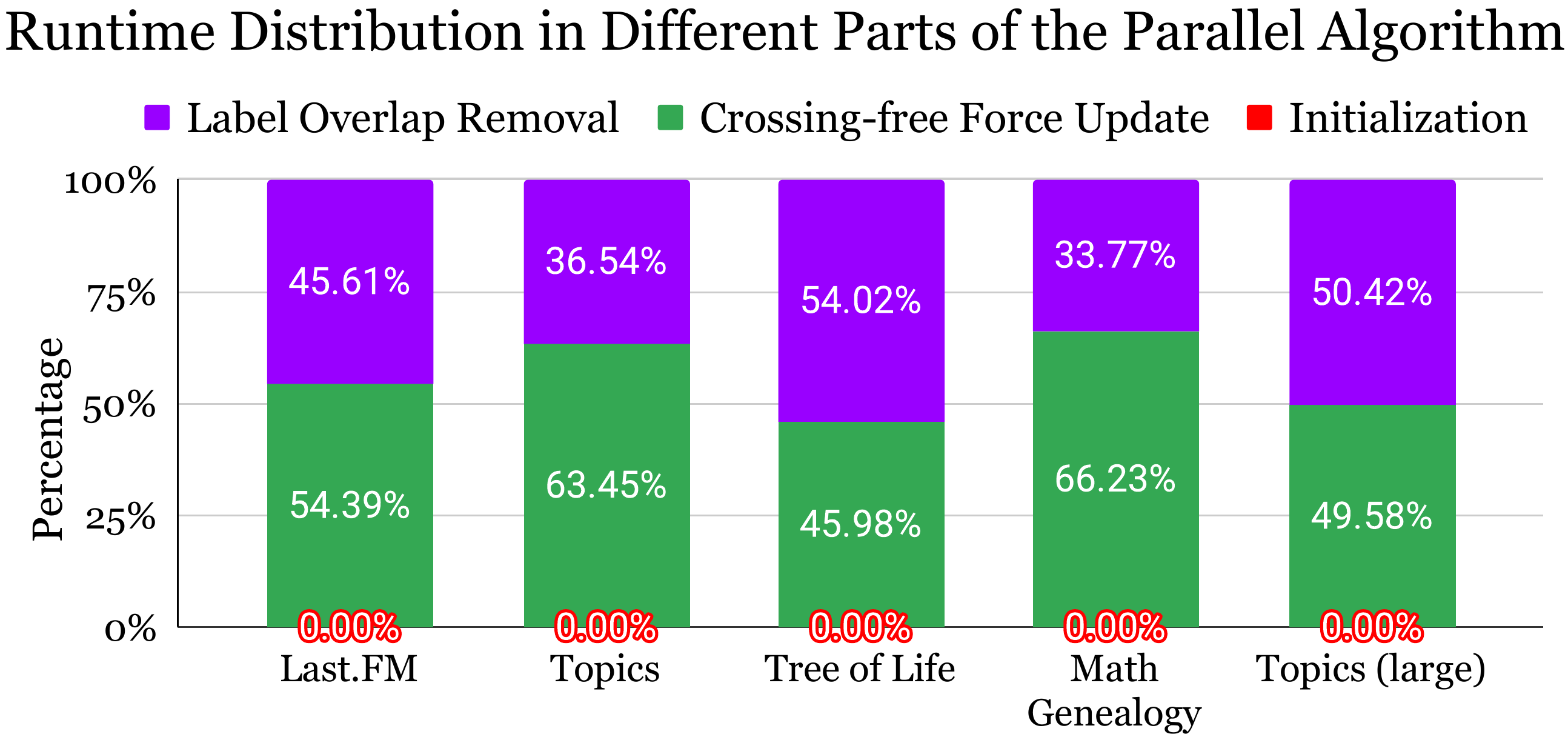}}
    \caption{Runtime distribution of different steps of the parallel \khaledsAlgoAcronym Algorithm for several input trees.}
    \label{fig:runtime_distribution}
\end{figure}

\medskip\noindent\textbf{Runtime Analysis of the \khaledsAlgoAcronym Algorithm}

There are three main steps in the \khaledsAlgoAcronym algorithm: (i) Initialization, (ii) Parallel Force-directed Improvement of Layout, and (ii) Parallel Label Overlap Removal. In Fig.~\ref{fig:runtime_distribution}, we report the percentage of total runtime spent in the different steps for several input trees. In all cases, we observe that the initialization step takes little time (less than a millisecond). This is expected since we run this step only once and the most time-consuming part of this step (finding center node by Alg.~{\fontfamily{lmtt}\selectfont PRT-Center-Node}) is fully parallelized. On the other hand, we iteratively run the crossing-free force update (parallel force-directed improvement) and label overlap removal steps multiple times. The force-directed update and label overlap removal steps consume almost 50\% of the total runtime for the large Topics tree, respectively.

\section{Limitations}
While we attempted to compare our readable tree layout framework with prior algorithms, we only compared against two. Similarly, we used only four real-world datasets and seven trees extracted from them for our evaluation. Further experiments with different types of trees, and with synthetically generated trees that test the limits of the prior and proposed methods (e.g., with respect to balance, degree distribution, diameter, etc.) are needed.

The utility of crossings-free, compact layouts that capture pre-specified edge lengths and show all node labels without overlaps needs to be evaluated. While intuitively these seem like  desirable features (non-overlapping labels make for readable layouts, non-crossing layouts help grasp the underlying structure, compact layouts require less panning and zooming), a human subjects study can further validate these goals.


We use a simple overlap removal technique to improve the compactness of the algorithm. Overlap removal is a well-studied problem and there are several advanced algorithms~\cite{marriott2003removing,gansner2009efficient,nachmanson2017node}. Modifying such  algorithms to ensure they do not change the topology of the current layout (e.g., by introducing edge crossings) remains an interesting open problem.

Since we optimize compactness, our algorithms might create zigzag-like paths,
making it difficult to quickly estimate paths lengths and to compare different paths. Balancing the need for compactness with such distortions requires further examination. 

Force directed algorithms often have a large parameter space.  We add on to this with parameters from two initial layouts and a final overlap step.  This gives us a very large parameter space.  While we have done some small-scale parameter space searches, a more careful and detailed analysis will likely yield better results.

\blue{Balancing multiple and contradictory optimization goals (e.g., compactness and edge length preservation) is difficult, and other optimization goals may create better layouts. We note that most of the ``squished'' layouts tend to result from the {\fontfamily{lmtt}\selectfont Edge-Length-Initialization} and can be avoided when selecting {\fontfamily{lmtt}\selectfont Compact-Initialization}.  
The quality of the layouts degrades in the parallel version (PRT) since some advanced force-directed features are not available in OpenMP~\cite{chandra2001parallel}.
A better implementation of the force-directed algorithms as well as consideration of other objectives besides desired edge length preservation and compactness remains a worthwhile future direction.
}


\section{Conclusions}
Both the quantitative evaluation and the visual analysis provide evidence of the utility of the proposed Readable Tree (RT) framework. The parallel version (PRT) makes the framework applicable to large instances, without much degradation in quality (edge length preservation and compactness).  

Comparing RT and PRT with \sfdpWithPrism and \yedCircular
also seems encouraging. 
On all 4 small networks
\reyansAlgoAcronym outperforms \sfdpWithPrism and \yedCircular in edge length preservation;
\mingweisAlgoAcronym outperforms \yedCircular and \sfdpWithPrism in compactness in most cases;
\khaledsAlgoAcronym is slower than both \sfdpWithPrism and \yedCircular, but not by much, especially given that \sfdpWithPrism introduces crossings in all layouts and \yedCircular has compactness scores that are orders of magnitude worse. 


Even though the layout of trees is a well-known and arguably solved problem, the \textit{readable tree layout problem} shows that there is more work to be done in this domain. We propose an algorithmic framework for creating readable tree layouts that guarantee crossings-free layouts with non-overlapping node labels. Our RT algorithm works well on smaller networks and can be executed on any computer, and our PRT algorithm  speeds up the computation making it applicable to larger networks but relies on more advanced hardware.
The utility of such algorithms goes beyond drawings of trees to help provide an interactive exploration of large networks, as illustrated by several examples and a video on the project website\footnote{
\url{https://tiga1231.github.io/zmlt/demo/overview.html}
}.  All source code, datasets, and analysis can be found at \url{https://github.com/abureyanahmed/multi_level_tree}.  

\section*{Acknowledgments}

This work was supported in part by NSF grants CCF-1740858, CCF-1712119, and DMS-1839274, a DOE grant DE-SC0022098.

\bibliographystyle{IEEEtranS}
\bibliography{references}

\begin{thebibliography}{10}
\providecommand{\url}[1]{#1}
\csname url@samestyle\endcsname
\providecommand{\newblock}{\relax}
\providecommand{\bibinfo}[2]{#2}
\providecommand{\BIBentrySTDinterwordspacing}{\spaceskip=0pt\relax}
\providecommand{\BIBentryALTinterwordstretchfactor}{4}
\providecommand{\BIBentryALTinterwordspacing}{\spaceskip=\fontdimen2\font plus
\BIBentryALTinterwordstretchfactor\fontdimen3\font minus
  \fontdimen4\font\relax}
\providecommand{\BIBforeignlanguage}[2]{{%
\expandafter\ifx\csname l@#1\endcsname\relax
\typeout{** WARNING: IEEEtranS.bst: No hyphenation pattern has been}%
\typeout{** loaded for the language `#1'. Using the pattern for}%
\typeout{** the default language instead.}%
\else
\language=\csname l@#1\endcsname
\fi
#2}}
\providecommand{\BIBdecl}{\relax}
\BIBdecl

\bibitem{Ahmed-JEA-19}
R.~Ahmed, P.~Angelini, F.~Sahneh, A.~Efrat, D.~Glickenstein, M.~Gronemann,
  N.~Heinsohn, S.~Kobourov, R.~Spence, J.~Watkins, and A.~Wolff, ``Multi-level
  {S}teiner trees,'' \emph{{ACM} Journal of Experimental Algorithmics},
  vol.~24, no.~1, pp. 2.5:1--2.5:22, 2019.

\bibitem{arleo2018distributed}
A.~Arleo, W.~Didimo, G.~Liotta, and F.~Montecchiani, ``A distributed multilevel
  force-directed algorithm,'' \emph{IEEE Transactions on Parallel and
  Distributed Systems}, vol.~30, no.~4, pp. 754--765, 2018.

\bibitem{Bachmaier2005}
C.~Bachmaier, U.~Brandes, and B.~Schlieper, ``Drawing phylogenetic trees,'' in
  \emph{ISAAC'05: Proceedings of the International Symposium on Algorithms and
  Computations}, ser. Lecture Notes in Computer Science, X.~Deng and D.~Du,
  Eds.\hskip 1em plus 0.5em minus 0.4em\relax Springer, 2005, pp. 1110--1121.

\bibitem{ballen2017walking}
C.~J. Ballen and H.~W. Greene, ``Walking and talking the tree of life: Why and
  how to teach about biodiversity,'' \emph{PLoS biology}, vol.~15, no.~3, p.
  e2001630, 2017.

\bibitem{bastian2009gephi}
M.~Bastian, S.~Heymann, and M.~Jacomy, ``Gephi: An open source software for
  exploring and manipulating networks,'' in \emph{International AAAI Conference
  on Web and Social Media}, 2009.

\bibitem{bavelas1950communication}
A.~Bavelas, ``Communication patterns in task-oriented groups,'' \emph{The
  journal of the acoustical society of America}, vol.~22, no.~6, pp. 725--730,
  1950.

\bibitem{blanch2015dendrogramix}
R.~Blanch, R.~Dautriche, and G.~Bisson, ``Dendrogramix: A hybrid tree-matrix
  visualization technique to support interactive exploration of dendrograms,''
  in \emph{2015 IEEE Pacific Visualization Symposium (PacificVis)}.\hskip 1em
  plus 0.5em minus 0.4em\relax IEEE, 2015, pp. 31--38.

\bibitem{bonacich2007some}
P.~Bonacich, ``Some unique properties of eigenvector centrality,'' \emph{Social
  networks}, vol.~29, no.~4, pp. 555--564, 2007.

\bibitem{doi:10.1177/1473871615594652}
K.~B\"{o}rner, A.~Maltese, R.~N. Balliet, and J.~Heimlich, ``Investigating
  aspects of data visualization literacy using 20 information visualizations
  and 273 science museum visitors,'' \emph{Information Visualization}, vol.~15,
  no.~3, pp. 198--213, 2016.

\bibitem{bostock2011d3}
M.~Bostock, V.~Ogievetsky, and J.~Heer, ``D$^3$ data-driven documents,''
  \emph{IEEE transactions on visualization and computer graphics}, vol.~17,
  no.~12, pp. 2301--2309, 2011.

\bibitem{bkk-mbs-05}
K.~W. Boyack, R.~Klavans, and K.~B{\"o}rner, ``Mapping the backbone of
  science,'' \emph{Scientometrics}, vol.~64, no.~3, pp. 351--374, 2005.

\bibitem{brandes2007eigensolver}
U.~Brandes and C.~Pich, ``Eigensolver methods for progressive multidimensional
  scaling of large data,'' in \emph{Graph Drawing}, Springer.\hskip 1em plus
  0.5em minus 0.4em\relax Springer, 2007, pp. 42--53.

\bibitem{burd2018graph}
R.~Burd, K.~Espy, I.~Hossain, S.~Kobourov, N.~Merchant, and H.~Purchase,
  ``{GRAM}: Global research activity map,'' in \emph{Intl.~Conference on
  Advanced Visual Interfaces (AVI)}.\hskip 1em plus 0.5em minus 0.4em\relax
  ACM, 2018, pp. 31:1--31:9.

\bibitem{chandra2001parallel}
R.~Chandra, L.~Dagum, D.~Kohr, R.~Menon, D.~Maydan, and J.~McDonald,
  \emph{Parallel Programming in OpenMP}.\hskip 1em plus 0.5em minus 0.4em\relax
  Morgan kaufmann, 2001.

\bibitem{chimani2011}
M.~Chimani, C.~Gutwenger, M.~J{\"u}nger, G.~W. Klau, K.~Klein, and P.~Mutzel,
  ``The open graph drawing framework {(OGDF)},'' \emph{Handbook of Graph
  Drawing and Visualization}, pp. 543--569, 2011.

\bibitem{phylodraw}
\BIBentryALTinterwordspacing
J.-H. Choi, H.-Y. Jung, H.-S. Kim, and H.-G. Cho, ``{PhyloDraw: a Phylogenetic
  Tree Drawing System },'' \emph{Bioinformatics}, vol.~16, no.~11, pp.
  1056--1058, 11 2000. [Online]. Available:
  \url{https://doi.org/10.1093/bioinformatics/16.11.1056}
\BIBentrySTDinterwordspacing

\bibitem{eades1984heuristic}
P.~Eades, ``A heuristic for graph drawing,'' \emph{Congressus Numerantium},
  vol.~42, pp. 149--160, 1984.

\bibitem{delnphard}
\BIBentryALTinterwordspacing
P.~Eades and N.~C. Wormald, ``Fixed edge-length graph drawing is np-hard,''
  \emph{Discrete Applied Mathematics}, vol.~28, no.~2, pp. 111--134, 1990.
  [Online]. Available:
  \url{https://www.sciencedirect.com/science/article/pii/0166218X9090110X}
\BIBentrySTDinterwordspacing

\bibitem{efrat2010}
\BIBentryALTinterwordspacing
A.~Efrat, D.~Forrester, A.~Iyer, S.~G. Kobourov, C.~Erten, and O.~Kilic,
  ``Force-directed approaches to sensor localization,'' \emph{ACM Trans. Sen.
  Netw.}, vol.~7, no.~3, oct 2010. [Online]. Available:
  \url{https://doi.org/10.1145/1807048.1807057}
\BIBentrySTDinterwordspacing

\bibitem{ellson2001graphviz}
J.~Ellson, E.~Gansner, L.~Koutsofios, S.~North, G.~Woodhull, S.~Description,
  and L.~Technologies, ``Graphviz — open source graph drawing tools,'' in
  \emph{Lecture Notes in Computer Science}.\hskip 1em plus 0.5em minus
  0.4em\relax Springer, 2001, pp. 483--484.

\bibitem{fruchterman1991graph}
T.~M.~J. Fruchterman and E.~M. Reingold, ``Graph drawing by force-directed
  placement,'' \emph{Software: Practice and Experience}, vol.~21, no.~11, pp.
  1129--1164, 1991.

\bibitem{GGK04}
P.~Gajer, M.~Goodrich, and S.~Kobourov, ``A fast multi-dimensional algorithm
  for drawing large graphs,'' \emph{Computational Geometry: Theory and
  Applications}, vol.~29, no.~1, pp. 3--18, 2004.

\bibitem{gansner2010gmap}
E.~R. {Gansner}, Y.~{Hu}, and S.~{Kobourov}, ``{GMap}: Visualizing graphs and
  clusters as maps,'' in \emph{2010 IEEE Pacific Visualization Symposium
  (PacificVis)}, 2010, pp. 201--208.

\bibitem{gansner2009putting}
\BIBentryALTinterwordspacing
E.~Gansner, Y.~Hu, S.~Kobourov, and C.~Volinsky, ``Putting recommendations on
  the map: Visualizing clusters and relations,'' in \emph{Proceedings of the
  Third ACM Conference on Recommender Systems}, ser. RecSys '09.\hskip 1em plus
  0.5em minus 0.4em\relax New York, NY, USA: Association for Computing
  Machinery, 2009, p. 345–348. [Online]. Available:
  \url{https://doi.org/10.1145/1639714.1639784}
\BIBentrySTDinterwordspacing

\bibitem{gansner2009efficient}
E.~R. Gansner and Y.~Hu, ``Efficient node overlap removal using a proximity
  stress model,'' in \emph{Graph Drawing}, I.~G. Tollis and M.~Patrignani,
  Eds.\hskip 1em plus 0.5em minus 0.4em\relax Berlin, Heidelberg: Springer,
  2009, pp. 206--217.

\bibitem{gansner1998improved}
\BIBentryALTinterwordspacing
E.~R. Gansner and S.~C. North, ``Improved force-directed layouts,'' in
  \emph{Proceedings of the 6th International Symposium on Graph Drawing}, ser.
  Graph Drawing '98.\hskip 1em plus 0.5em minus 0.4em\relax Springer, 1998, pp.
  364--373. [Online]. Available:
  \url{http://dl.acm.org/citation.cfm?id=647550.729069}
\BIBentrySTDinterwordspacing

\bibitem{goyal2017accurate}
P.~Goyal, P.~Doll{\'a}r, R.~Girshick, P.~Noordhuis, L.~Wesolowski, A.~Kyrola,
  A.~Tulloch, Y.~Jia, and K.~He, ``Accurate, large minibatch sgd: Training
  imagenet in 1 hour,'' \emph{arXiv preprint arXiv:1706.02677}, 2017.

\bibitem{hh-msadg-99}
R.~Hadany and D.~Harel, ``A multi-scale algorithm for drawing graphs nicely,''
  \emph{Discrete Applied Mathematics}, vol. 113, no.~1, pp. 3--21, 2001.

\bibitem{hadlak2015survey}
S.~Hadlak, H.~Schumann, and H.-J. Schulz, ``A survey of multi-faceted graph
  visualization.'' in \emph{EuroVis (STARs)}, 2015, pp. 1--20.

\bibitem{hu2005efficient}
Y.~Hu, ``Efficient, high-quality force-directed graph drawing,''
  \emph{Mathematica Journal}, vol.~10, no.~1, pp. 37--71, 2005.

\bibitem{hug2016new}
L.~A. Hug, B.~J. Baker, K.~Anantharaman, C.~T. Brown, A.~J. Probst, C.~J.
  Castelle, C.~N. Butterfield, A.~W. Hernsdorf, Y.~Amano, K.~Ise \emph{et~al.},
  ``A new view of the tree of life,'' \emph{Nature microbiology}, vol.~1,
  no.~5, pp. 1--6, 2016.

\bibitem{kittivorawongfast}
\BIBentryALTinterwordspacing
C.~Kittivorawang, D.~Moritz, K.~Wongsuphasawat, and J.~Heer, ``Fast and
  flexible overlap detection for chart labeling with occupancy bitmap,'' in
  \emph{IEEE VIS Short Papers}, 2020. [Online]. Available:
  \url{http://idl.cs.washington.edu/papers/fast-labels}
\BIBentrySTDinterwordspacing

\bibitem{kobak2019art}
D.~Kobak and P.~Berens, ``The art of using t-sne for single-cell
  transcriptomics,'' \emph{Nature communications}, vol.~10, no.~1, pp. 1--14,
  2019.

\bibitem{kobourov2014visualizing}
S.~Kobourov, S.~Pupyrev, and P.~Simonetto, ``{Visualizing Graphs as Maps with
  Contiguous Regions},'' in \emph{EuroVis - Short Papers}, N.~Elmqvist,
  M.~Hlawitschka, and J.~Kennedy, Eds.\hskip 1em plus 0.5em minus 0.4em\relax
  The Eurographics Association, 2014.

\bibitem{koren2002ace}
Y.~Koren, L.~Carmel, and D.~Harel, ``Ace: A fast multiscale eigenvectors
  computation for drawing huge graphs,'' in \emph{Information Visualization,
  2002. INFOVIS 2002. IEEE Symposium on}.\hskip 1em plus 0.5em minus
  0.4em\relax IEEE, 2002, pp. 137--144.

\bibitem{leow19GraphTSNE}
Y.~Y. Leow, T.~Laurent, and X.~Bresson, ``Graphtsne: A visualization technique
  for graph-structured data,'' in \emph{ICLR Workshop on Representation
  Learning on Graphs and Manifolds}, 2019.

\bibitem{itol}
\BIBentryALTinterwordspacing
I.~Letunic and P.~Bork, ``{Interactive Tree Of Life (iTOL) v5: an online tool
  for phylogenetic tree display and annotation},'' \emph{Nucleic Acids
  Research}, vol.~49, no.~W1, pp. W293--W296, 04 2021. [Online]. Available:
  \url{https://doi.org/10.1093/nar/gkab301}
\BIBentrySTDinterwordspacing

\bibitem{luboschik2008particle}
M.~Luboschik, H.~Schumann, and H.~Cords, ``Particle-based labeling: Fast
  point-feature labeling without obscuring other visual features,'' \emph{IEEE
  transactions on visualization and computer graphics}, vol.~14, no.~6, pp.
  1237--1244, 2008.

\bibitem{marriott2003removing}
K.~Marriott, P.~Stuckey, V.~Tam, and W.~He, ``Removing node overlapping in
  graph layout using constrained optimization,'' \emph{Constraints}, vol.~8,
  no.~2, pp. 143--171, 2003.

\bibitem{compact_measures}
M.~McGuffin and J.~Robert, ``Quantifying the space-efficiency of 2d graphical
  representations of trees,'' \emph{Information Visualization}, vol.~9, pp.
  115--140, 2010.

\bibitem{mote2007fast}
K.~Mote, ``Fast point-feature label placement for dynamic visualizations,''
  \emph{Information Visualization}, vol.~6, no.~4, pp. 249--260, 2007.

\bibitem{nachmanson2017node}
L.~Nachmanson, A.~Nocaj, S.~Bereg, L.~Zhang, and A.~Holroyd, ``Node overlap
  removal by growing a tree,'' pp. 33--43, 2016.

\bibitem{INKA}
\BIBentryALTinterwordspacing
Q.~H. Nguyen, ``{INKA:} an ink-based model of graph visualization,''
  \emph{CoRR}, vol. abs/1801.07008, 2018. [Online]. Available:
  \url{http://arxiv.org/abs/1801.07008}
\BIBentrySTDinterwordspacing

\bibitem{space_optimized}
Q.~V. Nguyen and M.~L. Huang, ``A space-optimized tree visualization,'' in
  \emph{IEEE Symposium on Information Visualization, 2002. INFOVIS 2002.},
  2002, pp. 85--92.

\bibitem{nobre2019state}
C.~Nobre, M.~Meyer, M.~Streit, and A.~Lex, ``The state of the art in
  visualizing multivariate networks,'' in \emph{Computer Graphics Forum},
  vol.~38, no.~3, 2019, pp. 807--832.

\bibitem{compactness}
\BIBentryALTinterwordspacing
H.~C. Purchase, D.~Carrington, and J.-A. Allder, ``Empirical evaluation of
  aesthetics-based graph layout,'' \emph{Empirical Software Engineering},
  vol.~7, no.~3, pp. 233--255, 2002. [Online]. Available:
  \url{https://doi.org/10.1023/A:1016344215610}
\BIBentrySTDinterwordspacing

\bibitem{rahman2020force2vec}
K.~Rahman, M.~Sujon, and A.~Azad, ``{Force2Vec: P}arallel force-directed graph
  embedding,'' in \emph{Intl.~Conference on Data Mining (ICDM)}.\hskip 1em plus
  0.5em minus 0.4em\relax IEEE, 2020, pp. 442--451.

\bibitem{rahman2020batchlayout}
M.~K. Rahman, M.~H. Sujon, and A.~Azad, ``{BatchLayout: A} batch-parallel
  force-directed graph layout algorithm in shared memory,'' in \emph{2020 IEEE
  Pacific Visualization Symposium (PacificVis)}.\hskip 1em plus 0.5em minus
  0.4em\relax IEEE, 2020, pp. 16--25.

\bibitem{Reingold1981}
E.~M. Reingold and J.~S. Tilford, ``Tidier drawings of trees,'' \emph{IEEE
  Transactions on Software Engineering}, vol. SE-7, no.~2, pp. 223--228, 1981.

\bibitem{shulz2011treevis}
H.~{Schulz}, ``Treevis.net: A tree visualization reference,'' \emph{IEEE
  Computer Graphics and Applications}, vol.~31, no.~6, pp. 11--15, Nov 2011.

\bibitem{overlap}
D.~Sun and K.~Wong, ``On evaluating the layout of {UML} class diagrams for
  program comprehension,'' in \emph{13th International Workshop on Program
  Comprehension (IWPC'05)}, 2005, pp. 317--326.

\bibitem{openlayers}
\BIBentryALTinterwordspacing
T.~O.~D. Team, \emph{OpenLayers}, 2020 (accessed December 22, 2020). [Online].
  Available: \url{https://openlayers.org/}
\BIBentrySTDinterwordspacing

\bibitem{van2008visualizing}
L.~Van~der Maaten and G.~Hinton, ``Visualizing data using t-sne.''
  \emph{Journal of machine learning research}, vol.~9, no.~11, 2008.

\bibitem{wiese2004yfiles}
R.~Wiese, M.~Eiglsperger, and M.~Kaufmann, ``yfiles visualization and automatic
  layout of graphs,'' in \emph{Graph Drawing Software}.\hskip 1em plus 0.5em
  minus 0.4em\relax Springer, 2004, pp. 173--191.

\bibitem{zager2008graph}
L.~A. Zager and G.~C. Verghese, ``Graph similarity scoring and matching,''
  \emph{Applied mathematics letters}, vol.~21, no.~1, pp. 86--94, 2008.

\end{thebibliography}

\textbf{Ryn Gray} is a current PhD student at the University of Arizona. They received their BS degree in Computer Science and Applied Mathematics from the University of Colorado at Boulder. Her research interests include network, data, and systems visualizations. Contact her at ryngray@arizona.edu\par

\textbf{Mingwei Li} is a postdoctoral researcher at the Department of Computer Science at Vanderbilt University. He received his PhD in Computer Science at the University of Arizona.
He received a BE degree in Electronics Engineering from the Hong Kong University of Science and Technology. 
His research interests include data visualization and machine learning. Contact him at mingwei.li@vanderbilt.edu\par

\textbf{Reyan Ahmed} is a visiting assistant professor at Colgate University, Hamilton, NY, USA. He received his PhD in Computer Science at the University of Arizona. He received BS and MS degrees in Computer Science and Engineering from Bangladesh University of Engineering and Technology. His research interests include graph algorithms, network visualization, and data science. Contact him at rahmed1@colgate.edu\par

\textbf{Md. Khaledur Rahman} is a research scientist at Meta Platforms Inc., Menlo Park, CA, USA. He received his PhD in Computer Science from Indiana University Bloomington. He received BS and MS degrees in Computer Science and Engineering from Bangladesh University of Engineering and Technology. His research interests include graph learning, recommendation systems, and high-performance computing. Contact him at khaledrahman@meta.com\par

\textbf{Ariful Azad} is an assistant professor of Intelligent Systems Engineering in the School of Informatics, Computing, and Engineering at Indiana University Bloomington. He obtained his Ph.D. from Purdue University and B.S. from Bangladesh University of Engineering and Technology.
His research interests include graph algorithms, high performance computing, and bioinformatics. Contact him at azad@iu.edu\par

\textbf{Stephen Kobourov} is a professor at the Department of Computer Science at the University of Arizona. He received a BS degree in Mathematics and Computer Science from Dartmouth College and MS and PhD
degrees from Johns Hopkins University. His research interests include information visualization, graph theory, and geometric algorithms. Contact him at kobourov@cs.arizona.edu\par

\textbf{Katy B\"{o}rner} is the Victor H. Yngve Professor of Information Science in the School of Informatics and Computing at Indiana University in Bloomington. She received an MS in Electrical Engineering from the University of Technology in Leipzig and a Ph.D. in Computer Science from the University of Kaiserslautern. Her research interests include network science, information visualization, and scientometrics. Contact her at katy@indiana.edu\par

\section{Supplementary Material}

In this section, we provide additional example layouts from the different datasets, and  that show a zoomed-in view of the images.
We also discuss topology preservation using an example.

\subsection{Evaluation}

\begin{figure}[t]
 	\centering
 	\subfloat[\yedCircular]{\label{fig:t8direct}\includegraphics[width=.6\columnwidth]{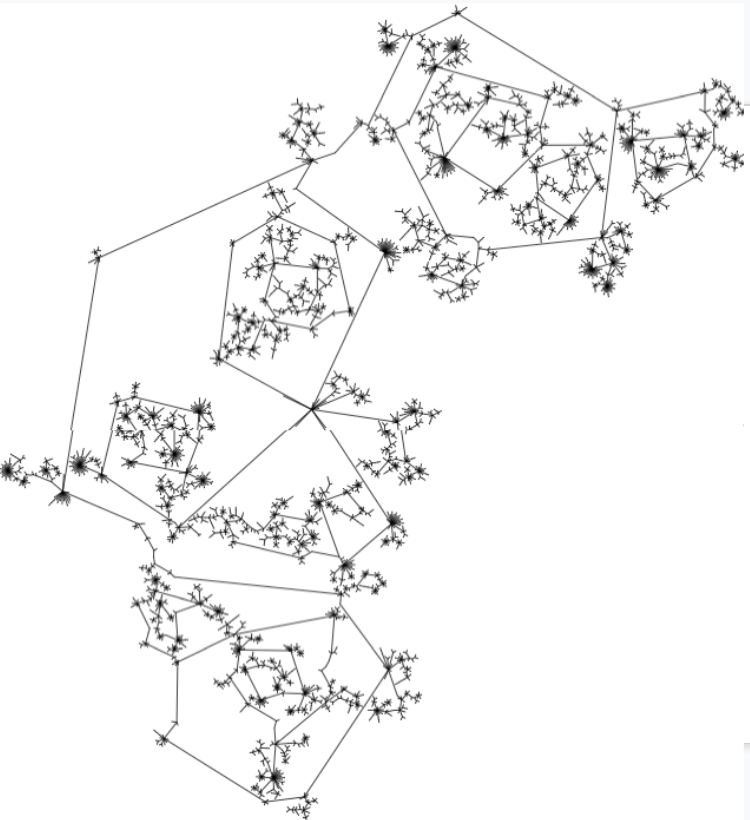}}\\
 	\subfloat[\sfdpWithPrism]{\label{fig:t8prism}\includegraphics[width=.8\columnwidth]{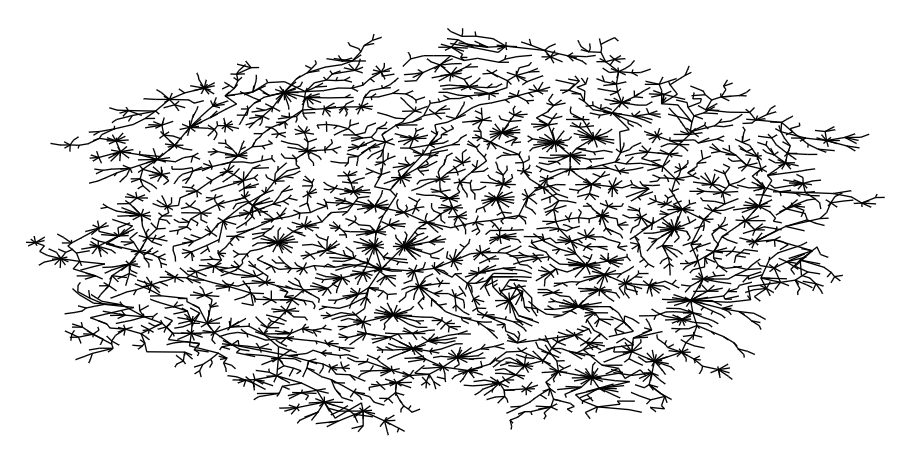}}\\
 	\subfloat[\reyansAlgoAcronym]{\label{fig:t8_angular}\includegraphics[width=.8\columnwidth]{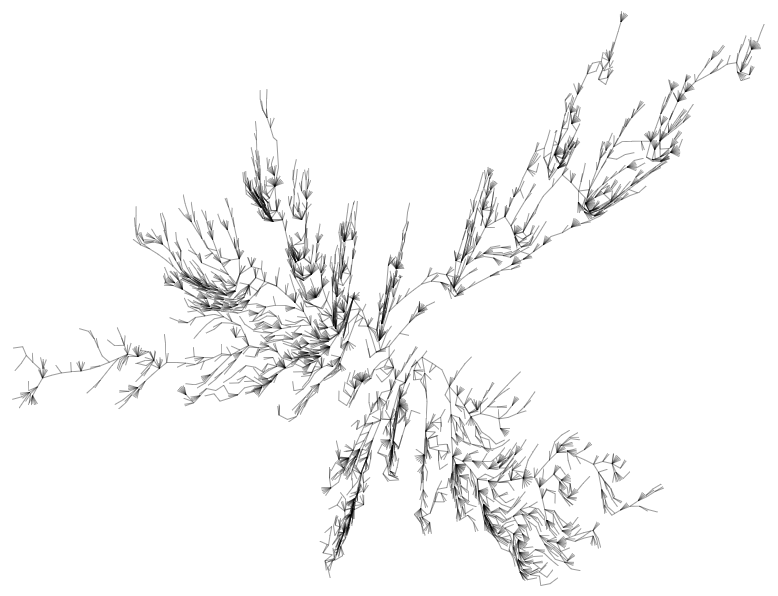}}\\
 	\subfloat[\mingweisAlgoAcronym]{\label{fig:t8mw}\includegraphics[width=.8\columnwidth]{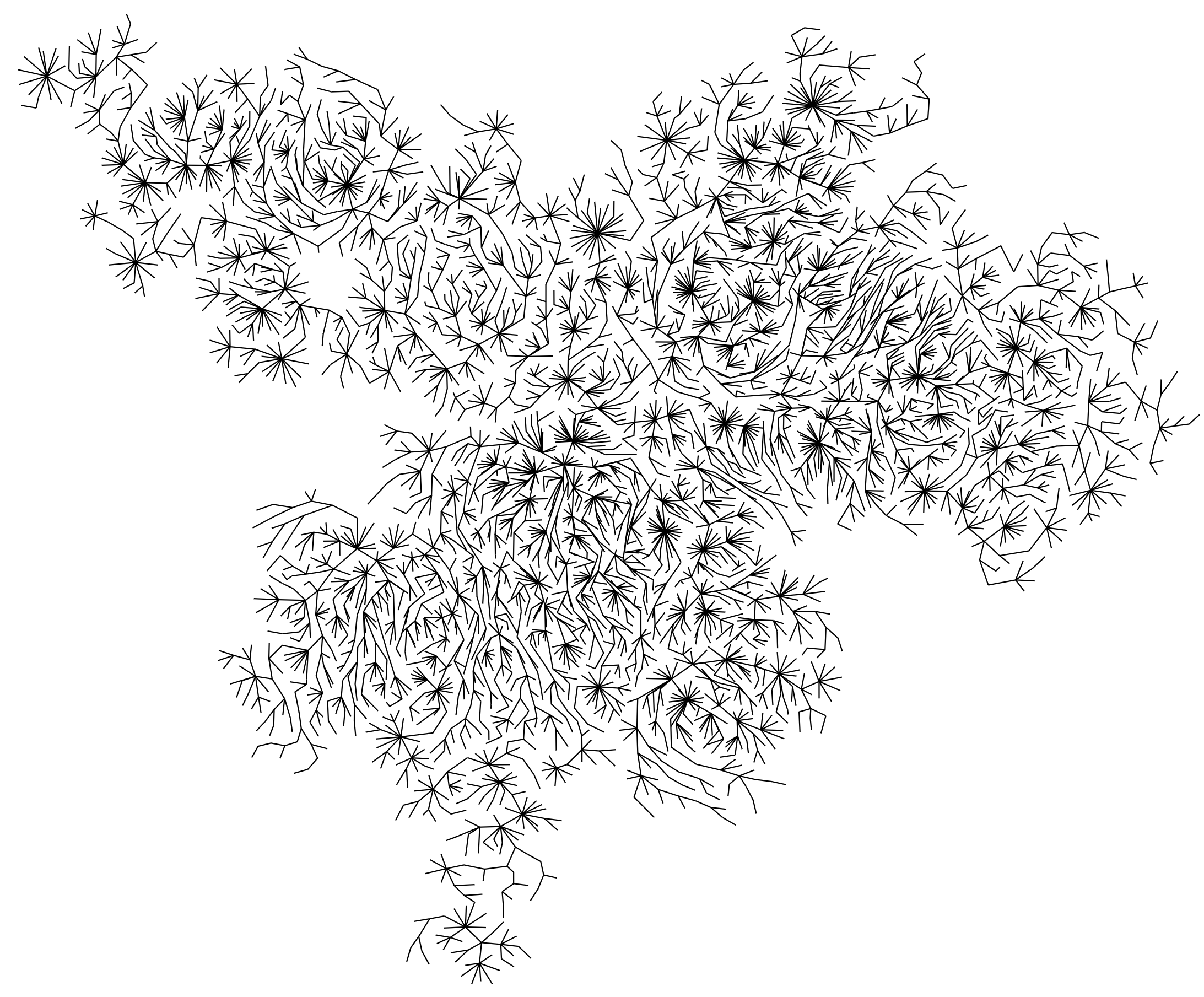}}\\
	
 	\caption{Comparison of the tree layout structure of the uniform Google Topics graph drawn with \yedCircular, \sfdpWithPrism, \reyansAlgoAcronym, and \mingweisAlgoAcronym}
 	\label{fig:compare-topics}
 \end{figure}

\begin{figure*}[thp]
\centering
\subfloat[\reyansAlgoAcronym]{
	\includegraphics[width=0.72\textwidth]{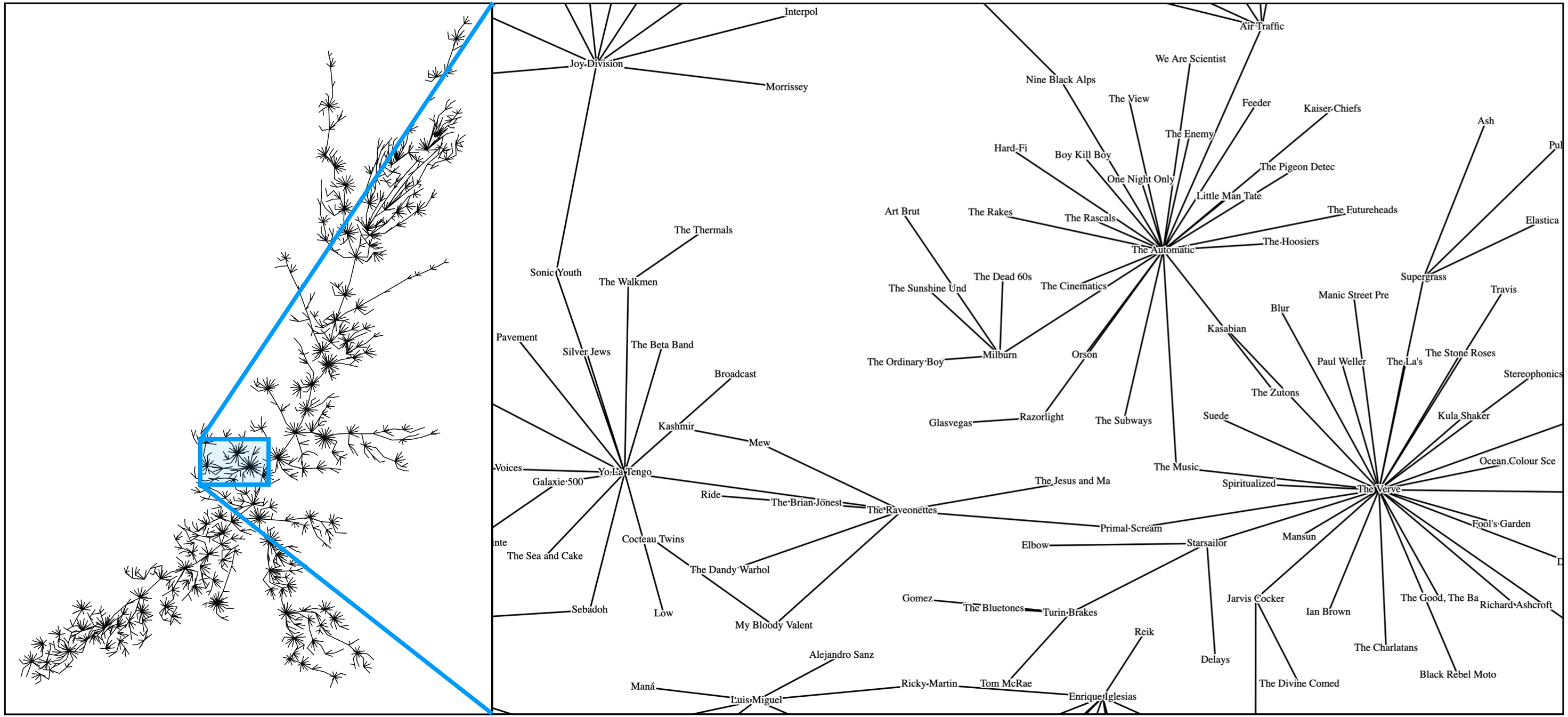}
}\\
\subfloat[\mingweisAlgoAcronym]{
	\includegraphics[width=0.72\textwidth]{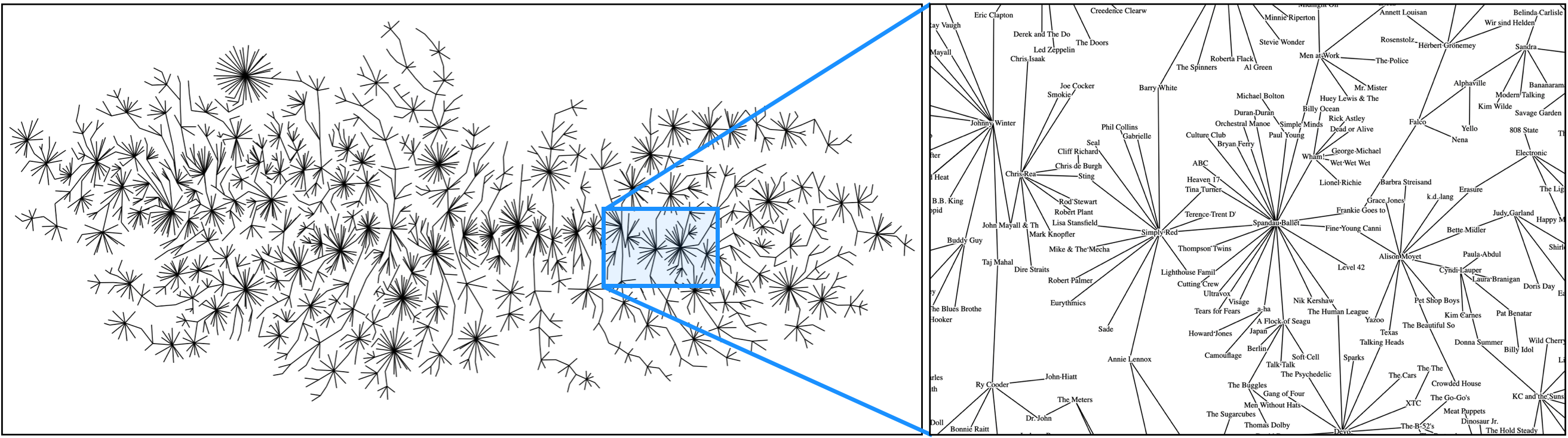}
}\\
\caption{Comparing Last.FM linear layouts drawn with our algorithms.}
\label{fig:lastfm_augmented}
\end{figure*}

\begin{figure*}[thp]
\centering
\subfloat[\reyansAlgoAcronym]{
	\includegraphics[width=0.72\textwidth]{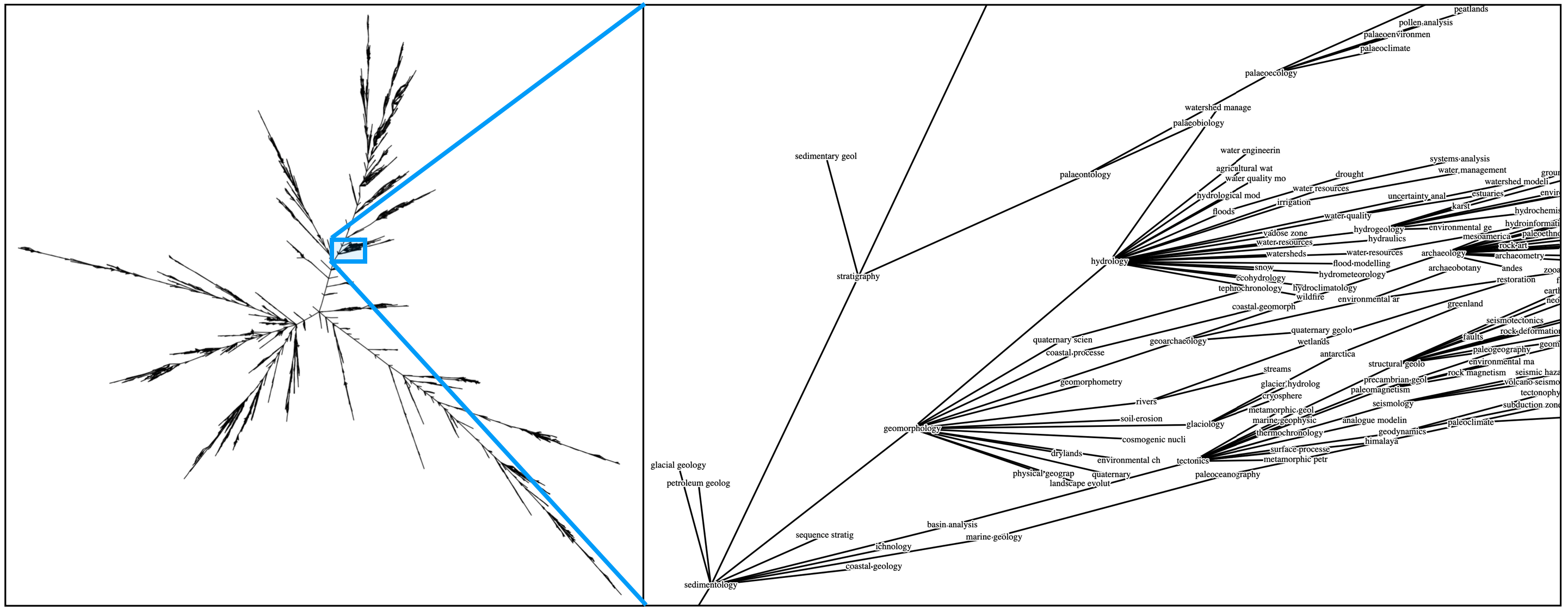}
}\\
\subfloat[\mingweisAlgoAcronym]{
	\includegraphics[width=0.72\textwidth]{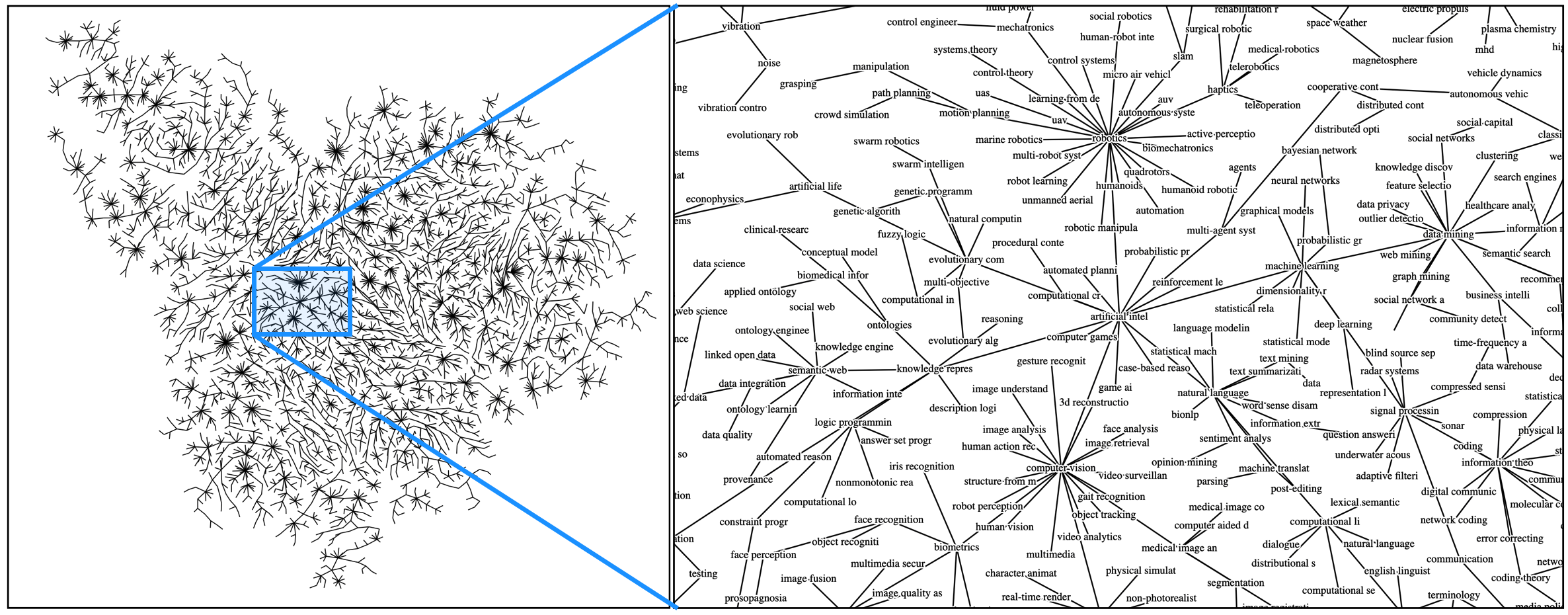}
}\\


\caption{Comparing Topics linear layouts drawn with our algorithms.}
\label{fig:topics_augmented}
\end{figure*}

\begin{figure*}[thp]
\centering
\subfloat[\reyansAlgoAcronym]{
	\includegraphics[width=0.65\textwidth]{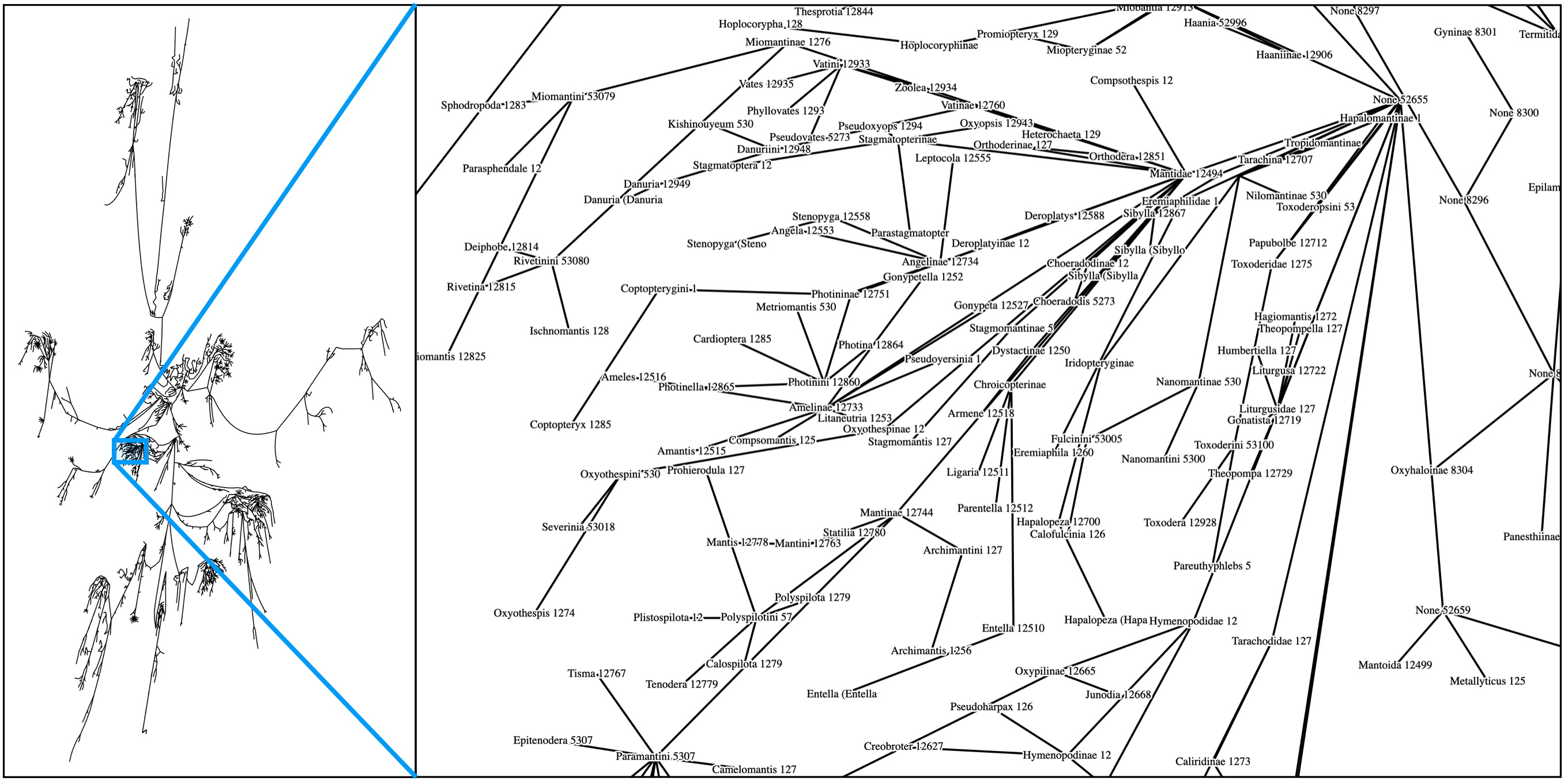}
}\\
\subfloat[\mingweisAlgoAcronym]{
	\includegraphics[width=0.65\textwidth]{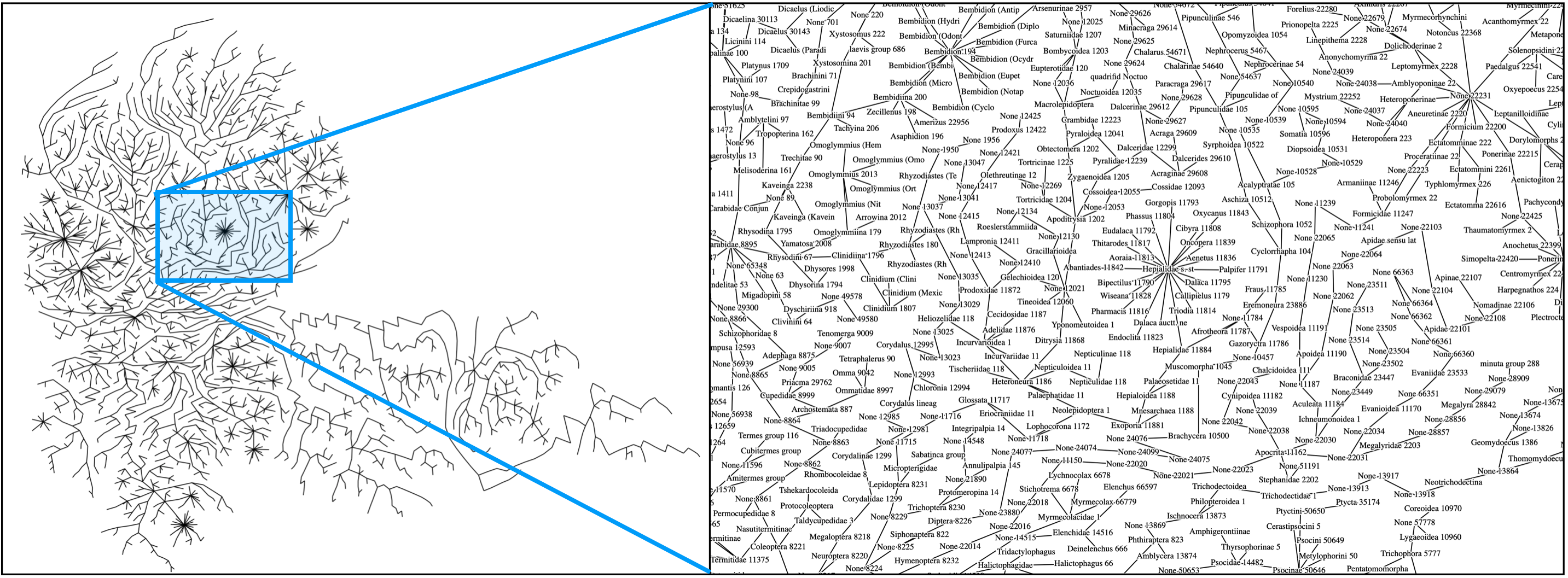}
}\\
\caption{Comparing Tree of Life linear layouts drawn with our algorithms.}
\label{fig:tol_augmented}
\end{figure*}

We now provide some images that show the outputs from different algorithms on different graphs.
In Fig.~\ref{fig:compare-topics}, we give layouts of uniform Topics Graph computed by two existing algorithms (CIR and sfdp+p) and ours (RT\_L and RT\_C). 
In Fig.~\ref{fig:lastfm_augmented}, we compare the layout of the Last.FM graph generated using RT\_L and RL\_C. 
Note that these overviews do not show details such as label overlaps, crossings, or compactness, but they do provide a feel for how the graphs are laid out.
For instance, it is easy to see that CIR does not preserve edge lengths while SFDP+P appears to do well. However, when zooming in we see many crossings and label overlaps in SFDP+P. 
The \reyansAlgoAcronym is able to preserve desired edge lengths more which helps to capture the overall topology of the layout. 
However, if we zoom in then we can see that the drawing is not compact: there are some free spaces among the labels. 
On the other hand, \mingweisAlgoAcronym focuses more on compactness and by zooming in the layout shows that relatively more labels are drawn compactly.
Similarly, \reyansAlgoAcronym preserve the desired edge lengths while \mingweisAlgoAcronym optimizes compactness for topics and tree of life graphs, see Fig.~\ref{fig:topics_augmented} and Fig.~\ref{fig:tol_augmented} respectively. 

\subsection{A Note on Topology Preservation}

\begin{figure}[thp]
	\centering
	\subfloat[]{	\includegraphics[width=.25\columnwidth,height=.37\columnwidth]{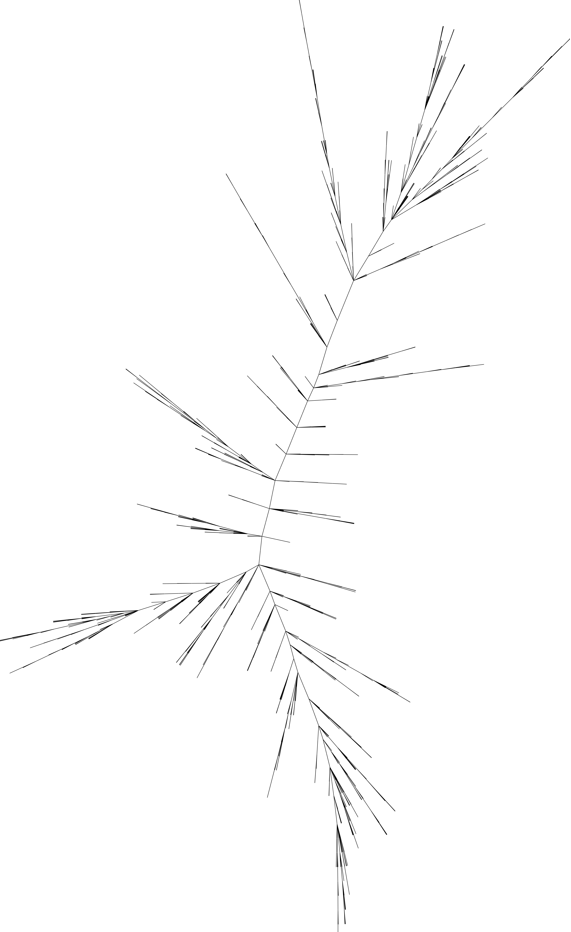}}\hspace{1.2cm}
	\subfloat[]{\includegraphics[width=.48\columnwidth]{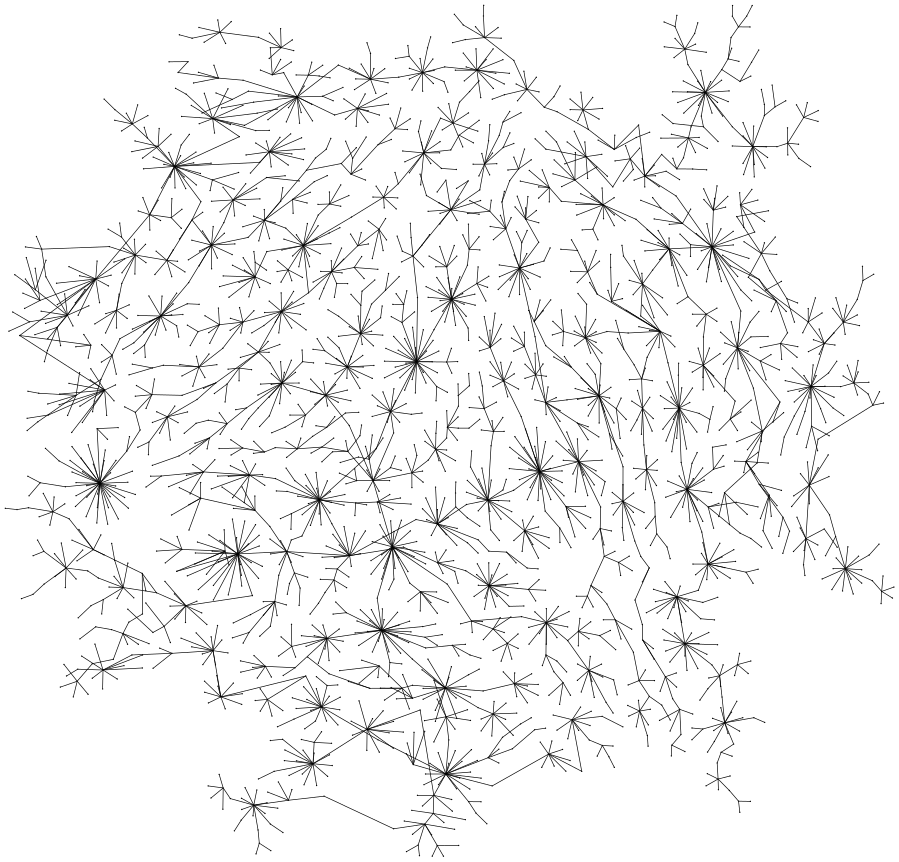}}\\
	\subfloat[]{\includegraphics[width=.44\columnwidth]{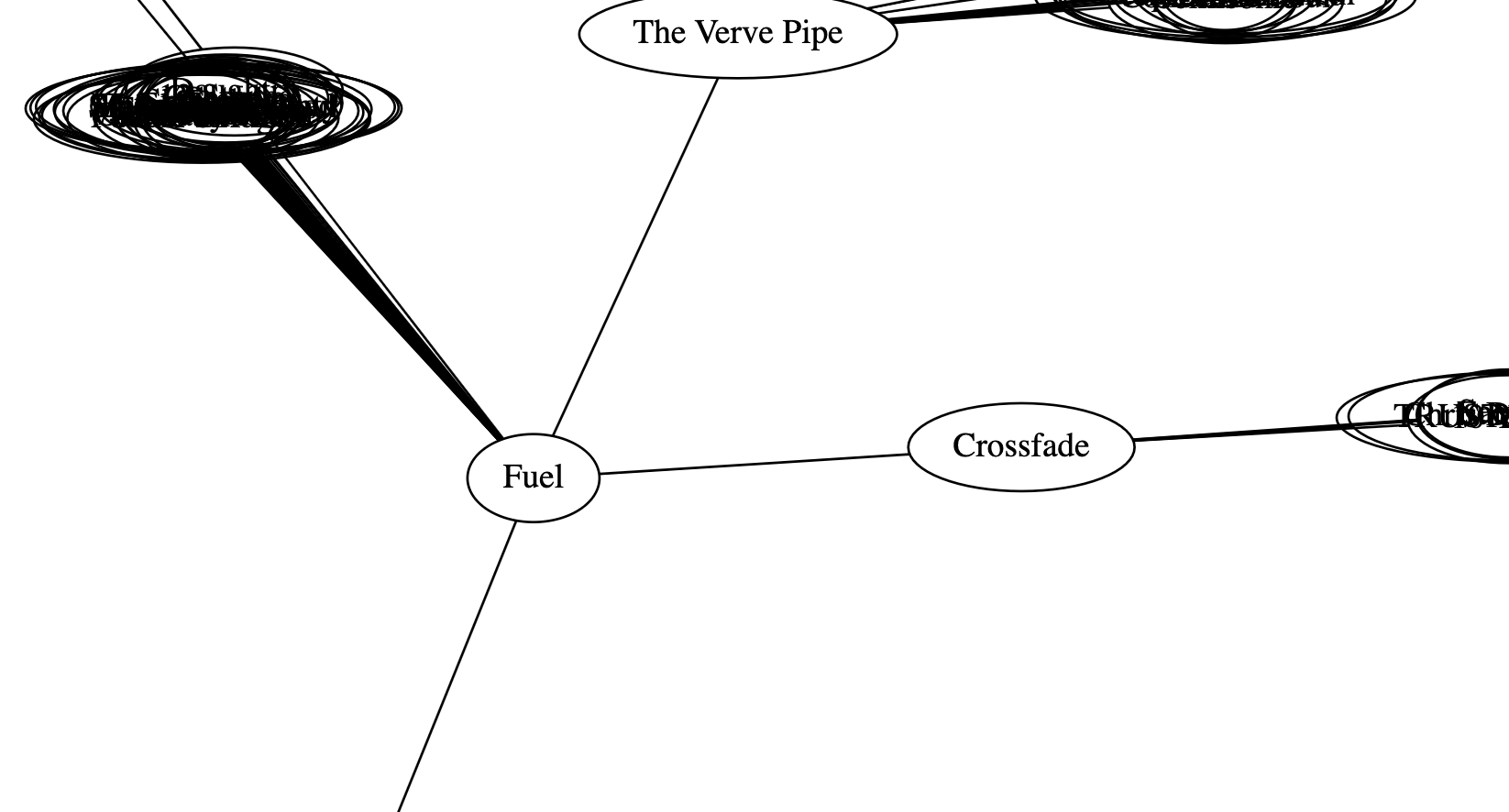}}
	\subfloat[]{\includegraphics[width=.52\columnwidth]{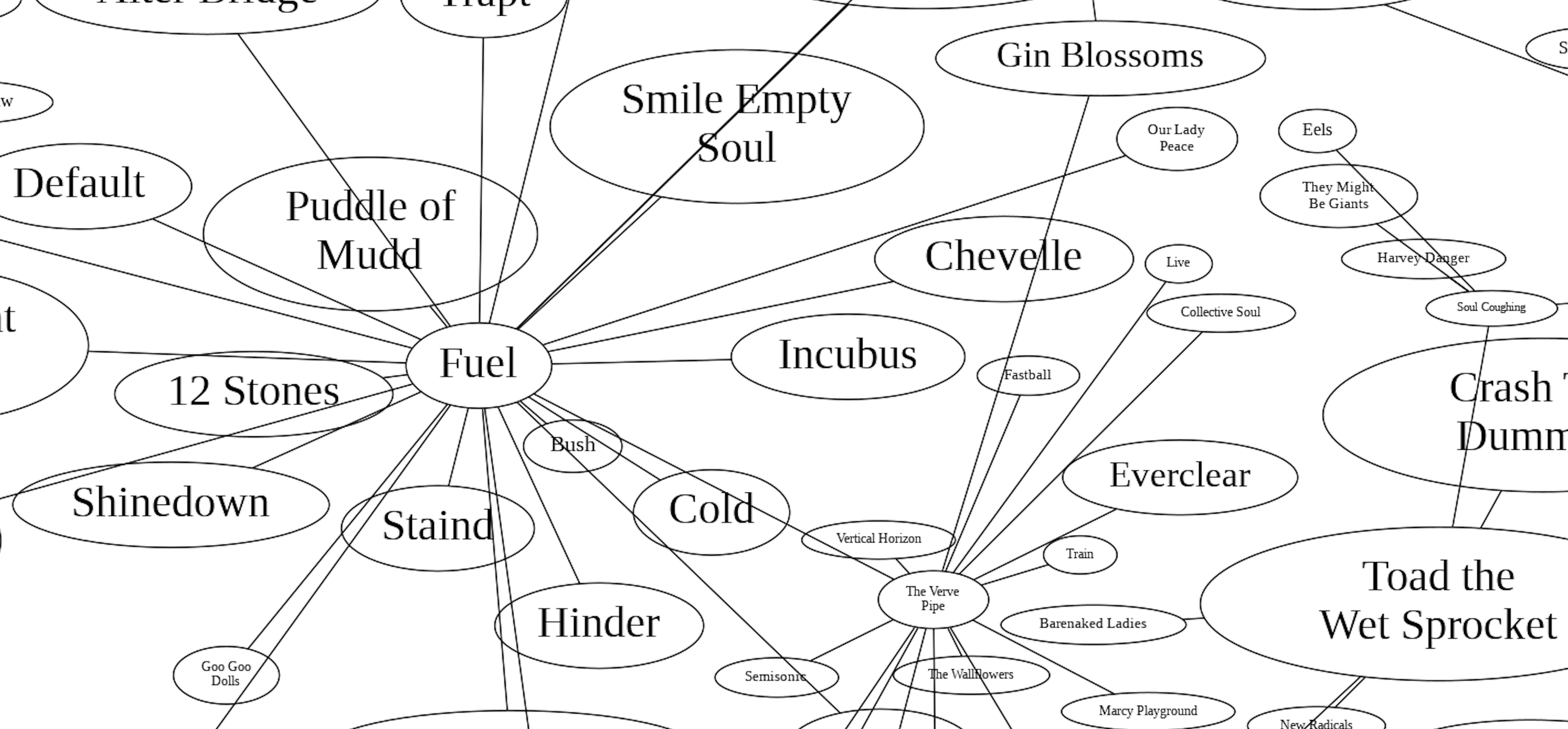}}
	\caption{\blue{
A crossings-free initial layout of the uniform last.FM network (a) and the corresponding output generated by sfdp (b). A zoomed in view in the two layouts in (a-b) shows that the many overlaps in the initial layout (c) were corrected at the expense of edge crossings (d).}}
	\label{fig:my_label}
\end{figure}

We discuss topology preservation in Sect~\ref{sec:topo_pres}, but here we illustrate the concept with an example. \blue{In Fig.~\ref{fig:my_label} we illustrate how prior methods that remove overlaps may do so at the expense of changing the topology of the underlying graph. Even then initialized with a crossings-free layout, \sfdpWithPrism removes the overlaps by introducing crossings and reordering node neighbors, thus changing the topology of the input layout. }

\end{document}